\documentclass[12pt]{article}
\usepackage{amsgen,amsmath,amstext,amsbsy,amsopn,amssymb,bbm, amsthm}
\usepackage{comment}
 \usepackage[pdftex]{hyperref}
\usepackage[utf8]{inputenc}
\usepackage[T1]{fontenc}	 
\usepackage{graphicx}
\usepackage{caption}
\usepackage{subcaption}
\usepackage{natbib}
\usepackage{fullpage}
\usepackage{authblk}
\usepackage{float}
\usepackage{xcolor}
\usepackage{setspace}
\usepackage{parskip}
\usepackage{mathtools, nccmath} 

\newcommand{\blind}{1} 

\DeclareMathOperator*{\argmax}{arg\,max}
\DeclareMathOperator*{\argmin}{arg\,min}

\def\N{\mathcal{N}}

\def\E{\mathbb{E}}
\def\by{\mathbf{y}}

\def\balpha{\boldsymbol{\alpha}}
\def\bbeta{\boldsymbol{\beta}}
\def\alphabar{\overline{\alpha}}
\def\balphabar{\overline{\balpha}}
\def\betabar{\overline{\beta}}

\def\btheta{\boldsymbol{\theta}}

\def\bz{\mathbf{z}}

\def\by{\mathbf{y}}
\def\bt{\mathbf{t}}
\def\bone{\mathbf{1}}
\def\bz{\mathbf{z}}
\def\bgamma{\boldsymbol{\gamma}}

\def\bw{\mathbf{w}}

\newcommand{\I}{\mathbf{I}}

\newcommand{\y}{\mathbf{y}}

\newcommand{\Balpha}{\boldsymbol{\alpha}}
\newcommand{\Btheta}{\boldsymbol{\theta}}
\newcommand{\Bbeta}{\boldsymbol{\beta}}

\newcommand{\Bgamma}{\boldsymbol{\gamma}}
\newcommand{\BGamma}{\boldsymbol{\Gamma}}

\newcommand{\1}{\mathbf{1}}
\newcommand{\0}{\mathbf{0}}
\DeclarePairedDelimiter{\nint}\lfloor\rceil
\newtheorem{theorem}{Theorem}

\newtheorem{proposition}[theorem]{Proposition}

\definecolor{PennRed}{RGB}{152, 30 50}
\definecolor{PennBlue}{RGB}{0, 44, 119}
\definecolor{PennGreen}{RGB}{94, 179,70}
\definecolor{PennViolet}{RGB}{141, 76, 145}
\definecolor{PennSkyBlue}{RGB}{14, 118, 188}
\definecolor{PennOrange}{RGB}{243, 117, 58}
\definecolor{PennBrightRed}{RGB}{223,82, 78}

\hypersetup{pdfborder = {0 0 0.5 [3 3]}, colorlinks = true, linkcolor = PennBrightRed, citecolor = PennSkyBlue}

\setcitestyle{semicolon}

\date{}

\begin{document}

\title{\bf Crime in Philadelphia: Bayesian Clustering with Particle Optimization}
  \author[$\star$]{Cecilia Balocchi\thanks{The authors gratefully acknowledge funding from the European Research Council (ERC), under the European Union Horizon 2020 research and innovation programme, Grant agreement No. 817257}}
\author[$\mathsection$]{Sameer K. Deshpande}
\author[$\ddagger$]{Edward I. George\thanks{The authors gratefully acknowledge funding from NSF grant DMS-1916245.}}
\author[$\ddagger$]{Shane T. Jensen}
\affil[$\star$]{{\small School of Mathematics, University of Edinburgh}}
\affil[$\mathsection$]{{\small Department of Statistics, University of Wisconsin--Madison}}
\affil[$\ddagger$]{{\small Department of Statistics, University of Pennsylvania}}
\maketitle
\vspace{-1.2cm}

\bigskip

\begin{abstract} 
Accurate estimation of the change in crime over time is a critical first step towards better understanding of public safety in large urban environments.
Bayesian hierarchical modeling is a natural way to study spatial variation in urban crime dynamics at the neighborhood level, since it facilitates principled ``sharing of information'' between spatially adjacent neighborhoods.
Typically, however, cities contain many physical and social boundaries that may manifest as spatial discontinuities in crime patterns. 
In this situation, standard prior choices often yield overly-smooth parameter estimates, which can ultimately produce mis-calibrated forecasts. 
To prevent potential over-smoothing, we introduce a prior that partitions the set of neighborhoods into several clusters and encourages spatial smoothness within each cluster.
In terms of model implementation, conventional stochastic search techniques are computationally prohibitive, as they must traverse a combinatorially vast space of partitions. 
We introduce an ensemble optimization procedure that simultaneously identifies several high probability partitions by solving one optimization problem using a new local search strategy. 
We then use the identified partitions to estimate crime trends in Philadelphia between 2006 and 2017. 
On simulated and real data, our proposed method demonstrates good estimation and partition selection performance. 
\end{abstract}


\onehalfspacing

\newpage
\section{Introduction}
\label{sec:introduction}
Beginning in the late 2000's and continuing through the 2010's, Philadelphia experienced population growth, underwent a rapid evolution in its built environment, and observed a generally decreasing trend in the total number of violent crimes reported in the city.
Although the \emph{total} number of violent crimes decreased city-wide, the decreasing trend was not experienced uniformly in all of the local neighborhoods making up the city.
In this paper, we carefully estimate both the neighborhood-level baseline levels and temporal trends in crime and separately cluster these estimates.
This enables us to identify clusters of neighborhoods whose trends in crime deviated markedly from the overall city-wide decrease as well as clusters of neighborhoods displaying substantially higher or lower baseline levels than surrounding neighborhoods. 

Accurate modeling of the neighborhood-level crime patterns benefits many stakeholders: urban planners can better understand how socioeconomic factors and the built environment affect crime, city officials can develop community programs and interventions to improve the overall quality of life for all residents, and law enforcement can more thoughtfully deploy resources to increase public safety. 
Accurate modeling of neighborhood trends is, however, complicated by several factors.
First, simple comparisons of the temporal trajectories of neighborhood-level counts fail to account for important differences in size, land use \& zoning, population, and the socio-demographic makeup of each neighborhood.
Consequently, in order to facilitate accurate comparisons across neighborhoods, we directly model the \emph{crime density} -- defined as the number of violent crimes divided by the land area (in square miles) -- of each neighborhood over time.
As we detail in Section~\ref{sec:data_model}, area is a more principled normalization than population for small urban areas. 

The second complication lies in the spatial modeling of crime density.
Bayesian hierarchical modeling is an attractive way to study crime at the neighborhood level as it allows us to ``borrow strength'' between spatially adjacent neighborhoods. 
Recently, \citet{Balocchi2019} demonstrated that Bayesian models that encourage spatial smoothing between neighborhoods with conditionally auto-regressive \citep[CAR;][]{besag1974spatial} priors yield more accurate neighborhood-level forecasts than fitting independent models to each neighborhood.
Though CAR models are an intuitive and popular way to introduce spatial dependence, they can produce overly smooth parameter estimates and forecasts that are at odds with the realities of complex urban environments.
In fact, as we will see in Section~\ref{sec:data_model}, while crime in Philadelphia displays considerable spatial correlation, there are also sharp discontinuities. 
While some discontinuities coincide with known landmarks like major streets, parks, and bodies of water, several occur along less obvious boundaries that may result from differences in socioeconomic conditions and disparate effects of public policy initiatives like community crime watches.

Specifying a CAR prior without accounting for potential discontinuities can lead to poor estimation of crime around these geographic or socioeconomic barriers. 
Directly modeling the locations of individual discontinuities between pairs of neighborhoods typically relies on heavily over-parametrized models; in fact, many such models introduce one latent parameter for each pair of adjacent neighborhoods.
We instead aim to identify \textit{clusters} of neighborhoods with similar crime dynamics. 

In this paper, we propose a ``CAR--within--clusters'' model where we introduce \emph{two latent spatial partitions} of neighborhoods in Philadelphia, one for the baseline levels of crime and one for the time trends.
We then specify separate CAR priors on the neighborhood-specific parameters within each cluster of each partition. 
Unlike many classical approaches to Bayesian spatial clustering, our proposed model allows (i) the spatial distribution of the baseline levels of crime to differ from that of the time trends and (ii) the parameters to vary both within and between clusters. 

Because we allow different model parameters to cluster differently, the combinatorial vastness of the underlying product space of partitions renders stochastic search techniques like Markov Chain Monte Carlo (MCMC) computationally prohibitive. 
We instead focus on posterior optimization.
However, rather than simply finding the \textit{maximum a posteriori} (MAP) partitions, we propose an extension of \citet{Rockova2018}'s ensemble optimization framework that simultaneously identifies multiple partitions with high posterior probability by solving a \emph{single} optimization problem.
Solving this problem is formally equivalent to finding a variational approximation of the discrete posterior distribution over the partitions.
We introduce a new local search strategy that, at a high level, runs several greedy searches that are made ``mutually aware'' through an entropy penalty.
This penalty promotes diversity among the estimated partitions by discouraging different search paths from visiting the same point in the latent discrete space. 

Here is an outline for the rest of the paper.
In Section~\ref{sec:data_model} we describe our crime data and introduce our ``CAR--within--clusters'' model. 
Then, in Section~\ref{sec:lit_review}, we highlight important similarities and differences between our proposed model and existing Bayesian spatial clustering methods.
We introduce our variational approximation and local optimization strategy in Section~\ref{sec:variational_approximation} before demonstrating its use on synthetic data in Section~\ref{sec:illustration}.
We then apply our method to the Philadelphia crime data in Section~\ref{sec:philly_example}.
Section~\ref{sec:discussion} concludes with a discussion of our results and an outline of potential future directions. 
\if1\blind
{An \texttt{R} package implementing our method and all code and data to replicate the results in this paper are available at \url{github.com/cecilia-balocchi/particle-optimization}.
  } \fi
\if0\blind
{An \texttt{R} package implementing our method and all code and data to replicate the results in this paper are available at \url{https://anonymous.4open.science/r/particle-optimization-663E} (link and repository have been anonymized for double-blindness).
} \fi

\section{Data and the ``CAR--within--clusters'' Model}
\label{sec:data_model}
Our crime data comes from \url{opendataphilly.org}, where the Philadelphia Police Department publicly releases the location, time, and type of each reported crime in the city. 
We focus on \emph{violent} crimes, which include  homicides, rapes, robberies, and aggravated assaults \citep{UCR}, aggregated at the census tract level.
Philadelphia is divided into $N = 384$ census tracts, which range in size from 0.04 to 6.65 square miles and whose populations range from 0 to 8300, with mean of 4000, according to the 2010 Census.

When comparing crime incidence across heterogeneous regions, it is very common for criminologists to work with per-capita crime rates (e.g. crime rates per 100,000 inhabitants). 
However, for small geographic regions within urban environments, \citet{zhang2007spatial} point out that frequently neither criminals nor victims are from the same neighborhood as the crime location.
They instead recommend normalizing crime counts by the area of the neighborhood rather than the population.

To this end, let $c_{i,t}$ be the total number of violent crimes reported in neighborhood $i$ in year $t,$ with $t = 0$ corresponding to the year 2006 and $t = 11$ corresponding to the year 2017.
Additionally, let $A_{i}$ be the area of neighborhood $i$ in square miles.
Since the distribution of \textit{crime density} $c_{i,t}/A_{i}$ is quite skewed, we transform the densities using the inverse hyperbolic sine transformation: $y_{i,t} = \log\left(c_{i,t}/A_{i} + (1 + (c_{i,t}/A_{i})^{2}))^{1/2}\right) -\log (2).$
This transformation behaves quite similarly to the logarithmic transform but is well-defined at zero \citep{burbidge1988alternative}.

At a high-level, we could model, independently for each census tract $i = 1, \ldots, N$ and $t = 0, \ldots, 11,$ $y_{i,t} \sim \mathcal{N}(f_{i,0}(x_{t}), \sigma^{2}),$
where $x_{t}$ is the time index standardized to have mean zero and unit variance and the unknown regression function $f_{i,0}$ measures the expected transformed crime density in neighborhood $i.$
Rather than attempt to estimate each $f_{i,0}$ exactly -- a task made complicated by the fact that we only have 12 observations per census tract -- we instead focus on \textit{approximating} $f_{0}$ up to the first order.
That is, we introduce parameters $\alpha_{i}$ and $\beta_{i}$ for each census tract and fit the approximate model
\begin{equation}
\label{eq:tract_model}
y_{i,t} \sim \mathcal{N}(\alpha_{i} + \beta_{i}x_{t}, \sigma^{2}).
\end{equation}
The parameters $\alpha_{i}$ and $\beta_{i}$ respectively approximate the mean level and time trend of the transformed crime density in census tract $i.$
Although the approximation in Equation~\eqref{eq:tract_model} may not perfectly characterize the temporal evolution of crime, such linear approximations are routinely used in limited data settings like ours, which includes only 12 observations per census tract \citep{bernardinelli1995bayesian, anderson2017spatial}.
Moreover, an exploratory analysis (see Section S4.1 in the Supplementary Materials) did not reveal strong suggestions of non-linearity in the $y_{i,t}.$
While such a first order approximation is justified in our setting, we note that the methods we develop in Section~\ref{sec:variational_approximation} to estimate and cluster the $\alpha_{i}$'s and $\beta_{i}$'s may be extended to higher-order approximations of the $f_{0,i}$'s that can capture non-linearities; we defer discussion of such extensions of Section S4.2 in the Supplementary Materials.




Although the raw crime counts $c_{i,t}$ were not of primary interest in our analysis, we can, nevertheless, elaborate the model in~\eqref{eq:tract_model} to directly model the observed crime counts.
For instance, following \citet{KowalCanale2020_star}, we can model crime counts as a suitably rounded and transformed latent Gaussian random variable.
Specifically, we can model $c_{i,t} = \nint{A_{i}\sinh{(\log(2) + y^{\star}_{it})}}$ where $\nint{u}$ is the nearest integer to $u$ and $y^{\star}_{i,t} \sim \mathcal{N}(\alpha_{i} + \beta_{i}x_{t}, \sigma^{2})$ is a latent variable.
Recent work \citep{KowalCanale2020_star, KowalWu2021_count} demonstrates that such transformed and truncated latent Gaussian models are more accurate and flexible than conventional Poisson and negative binomial models of counts. 

\subsection{Related work}
\label{sec:lit_review}

The model in Equation~\eqref{eq:tract_model} is a linear model with spatially-varying coefficients.
Geographically weighted regression (GWR; \citet{Fotheringham2002}) is a popular method that estimates the parameters $(\alpha_{i}, \beta_{i})$ in region $i$ using data from region $i$ and weighted data from nearby regions, with regions closer to $i$ receiving higher weights.
Within the Bayesian literature, Gaussian processes (GPs) are a common prior choice to induce spatial dependence between the parameters estimated at different locations.
In the presence of \textit{a priori} unknown spatial discontinuities, however, both GWR and GP-based methods can introduce an inappropriate amount of smoothness, resulting in substantial bias in the parameter estimates and downstream crime forecasts.

There is a vast literature on data-adaptive strategies for detecting such discontinuities.
One approach involves first fitting a simple model that does not account for potential discontinuities and then identifying jumps in the fitted values 
(see, e.g., \citet{boots2001using}, \citet{li2011mining}, \citet{banerjee2012bayesian}, \citet{lu2005bayesian}, and \citet{lee2013locally}).
Alternatively, many authors directly model uncertainty about which borders correspond to sharp discontinuities within larger Bayesian hierarchical models (see, e.g., \citet{lee2012boundary}, \citet{lu2007bayesian}, and \citet{Balocchi2019}).
Despite their intuitive appeal, these latter models are heavily over-parametrized, often introducing one latent parameter for every pair of adjacent neighborhoods. 

Rather than looking for individual discontinuities between pairs of neighborhoods, several authors instead look for clusters of neighborhoods with similar parameter values.
Such clustered models are much more parsimonious and readily interpreted than models of pairwise discontinuities. 
Our goal in this work is to cluster and estimate the mean-levels $\alpha_{i}$ and time-trends $\beta_{i}$ in Equation~\eqref{eq:tract_model}.
Because there is no \textit{a priori} reason to expect the spatial distribution of the $\alpha_{i}$'s to match the spatial distribution of the $\beta_{i}$'s, we require a method that can estimate separate clusters for both sets of parameters.
We further require a method that can recover \textit{spatial partitions}, whose clusters consist of neighborhoods which are spatially adjacent. 
Unfortunately, most existing methods for spatially clustered regression fail to meet both criteria.

Two important early works on Bayesian spatial clustering are \citet{knorr2000bayesian} and \citet{denison2001bayesian}.
Both fit intercept-only Poisson regression models in which the intercepts are constant within clusters.
In both methods, clusters are defined with respect to carefully chosen distance metrics.
\citet{gangnon2000bayesian} and \citet{green2002hidden} also fit Poisson regression models with clustered intercepts.
They also assumed that parameters were constant within clusters but specified priors on the partitions that respectively depended on the cluster's shape and number of adjacent units within each cluster.
In general, neither of their procedures recover spatial partitions.

\citet{feng2016spatial} substantially expand \citet{knorr2000bayesian}'s model by incorporating additional covariates with spatially clustered effects.
Despite producing spatial clusters, their extension assumes that (i) covariate effects are constant within clusters and (ii) the effects of every covariate are clustered identically.
\citet{PageQuintana2016}'s spatial product partition model makes the same assumptions but is further limited by the fact that it places positive prior probability on non-spatial partitions.

\citet{anderson2017spatial} study the spatiotemporal variation in respiratory disease risk by fitting a Poisson regression model in each areal unit.
Like us, their goal is to estimate an intercept and time-trend within each areal unit and, unlike \citet{feng2016spatial}, they do not assume that the intercepts and time-trends are clustered identically.
However, they do not restrict attention to spatial clusters and their computational procedure requires the user to specify, \textit{a priori}, the maximum number of clusters for both the intercept and time-trend.

Recently \citet{Teixeiria2019} and \citet{LiSang2019} proposed elegant methods for spatial clustering based on spanning tree representations of the adjacency structure of the neighborhoods.
Whereas \citet{LiSang2019}'s method is based on solving a fused lasso problem corresponding to a fixed spanning tree, \citet{Teixeiria2019} instead developed a Gibbs sampler that sequentially samples a spanning tree, spatial partition, and regression parameters.
Despite the elegance of their framework, \citet{Teixeiria2019} noted that sampling a spanning tree from its exact posterior conditional distribution is computationally prohibitive.
They instead sampled from an approximate conditional and the extent to which their approximation affected their downstream inference is unclear. 

\subsection{Model}
\label{sec:model}
\begin{figure}[h!]
\begin{subfigure}[b]{0.495\textwidth}
\centering
\includegraphics[width=\textwidth]{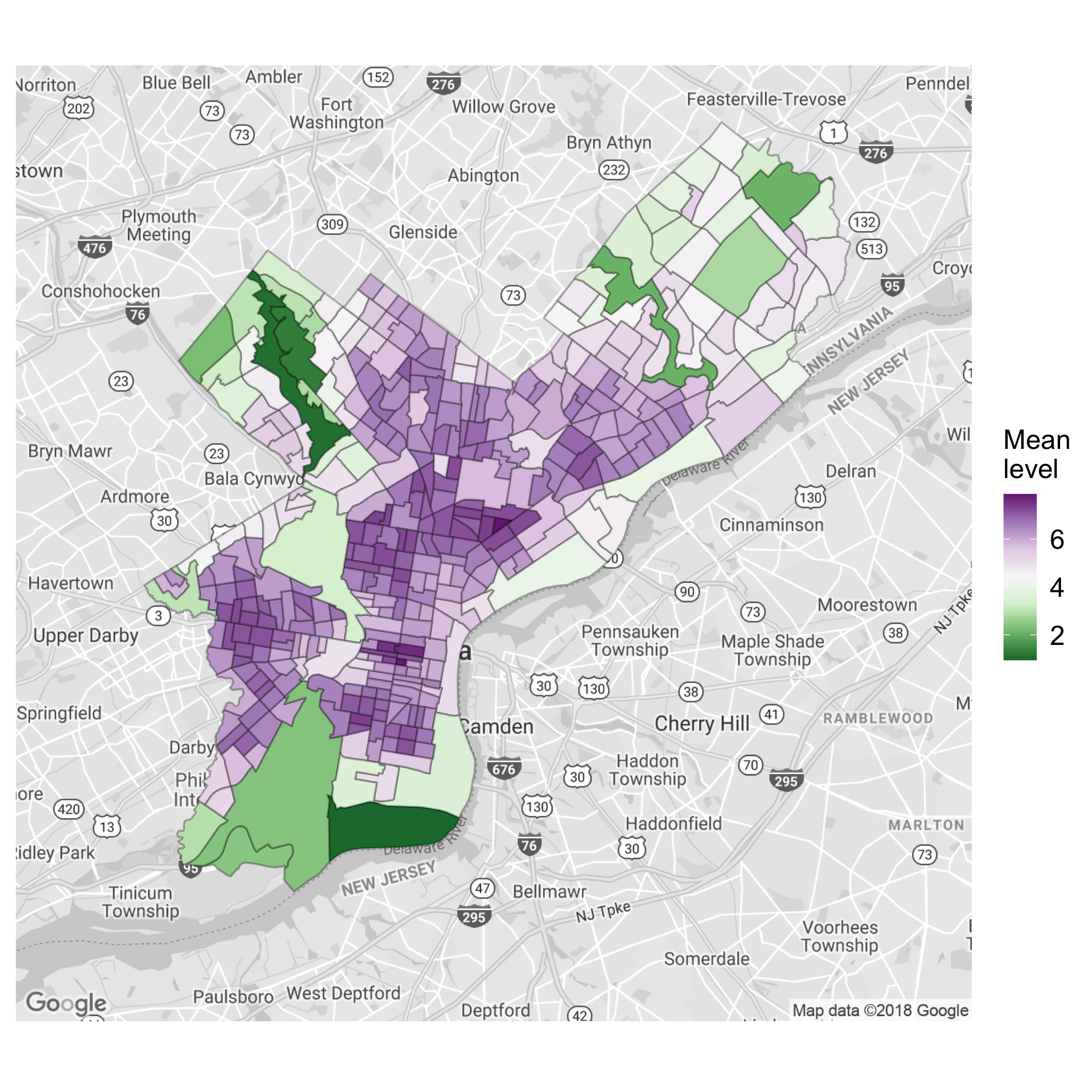}
\end{subfigure}
\begin{subfigure}[b]{0.495\textwidth}
\centering
\includegraphics[width=\textwidth]{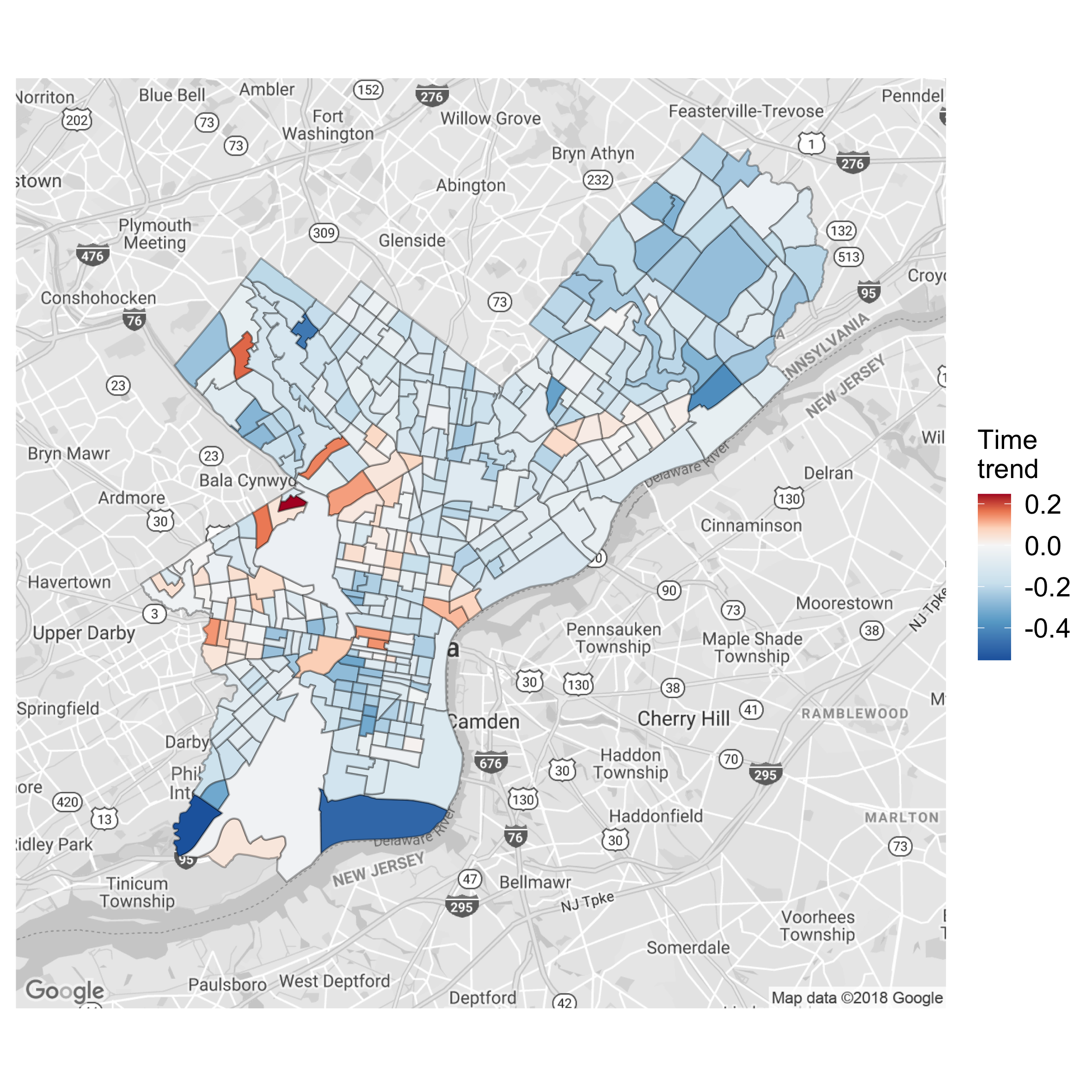}
\end{subfigure}
\caption{\small{Visualization of the maximum likelihood estimates of the tract-level intercepts $\balpha$ (left panel) and time-trends $\bbeta$ (right panel) for the model defined in Section~\ref{sec:model}}}
\label{fig:mle}
\end{figure}

To motivate our proposed CAR--within--cluster model, we first note that the tract-specific maximum likelihood estimates (MLEs) of $\alpha_{i}$ and $\beta_{i}$ suggest that the crime may not have decreased uniformly across the city (Figure~\ref{fig:mle}).
In fact, in a small number of neighborhoods, crime has actually increased over the last decade.

We further observe that, with a few notable exceptions, spatially adjacent neighborhoods tend to have similar MLEs, suggesting a high degree of spatial correlation in the neighborhood-level crime dynamics.
We take a hierarchical Bayesian approach in order to ``borrow strength'' between neighborhoods that involves specifying a prior distribution on the parameters $\balpha = (\alpha_{1}, \ldots, \alpha_{N})$ and $\bbeta = (\beta_{1}, \ldots, \beta_{N}).$ 
Because we expect the tract-specific parameters to display some spatial continuity, we use priors that explicitly introduce dependence between parameters from neighboring tracts. 

Conditionally autoregressive (CAR) models are a popular class of such priors and we use a version introduced in \citet{leroux2000estimation}.
Letting $W = (w_{i,j})$ be a binary adjacency matrix with $w_{i,j} = 1$ if and only if neighborhoods $i$ and $j$ share a border, we say that the vector $\btheta = (\theta_{1}, \ldots, \theta_{n})$ follows a CAR model with grand mean $\overline{\theta}$ and variance scale $\tau^{2}$ if and only if all of the full conditional distributions have the form
$$
\theta_{i} \mid \btheta_{-i}, \overline{\theta}, \tau^{2} \sim N\left(\frac{(1 - \rho) \overline{\theta} + \rho\sum_{j}{w_{i,j}\theta_{j}}}{1 - \rho + \rho\sum_{j}{w_{i,j}}}, \frac{\tau^{2}}{1 - \rho + \rho \sum_{j}{w_{i,j}}}\right).
$$

In this CAR model, the conditional mean of $\theta_{i} \mid \btheta_{-i}$ is a weighted average of the grand mean $\overline{\theta}$ and the average of the $\theta_{j}$'s from the neighborhoods that border neighborhood $i.$
The degree to which $\theta_{i}$ is shrunk toward either of these targets is governed by a parameter $\rho,$ which is typically set by the analyst, and the number of neighbors.
These full conditionals uniquely determine the joint distribution $\btheta \sim N(\overline{\theta}\bone_{n}, \tau^{2}\Sigma_{\rm{CAR}})$ where
$$
\Sigma_{\rm{CAR}} = 
\begin{cases}
\left[\rho W^{\star} + (1 - \rho)I_{n}\right]^{-1} & \text{if $n \geq 2$} \\
\frac{1}{1 - \rho} & \text{if $n = 1$}
\end{cases},
$$ 
$\bone_{n}$ is the $n$-vector of ones, and $W^{\star}$ is the unweighted graph Laplacian of the adjacency matrix $W$.
For compactness, we will write $\btheta \mid \overline{\theta}, \tau^{2} \sim {\rm CAR}(\overline{\theta}, \tau^{2}, W).$

Cities typically contain many geographic and social barriers like rivers and highways that manifest in sharp spatial discontinuities.
In the presence of these discontinuities, a na\"{i}vely specified CAR model can induce a level of spatial smoothness among the parameters at odds with the data.
To avoid this behavior, we seek \textit{clusters} of parameters that demonstrate considerable spatial continuity within but not between clusters.
We introduce two latent partitions of the neighborhoods, $\gamma^{(\alpha)}$ and $\gamma^{(\beta)}$, where $\gamma^{(\cdot)} = \{S_{1}^{(\cdot)}, \ldots, S_{K^{(\cdot)} }^{(\cdot)}\}$.
We refer to the sets $S^{(\cdot)}_{k}$ as \textit{clusters} and restrict attention to partitions consisting of clusters of spatially connected neighborhoods. 
We denote the set of all such partitions by $\mathcal{SP}$ and let $\bgamma := (\gamma^{(\alpha)}, \gamma^{(\beta)})$ be the pair of latent spatial partitions underlying the mean levels and the time trends of crime across neighborhoods.
In what follows, we will refer to $\bgamma$ as a \textit{particle}.

To simplify our presentation, we describe only the prior over the mean levels of crime $\balpha$; we place an analogous prior on the time trends $\bbeta.$
We place independent CAR priors on the collections $\balpha_{k} = \{\alpha_{i}: i \in S^{(\alpha)}_{k}\}$, so that the joint prior density $\pi(\balpha \mid \gamma^{(\alpha)}, \sigma^{2})$ factorizes over the collection of all clusters:
$\pi(\balpha \mid \gamma^{(\alpha)}, \sigma^{2}) = {\prod_{k = 1}^{K^{(\alpha)}}{\pi(\balpha_{k} \mid \sigma^{2})}}$. 
To this end, we introduce a collection of grand cluster means $\balphabar = \left\{\alphabar_{1}, \ldots, \alphabar_{K^{(\alpha)}}\right\}$ and model $\balpha_{k} \mid \alphabar_{k}, \sigma^{2} \sim {\rm CAR}(\alphabar_{k}, a_{1}\sigma^{2}, W^{(\alpha)}_{k}),$ where $W^{(\alpha)}_{k}$ is the sub-matrix of $W$ whose rows and columns are indexed by the cluster $S_{k}^{(\alpha)}.$
We further place independent $N(0, a_{2}\sigma^{2})$ priors on the grand cluster means $\alphabar_{k}.$

We place independent truncated Ewens-Pitman priors on the two latent spatial partitions $\gamma^{\alpha}$ and $\gamma^{\beta}.$
The Ewens-Pitman distribution on partitions is equivalent to the exchangeable partition probability function (EPPF) of a Chinese Restaurant Process \citep{aldous1985exchangeability,pitman2002combinatorial}. It is characterized by favoring a small number of large clusters and asymptotically it induces an average of $\eta \log(N)$ clusters.
To recover partitions with connected clusters, we truncate this distribution to the set of spatial partitions $\mathcal{SP}$.
The probability mass function of this prior is given by
\begin{equation}
\label{eq:ep_prior}
\pi(\gamma) \propto \eta^{K}\prod_{k = 1}^{K}{(n_{k} - 1)!} \times \mathbf{1}(\gamma \in \mathcal{SP})
\end{equation}
and we denote it by $\mathcal{T\mbox{-}EP}$.
We complete our hierarchical prior with an Inverse Gamma prior on the residual variance $\sigma^{2} \sim \text{IG}\left(\frac{\nu_{\sigma}}{2}, \frac{\nu_{\sigma}\lambda_{\sigma}}{2}\right).$
To summarize, our model is
\begin{align}
\label{eq:general_car_cluster_model}
\begin{split}
\gamma^{(\alpha)}, \gamma^{(\beta)} &\overset{iid}{\sim} \mathcal{T\mbox{-}EP} \\
\sigma^{2} &\sim \text{IG}\left(\frac{\nu_{\sigma}}{2}, \frac{\nu_{\sigma}\lambda_{\sigma}}{2}\right) \\
\alphabar_{1}, \ldots, \alphabar_{K_{\alpha}} \mid \gamma^{(\alpha)}, \sigma^{2} &\overset{iid}{\sim} N(0, a_{2}\sigma^{2}) \\
\betabar_{1}, \ldots, \betabar_{K_{\beta}} \mid \gamma^{(\beta)}, \sigma^{2} &\overset{iid}{\sim} N(0, b_{2}\sigma^{2}) \\
\balpha_{k} \mid \alphabar_{k}, \sigma^{2}, \gamma^{(\alpha)} &\sim \text{CAR}(\alphabar_{k}, a_{1}\sigma^{2}, W^{(\alpha)}_{k}) \quad \text{for $k = 1, \ldots, K_{\alpha}$} \\
\bbeta_{k'} \mid \betabar_{k'}, \sigma^{2}, \gamma^{(\beta)} &\sim \text{CAR}(\betabar_{k'}, b_{1}\sigma^{2}, W^{(\beta)}_{k'}) \quad \text{for $k' = 1, \ldots, K_{\beta}$} \\
y_{i,t} \mid \balpha, \bbeta, \sigma^{2} &\sim N(\alpha_{i} + \beta_{i}x_{t}, \sigma^{2})
\end{split}
\end{align}

The high degree of conditional conjugacy in~\eqref{eq:general_car_cluster_model} enables us to derive analytic expressions for quantities such as the marginal likelihood $p(\by \mid \bgamma)$ as well as the conditional posterior expectations $\E[\balpha,\bbeta \mid \bgamma, \by].$ 
The availability of these expressions will be crucial for the posterior exploration strategy we develop below. 

Given the residual variance $\sigma^{2}$ and latent partitions $\gamma^{(\alpha)}$ and $\gamma^{(\beta)},$ parameters in different clusters are conditionally independent.
In other words, our model falls with the class of conditional product partition models (PPMs) that have been widely used in Bayesian spatial statistics (see, e.g., \citet{knorr2000bayesian}, \citet{denison2001bayesian}, and \citet{feng2016spatial}).
Unlike these papers, however, we are interested in recovering two latent partitions, one each for the mean levels and time-trends within each census tract.
Our model is perhaps most similar to \citet{anderson2017spatial}, who also seek to estimate two distinct partitions, one for the intercepts and one for the slopes. 
However, they limit attention to partitions containing five or fewer clusters for computational simplicity, whereas we will not need to impose any \textit{a priori} restriction on the number of clusters. 

Our CAR--within--cluster model depends on several hyperparameters, including $\rho$, which regulates the amount of within-cluster spatial autocorrelation, and $\eta,$ which regulates the expected number of clusters.
Throughout our analysis, we fix these hyperparameters and set the remaining hyperparameters $a_{1}, a_{2}, b_{1}, b_{2}, \nu_{\sigma}$ and $\lambda_{\sigma}$ in a data-dependent fashion, as described later in Section~\ref{sec:particle_optimization}. 

\section{Posterior Exploration and Summarization}
\label{sec:variational_approximation}
In fitting our model, we have three simultaneous aims: (i) identify the latent pair of partitions $\bgamma = (\gamma^{(\alpha)}, \gamma^{(\beta)}),$ (ii) estimate the approximate baseline levels $\balpha$ and time trends $\bbeta$ of crime in each neighborhood, and (iii) make predictions about future incidents of crime in each neighborhood.
These latter two tasks can generally be expressed as evaluating posterior expectations of the form $\E[g(\balpha, \bbeta) \mid \by]$ where $g$ is a functional of interest.
For instance, we can obtain point estimates of individual mean levels with $g(\balpha, \bbeta) = \alpha_{i}$ and future crime forecasts at some future time with $g(\balpha, \bbeta) = \alpha_{i} + \beta_{i}x^{\star}.$
The combinatorial vastness of the space $\mathcal{SP}^{2},$ which contains all possible pairs of spatial partitions, renders it impossible to enumerate all $\bgamma$'s for even small values of $N.$
As a result, we cannot compute the posterior probability $\pi(\bgamma \mid \by)$ exactly. 

It is tempting to resort to Markov Chain Monte Carlo (MCMC) simulations to approximate expectations $\E[g(\balpha, \bbeta) \mid \by].$
Unfortunately, due to the vastness of $\mathcal{SP}^{2},$ such MCMC simulations may require a prohibitive amount of time to mix.
To get around this difficulty, \citet{anderson2017spatial} arbitrarily restricted attention to partitions with no more than five clusters each. 
Even with such a restriction, which we will not impose, it is still quite difficult to distill the thousands of resulting draws of $\bgamma$ into a single point estimate and to quantify parameter and partition uncertainty in an interpretable fashion \citep[see, e.g.,][]{wade2017bayesian}.

A popular alternative approach is posterior optimization, which usually focuses on identifying the \textit{maximum a posteriori} (MAP) pair of partitions $\bgamma^{(1)}$ or some other decision-theoretic optimal point estimate (see, e.g., \citet{Lau2007} and \citet{rastelli2018optimal}).
One then estimates the marginal expectation $\E[g(\balpha, \bbeta) \mid \by]$ with a ``plug-in'' estimator $\E[g(\balpha, \bbeta) \mid \by, \bgamma^{(1)}].$
Though this procedure might be substantially faster than MCMC, it completely eschews exploration of the uncertainty about $\bgamma.$
As a result, the MAP plug-in estimator may result in over-confident inference about the functional.
Essentially, using the MAP plug-in amounts to approximating the full posterior distribution over $\mathcal{SP}^{2}$ with a single point mass.

Instead of identifying only a single $\bgamma$ with high posterior probability, what if we could identify $\Gamma_{L} = \{\bgamma^{(1)}, \ldots, \bgamma^{(L)}\},$ the set of $L$ pairs of partitions with largest posterior mass? 
Given $\Gamma_{L},$ a natural approximation is
$$
\E[g(\balpha, \bbeta) \mid \by, \bgamma] \approx \sum_{\bgamma \in \Gamma_{L}}{\tilde{\pi}_{L}(\bgamma^{(\ell)} \vert \by)\E[g(\balpha, \bbeta) \vert \bgamma^{(\ell)}, \by]},
$$
where $\tilde{\pi}_{L}(\cdot \vert \by)$ is the truncation of $\pi(\bgamma \vert \by)$ to the set $\Gamma_{L}.$
Because this approximation averages over more of the posterior uncertainty about $\bgamma,$ we might reasonably expect it to be more accurate than the MAP plug-in.

\subsection{A Variational Approximation}

It turns out that we can identify $\Gamma_{L}$ by finding a variational approximation to the full posterior distribution. 
Before proceeding, we introduce some more notation.
For any collection of $L$ particles $\Gamma = \left\{\bgamma_{1}, \ldots, \bgamma_{L}\right\}$ and vector $\bw = (w_{1}, \ldots, w_{L})$ in the $L$-dimensional simplex, let $q(\cdot \mid \Gamma, \bw)$ be the discrete distribution that places probability $w_{\ell}$ on the particle $\bgamma_{\ell}.$
Following \citet{Rockova2018}, we will refer to individual pairs of partitions $\bgamma$ as \textit{particles}, the collection $\Gamma$ as a \textit{particle set} and $\bw$ as \textit{importance weights}.
Let $\mathcal{Q}_{L}$ be the collection of all such distributions supported on at most $L$ particles.
Finally, for each $\lambda > 0$, let $\Pi_{\lambda}$ be the tempered marginal posterior with mass function $\pi_{\lambda}(\bgamma) \propto \pi(\bgamma \mid \by)^{\frac{1}{\lambda}}.$
Note that the particles in $\Gamma_{L},$ which are the $L$ particles with largest posterior mass, are also the $L$ particles with largest tempered posterior mass for all $\lambda.$

The following proposition provides the foundation for estimating $\Gamma_{L}.$
\begin{proposition}
Suppose that $\pi(\bgamma \mid \by)$ is supported on at least $L$ distinct particles and that $\pi_{\lambda}(\bgamma) \neq \pi_{\lambda}(\bgamma')$ for $\bgamma \neq \bgamma'.$ 
Let $q^{\star}_{\lambda}(\cdot | \Gamma^{\star}(\lambda), \bw^{\star}(\lambda))$ be the distribution in $\mathcal{Q}_{L}$ that is closest to $\Pi_{\lambda}$ in a Kullback-Leibler sense:
$$
q^{\star}_{\lambda} = \argmin_{q \in \mathcal{Q}_{L}}\left\{\sum_{\bgamma}{q(\bgamma)\log{\frac{q(\bgamma)}{\pi_{\lambda}(\bgamma)}}}\right\}.
$$
Then $\Gamma^{\star}(\lambda) = \Gamma_{L}$ and for each $\ell = 1, \ldots, L,$ $w_{\ell}^{\star}(\lambda) \propto \pi(\bgamma^{(\ell)} | \by)^{\frac{1}{\lambda}}$
\end{proposition}

The proof of Proposition 1 is in Section S1 of the Supplementary Materials and follows from the definition of the Kullback-Leibler divergence.

In light of Proposition 1, we can recover $\Gamma_{L}$ by finding an approximation of any tempered posterior $\Pi_{\lambda}.$
This is equivalent to solving 
\begin{equation}
\label{eq:particle_optimization}
\left(\Gamma^{\star}(\lambda), \bw^{\star}(\lambda)\right) = \argmax_{\left(\Gamma, \bw\right)}\left\{\sum_{\ell = 1}^{L}{w_{\ell}\log{p(\by, \bgamma_{\ell})}} + \lambda H(\Gamma, \bw) \right\},
\end{equation}
where $H(\Gamma, \bw) = -\E_{q}[\log{q(\cdot | \Gamma, \bw)}]$ is the entropy of the approximating distribution $q(\cdot | \Gamma, \bw).$

Before proceeding, we stress that we are not finding a variational approximation of $\pi(\balpha, \bbeta, \sigma^{2} \vert \by),$ the marginal posterior distribution of the continuous parameters of interest.
Instead, we are approximating the discrete posterior distribution $\pi(\bgamma \mid \by),$ which places positive probability over all particles $\bgamma = (\gamma^{\alpha}, \gamma^{\beta}),$ with another discrete distribution $q^{\star}$ that places positive probability on only $L$ particles.

We pause briefly to reflect on the two terms in Equation~\eqref{eq:particle_optimization}.
The first term is, up to an additive constant depending only on $\by$, the $\bw$-weighted average of the height of the log-posterior at each particle in the particle set $\Gamma.$
This term is clearly maximized when all of the particles in $\Gamma$ are equal to the MAP.
On the other hand, the entropy $H(\Gamma, \bw)$ of the approximating distribution is maximized when all of the particles in $\Gamma$ are distinct and each $w_{\ell} = L^{-1}.$
The penalty term $\lambda$ balances these two opposing forces.

Finally, we note that \citet{Rockova2018} introduced essentially the same family of optimization problems to identify sparse high-dimensional linear regression models and described a coordinate ascent strategy that iteratively updated $\bw$ and $\Gamma.$
In that work, $\bgamma$ was a binary vector indicating which variables to include in the model.

\subsection{Particle Optimization}
\label{sec:particle_optimization}

Finding the global optimum of~\eqref{eq:particle_optimization} exactly is practically impossible, given the enormous size of the set of all possible particle sets $\Gamma.$ 
Instead, like \citet{Rockova2018}, we deploy a coordinate ascent strategy: starting from an initial particle set $\Gamma$ and initial weight vector $\bw,$ we iteratively update one of $\bw$ and $\Gamma$ until we reach a stationary point.

To update the particle set $\Gamma,$ we sweep over the individual particles, sequentially updating each one while holding the others fixed.
Finding the globally optimum single particle update is practically impossible, since doing so would require searching over the space of all possible pairs of spatial partitions.
We instead update each $\bgamma_{\ell}$ by updating each of its constituent partitions one at a time.
Even updating a single partition within a single particle is practically impossible, given the vastness of the space of spatial partition $\mathcal{SP}.$
For that reason, we must rely on a local algorithm to search over $\mathcal{SP}.$
That is, we update a single partition by first enumerating a large number of candidate partitions and moving to the candidate that maximizes the objective function Equation~\eqref{eq:particle_optimization}.

Perhaps the simplest candidate set consists of all partitions that can be formed by reallocating a single neighborhood to a new or existing cluster.
Such one-neighborhood updates directly parallel conventional Gibbs samplers for Dirichlet process mixture models. 
In our optimization setting, such a restrictive search strategy results in premature termination at a sub-optimal particle set $\Gamma.$
Instead, a more promising strategy for navigating the space of partitions is to allow multiple neighborhoods to be reallocated at once. 

\begin{figure}[h]
\centering
\includegraphics[width = 0.5\textwidth]{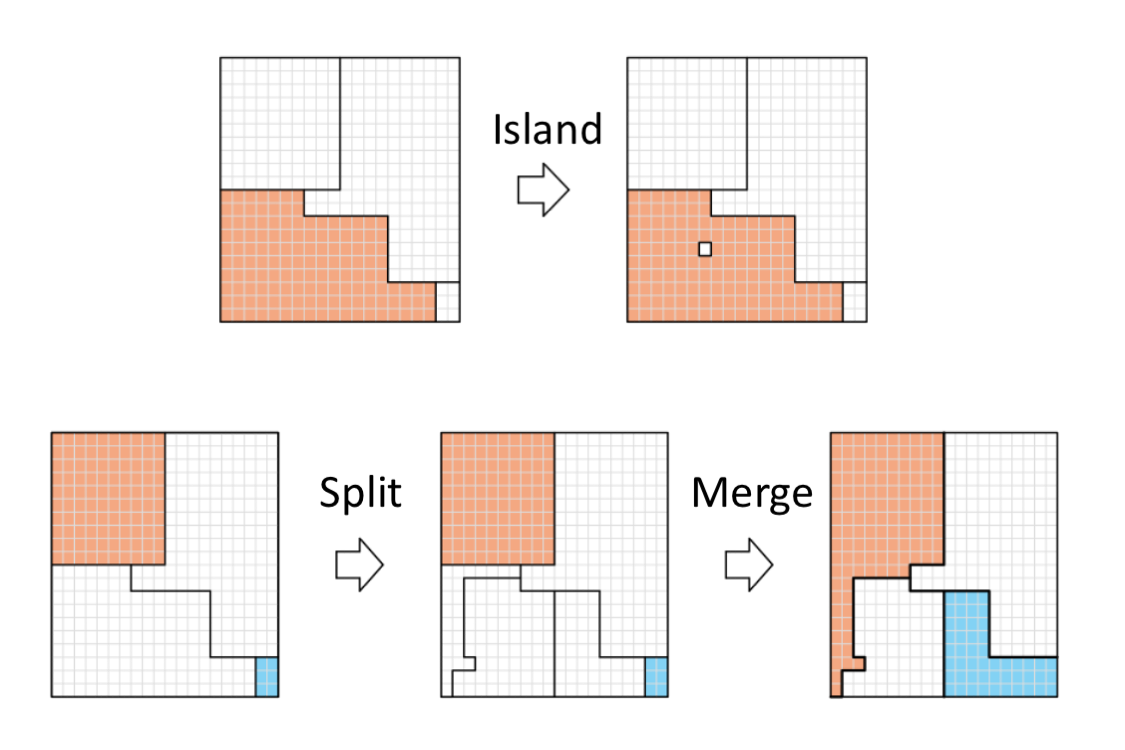}
\caption{Two broad types of transitions that we consider. An ``island'' move (top) removes a single neighborhood from an existing cluster (the lower left orange cluster) and creates a new singleton cluster. A ``split and merge'' move (bottom) first splits an existing cluster (the bottom left cluster) into multiple sub-clusters and then merges some or all of the new sub-clusters into existing clusters}
\label{fig:transitions}
\end{figure}

We build a candidate set using both \textit{fine} transitions, which reallocate a single neighborhood to a new or existing cluster, and \textit{coarse} transitions, which simultaneously reallocate multiple neighborhoods.
There are two types of fine transitions, ``island'' moves, in which a neighborhood is remove from its current cluster and reallocated to a new singleton cluster, and ``border'' moves, which move a single neighborhood located at the boundary between two clusters across the boundary.
We also consider three types of coarse transitions: (i) ``merge'' moves, which combine two adjacent clusters into a single clusters; (ii) ``split'' moves, which divide an existing cluster into multiple smaller sub-clusters; and (iii) ``split-and-merge'' moves, which first split an existing clusters into multiple pieces and then merge some or all of those pieces with other existing clusters.
Figure~\ref{fig:transitions} illustrates these moves.
We note that sometimes, removing a single neighborhood from a cluster leaves the resulting cluster spatially disconnected.
When this happens, we treat the resulting components as individual clusters.

\textbf{Local search heuristics}. In our local search algorithm, it is not practical to enumerate all possible coarse and fine transitions. 
We instead use a number of heuristics to limit the number of possible moves considered in each step.
For brevity, we describe these heuristics for transitions for $\gamma^{(\alpha)}$; we use exactly the same heuristics for $\gamma^{(\beta)}.$

The conditional conjugacy of our ``CAR--within--cluster'' model allows us to quickly compute $\E[\alpha_{i} \mid \bgamma, \by]$ and $\E[\alphabar_{k} \mid \gamma, \by].$ 
We use these conditional means as running estimates with which to propose transitions. 
Rather than attempting to merge an existing cluster $k$ with each one of its neighboring clusters, we only attempt to merge $k$ with its neighbor $k'$ whose grand cluster mean $\overline{\alpha}_{k'}$ is closest to $\overline{\alpha}_{k}.$
Moreover, when attempting split-and-merge moves, we cap the number of new sub-clusters at five. 
Finally, for island moves, we initially only attempt to remove neighborhood $i$ from its current cluster and move it to a new singleton if the estimated $\alpha_{i}$ is in the top or bottom 5\% of the distribution of estimates within the cluster.
During our coordinate ascent algorithm, if we find that none of these transitions are accepted, we try all $N$ island moves.
This last check ensures that our algorithm converges locally in the sense that no one-neighborhood update to an individual partition will result in a higher objective.

These heuristics are admittedly ad hoc but we have found that they strike a good balance between solution optimality and computational speed.
We found, for instance, that merge moves that tried to merge clusters with dissimilar estimated grand means were rarely accepted.
We also found that if we restricted split moves to create only two new sub-clusters at a time, it was more difficult to identify partitions containing many small clusters.

\textbf{Choice of hyperparameters}.
Our prior depends on several hyperparameters, some of which we recommend fixing and some of which we set in a data-dependent manner.
Specifically, we fix $\rho = 0.9$ so that our prior concentrates on clusters with relatively high levels of spatial autocorrelation. 
This choice of $\rho$ also ensures prior propriety and results in somewhat more numerically stable posterior computations than values of $\rho$ much closer to 1.
Although the posterior distribution over $\bgamma$ is somewhat sensitive to $\rho$, resulting estimates of $\balpha$ and $\bbeta$ are somewhat less sensitive to misspecification of $\rho$ (see Section S3.4 of the Supplementary Materials for a detailed sensitivity analysis). 

In our data analysis and simulations, the number of spatial units $N$ is approximately 400.
For a non-truncated Ewens-Pitman prior, the expected number of clusters grows at the rate $\eta \log{(N)}.$
To us $\log{(400)} \approx 6$ seemed like a reasonably adequate level of summarization of crime patterns; accordingly, we fixed $\eta = 1$ throughout. 

The remaining hyperparameters are set in a data-dependent fashion.
At a high-level, to set $\nu_{\sigma}$ and $\lambda_{\sigma},$ we first fit separate linear models to the data in each neighborhood to get an empirical distribution of the residual variance.
We then set the hyperparameters of the $\text{Inv. Gamma}(\nu_{\sigma}/2, \nu_{\sigma}\lambda_{\sigma}/2)$ prior on $\sigma^{2}$ to match the first and second moments of this empirical distribution.
To set $a_{1}, a_{2}, b_{1}$, and $b_{2},$ we use an empirical Bayes approach based on computing the MAP estimate of $\bgamma.$
We provide full details of this strategy to Section S2 of the Supplementary Materials. 
In our experiments, we have found this strategy to work well and so have automated its implementation as a default option in our R package.

\textbf{Initialization and choice of $\lambda$}.
To run our local search algorithm, one must additionally specify (i) the number of particles $L$, (ii) the initial locations of the particles, and (iii) the inverse temperature $\lambda > 0.$
The choice of $L$ is largely dependent on the computational budget, as the per-iteration complexity of our algorithm scales linearly in $L.$
For problems of our size, with around $N = 400$ spatial units, we have found that setting $L = 10$ or $L = 20$ strikes a good balance between posterior exploration and computational effort. 

We initialize the particle set by randomly drawing particles $(\hat{\gamma}^{(\alpha)}_{K}, \hat{\gamma}^{(\beta)}_{K'})$ with replacement where $\hat{\gamma}^{(\alpha)}_{K}$ is the partition obtained by running k-means on the maximum likelihood estimates of $\balpha$ with $k = K$ clusters.
We let $K,K' = 1, \ldots, \lfloor \log(N)\rfloor.$
In this initialization, the probability of drawing particle $(\hat{\gamma}^{(\alpha)}_{K}, \hat{\gamma}^{(\beta)}_{K'})$ is proportional to its marginal posterior probability. 
Our initialization allows our algorithm to pursue several search directions simultaneously but also allows for some redundancy in the initial particle set.
In regions of high posterior probability, such redundancy allows multiple particles to search around a dominant mode, providing a measure of local uncertainty.

Finally, it is important to stress that $\lambda$ is not a model parameter; neither the data likelihood nor the prior distribution depends on it.
Moreover, although $\lambda$ plays the role of a penalty parameter in Equation~\eqref{eq:particle_optimization}, Proposition 1 guarantees that the \textit{global} solution to that optimization problem is the same for all $\lambda > 0.$
This is in marked contrast to most penalized likelihood procedures, whose solutions are highly dependent on the choice of penalty parameter.
In practice, however, the solution obtained by our \textit{local} search algorithm is somewhat sensitive to the choice of $\lambda.$

In our experiments, we have found that setting $\lambda = 1$ often results in all $L$ particles collapsing onto a single point.
On further inspection, we found that changes in the entropy, being upper bounded by $\log{(L)},$ are often an order of magnitude smaller than changes in the $\bw$-weighted log-posterior. 
To offset this, we recommend running our local search algorithm with large $\lambda.$
Typically, with larger values of $\lambda,$ particles are encouraged to drift to regions of lower posterior probability more forcefully than with lower values of $\lambda.$
In our experiments, we have found that $\lambda = 10$ and $\lambda = 100$ often produce good results. 

\section{Synthetic Data Evaluation}
\label{sec:illustration}

\begin{figure}[t]
\centering
\includegraphics[width = 0.65\textwidth]{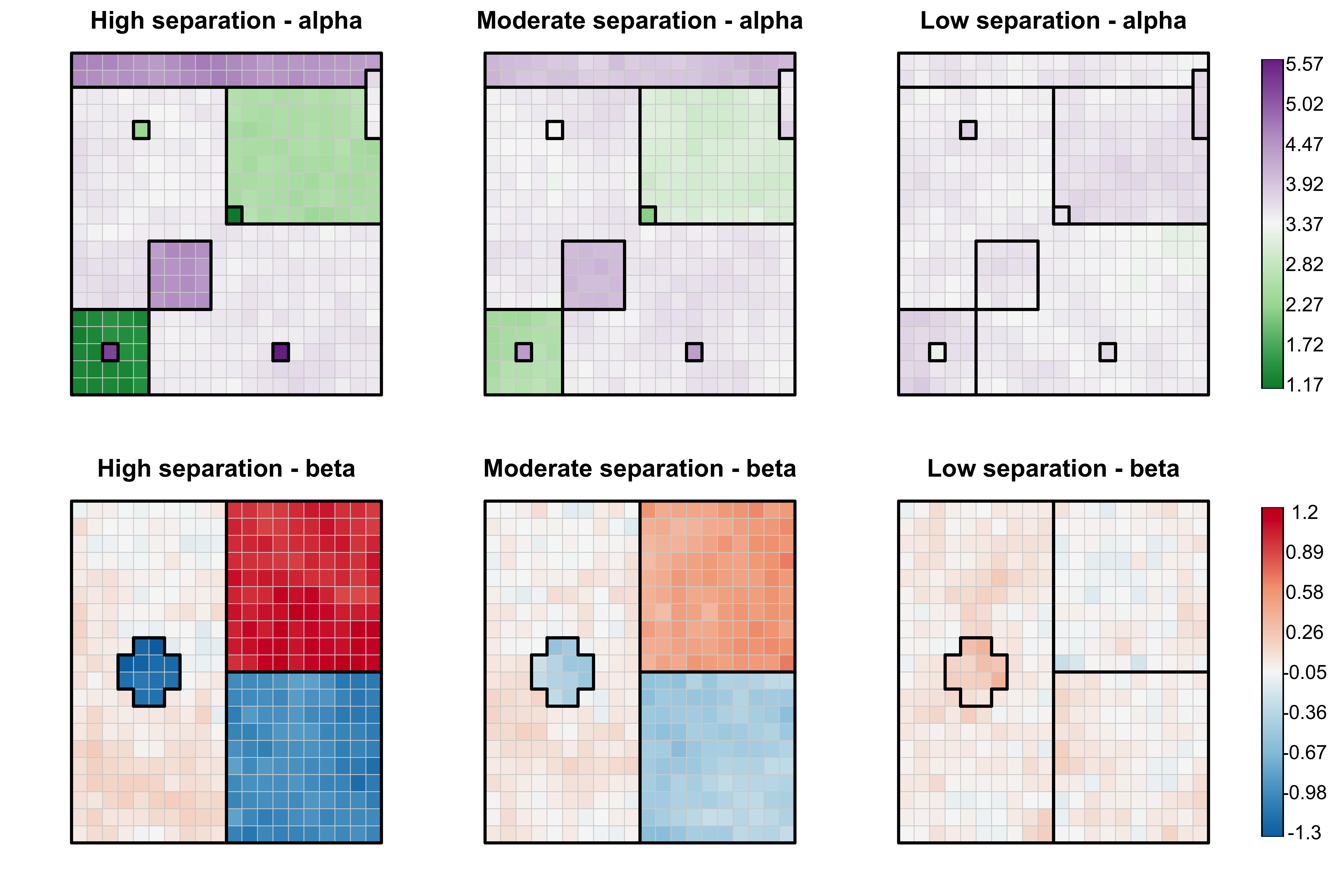}
\caption{The spatial partitions used to generate our synthetic data and three different settings of $\balpha$ and $\bbeta$ values. The color of each square corresponds to the value of the parameter. Different color scales have been used for $\balpha$ and $\bbeta.$ Going from left to right, the distances between the cluster means gets progressively small and the overall cluster structure is less visibly apparent.}
\label{fig:example_partitions}
\end{figure}

To investigate the behavior of our proposed particle optimization procedure, which we will refer to as \texttt{PartOpt} throughout this section, we conducted several experiments using synthetic data generated over a $20 \times 20$ grid of equally-sized spatial units.
Throughout our experiments, we fix the true spatial partitions $\tilde{\gamma}^{\alpha},$ which contains 10 clusters ranging in size from one unit to 237 units, and $\tilde{\gamma}^{\beta},$ which contains four clusters of sizes 188, 100, 100, and 12.
Without losing any generality, we assumed throughout our simulation study that each region had unit area $A_{i} = 1.$ 
Given the true spatial partitions, we generated synthetic data according to the model in Equation~\eqref{eq:tract_model} where the $\alpha_{i}$'s and $\beta_{i}$'s were drawn from CAR--within--cluster distributions.
We considered three different settings, corresponding to varying levels of separation between the means $\overline{\alpha}_{k}$ and $\overline{\beta}_{k'}$ associated with each cluster of $\tilde{\gamma}^{\alpha}$ and $\tilde{\gamma}^{\beta}.$
This allowed us to investigate how \texttt{PartOpt} works in settings where the cluster structure is readily apparent and in settings where the cluster structure is less well-defined.
For simplicity, although $\tilde{\gamma}^{\alpha}$ contains ten distinct clusters, we fixed the corresponding cluster means $\overline{\alpha}_{k}$ in such a way that there were only five distinct cluster means. 
Figure~\ref{fig:example_partitions} shows the true spatial partitions we used as well as a single draw of $\balpha$ and $\bbeta$ in each of the three settings we considered. 
Full details for generating our synthetic data may be found in Section~S3.1 of the Supplementary Materials. 

We compared \texttt{PartOpt} method with the following competitors: (i) \citet{anderson2017spatial}'s method for finding clusters of slopes and intercepts in spatial Poisson regression (hereafter \texttt{And}); (ii) \citet{LiSang2019}'s spatially clustered coefficient procedure (hereafter \texttt{SCC}); (iii) separately running K-means on the MLEs of $\balpha$ and $\bbeta$ (hereafter \texttt{KM}); and (iv) separately running spectral clustering \citep{ng2002spectral} on the MLEs of $\balpha$ and $\bbeta$ (hereafter \texttt{SC}).

In our experiments and data analysis, we placed independent truncated Ewens-Pitman prior \eqref{eq:ep_prior} on the latent partitions $\gamma^{\alpha}$ and $\gamma^{\beta}$ with $\eta = 1.$
We ran \texttt{PartOpt} with $L = 10$ particles and three different penalties $\lambda \in \{1, 10, 100\}.$
We further fixed $\rho = 0.9$ in the CAR--within--clusters prior and defer a sensitivity analysis of this choice of $\rho$ to Section~S3.4 of the Supplementary Materials.
We moreover set the remaining hyperparameters $a_{1}, a_{2}, b_{1},$ and $b_{2}$, together with $\nu_{\sigma}$ and $\lambda_{\sigma}$, in a data-dependent way.
At a high level, we first use a heuristic based on the MLE estimate of $\balpha$ and $\bbeta$ and on the expected number of clusters to estimate temporary values for the hyperparameters, with which we find the MAP; we then find an empirical Bayes estimate of the hyperparameters given the MAP and run our full procedure to recover the particle set.
For further details, see Section~S2 of the Supplementary Materials.

Note that our synthetic transformed crime densities $y_{i,t}$ are generated from the normal model in Equation~\eqref{eq:tract_model}, while \texttt{And} is designed for count-valued data.
In order to facilitate a comparison between \texttt{PartOpt} and \texttt{And}, we first transformed our $y_{i,t}$'s into crime counts $c_{i,t}$ using the formula $c_{i,t} = \nint{\sinh(y_{i,t} + \log{2})}$ where $\nint{x}$ is the nearest integer to $x.$
As noted earlier, \citet{anderson2017spatial} restricts attention to partitions with a pre-specified number of clusters.
Following their recommendations we set the maximum number of clusters per partition to be five when we ran \texttt{And},.
In Section~S3.3 of the Supplementary Materials, we consider an additional comparison between \texttt{PartOpt} and \texttt{And} in which we first generate crime counts $c_{i,t}$ from a Poisson regression model whose slopes and intercepts are drawn from a CAR--within--clusters prior, and then compute the transformed crime densities $y_{i,t}$ using the inverse hyperbolic sine transformation. 
 
When running \texttt{KM} on the MLEs of $\balpha$ and $\bbeta$, we varied the initial number of clusters from one to five and selected the final number of clusters using the silhouette method \citep{Aranganayagi2007}.
Since \texttt{KM} and \texttt{And} do not generally return spatially connected clusters, we post-processed the identified clusters by splitting them into their connected components.
For \texttt{SC}, we varied the number of clusters from one to 10 and used the silhouette method to select the number of clusters.
Unlike \texttt{PartOpt}, \texttt{And}, and \texttt{SCC}, which all attempt to learn the latent partitions jointly, \texttt{KM} and \texttt{SC} estimate $\gamma^{\alpha}$ and $\gamma^{\beta}$ essentially independently.
For both \texttt{KM} and \texttt{SC}, we estimated the tract-level parameters $\alpha_{i}$ and $\beta_{i}$ using the posterior mean conditional on the estimated partitions. 

We compared the parameter estimation and predictive performance of these methods as well as their abilities to recover the true spatial partitions.
We assessed parameter estimation using the root mean square error (RMSE) for estimating the concatenated vector of parameters $(\balpha^{\top}, \bbeta^{\top})^{\top}$.
We assessed predictive performance using the RMSE  for predicting the vector of one-step-ahead out-of-sample observations $\y_{T+1} = (y_{i,T+1})_{i}$ where $y_{i,T+1} \sim N(\alpha_i + \beta_i x_{T+1}, \sigma^2)$.
\texttt{SCC}, \texttt{KM}, and \texttt{SC} all estimate a single pair of partitions.
We measured these methods' ability to recover the latent spatial partitions by computing the adjusted Rand indices \citep{hubert1985comparing} between (i) the estimated intercept partition $\hat{\gamma}^{\alpha}$ and the true partition $\tilde{\gamma}^{\alpha}$ and (ii) the estimated slope partition $\hat{\gamma}^{\beta}$ and the true partition $\tilde{\gamma}^{\beta}.$
The Rand index \citep{rand1971} between two partitions is the proportion of pairs of items that are clustered together in both partitions.
Values close to one indicate a high a degree of similarity between partitions.
The adjusted Rand index corrects the Rand index for chance agreement.

For \texttt{And}, we first computed the adjusted Rand indices between each sampled partition and the corresponding true partition.
We then approximated the posterior mean adjusted Rand index for each of $\gamma^{\alpha}$ and $\gamma^{\beta}.$
We similarly \textit{approximated} the posterior mean adjusted Rand indices using the particle set identified by \texttt{PartOpt}.
Large values of the (approximate) posterior mean adjusted Rand index indicates that the (approximate) posterior places more of its mass near the true partitions that generated the data. 

\begin{figure}[t!]
\centering
\includegraphics[width = 0.7\textwidth]{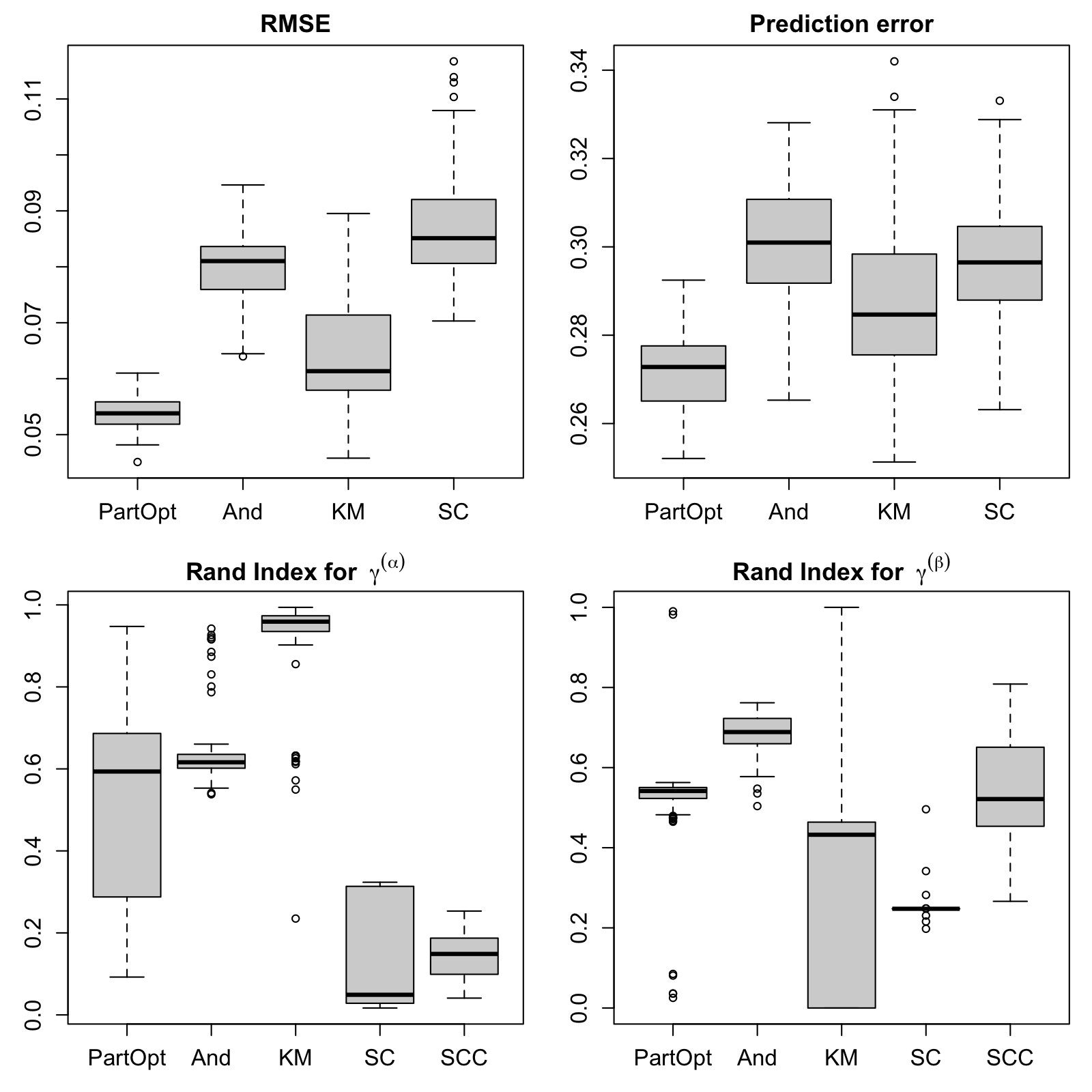}
\caption{The estimation and partition selection performance of our method run with $\lambda = 100$ and several competitors in the moderate cluster separation setting. Across all simulations, the estimation and prediction error of \texttt{SCC} was substantially greater than those of other methods and are not shown.}
\label{fig:sim_medsep_boxplot}
\end{figure}

For each of the three settings of cluster separation, we generated 100 synthetic datasets.
Figure~\ref{fig:sim_medsep_boxplot} shows the overall performance of the methods considered in the medium separation setting (analogous figures for the high and low separation setting may be found in Section~S3.2 of the Supplementary Materials).

Across our simulations, \texttt{PartOpt} obtained consistently lower RMSE and prediction error than the competitors in each of the high, medium, and low separation settings. 
In all of our experiments, the estimation and prediction error of \texttt{SCC} was several times larger than that of every other method; indeed, in the medium separation setting, the estimation error of \texttt{SCC} was between 2.7 and 3.0 while the prediction error was between 4.3 and 5.4.
On further inspection, we found that the clusters recovered by \texttt{SCC} were quite different from the clusters identified by any of the other methods.
Often, \texttt{SCC} would cluster together spatial units with very different true parameter values, introducing substantial bias in the parameter estimates.
Interestingly, while \texttt{KM} achieved smaller estimation and prediction than \texttt{SC} in both the high and medium separation setting, \texttt{SC} performed better in the low separation setting, when the cluster structure was much less distinctive.
As the data in our main simulation experiments was generated from a Gaussian likelihood, the Poisson regression model fit by \texttt{And} is mis-specified.
As a result, its parameter estimates and one-step-ahead predictions are somewhat worse than those of \texttt{PartOpt}, which is well-specified.
Interestingly, in our second set of experiments, where we generated crime counts from a Poisson regression model and then ran \texttt{PartOpt} on the transformed densities, \texttt{PartOpt} often outperformed the correctly specified \texttt{And} (see Section~S3.3 of the Supplementary Materials). 

In the high separation setting, each of \texttt{And}, \texttt{KM}, and \texttt{PartOpt} consistently recovered the true partitions.
This is, in-and-of-itself, not especially surprising: in the high separation setting, the cluster structure is obvious upon visual inspection of the MLEs of $\balpha$ and $\bbeta.$
Nevertheless, it is interesting to note that both \texttt{SC} and \texttt{SCC} perform quite poorly in terms of recovering the true partitions in this setting (see Figure~S2 in the Supplementary Materials).
In the medium and low separation setting, when the true cluster structure is less distinctive, the approximate posterior distribution over $\bgamma$ identified by \texttt{PartOpt} tends not to place much mass near the true partitions.
On further inspection, we found that in these two settings, the true partitions that generated the data had considerably less posterior probability than the majority of the particles identified by \texttt{PartOpt}.
This is also not especially surprising: when the data-generating parameters formed clusters which were all quite similar, the posterior favored forming a single big cluster rather than many smaller clusters with similar parameter values.

\section{Clustering Crime Dynamics in Philadelphia}
\label{sec:philly_example}

As described in Section~\ref{sec:data_model}, we model the transformed density of violent crimes $y_{i,t}$ in neighborhood $i$ at time $t$ as $y_{i,t} \sim \mathcal{N}(\alpha_{i} + \beta_{i}x_{t}, \sigma^{2}).$
We further wish to identify two partitions of neighborhoods: one, $\gamma^{(\alpha)},$ that clusters together neighborhoods with similar mean levels of crime $\alpha_{i}$, and the other, $\gamma^{(\beta)},$ that clusters together neighborhoods with similar time trends $\beta_{i}.$ 
For our analysis of the Philadelphia crime data, we ran \texttt{PartOpt} using $L = 20$ particles and with $\lambda = 100.$
Like in our simulation experiments, we fixed the hyperparameter $\eta = 1$ in the truncated Ewens-Pitman priors on $\gamma^{(\alpha)}$ and $\gamma^{(\beta)}.$

\begin{figure}[h]
\centering
\includegraphics[width = 0.455\textwidth]{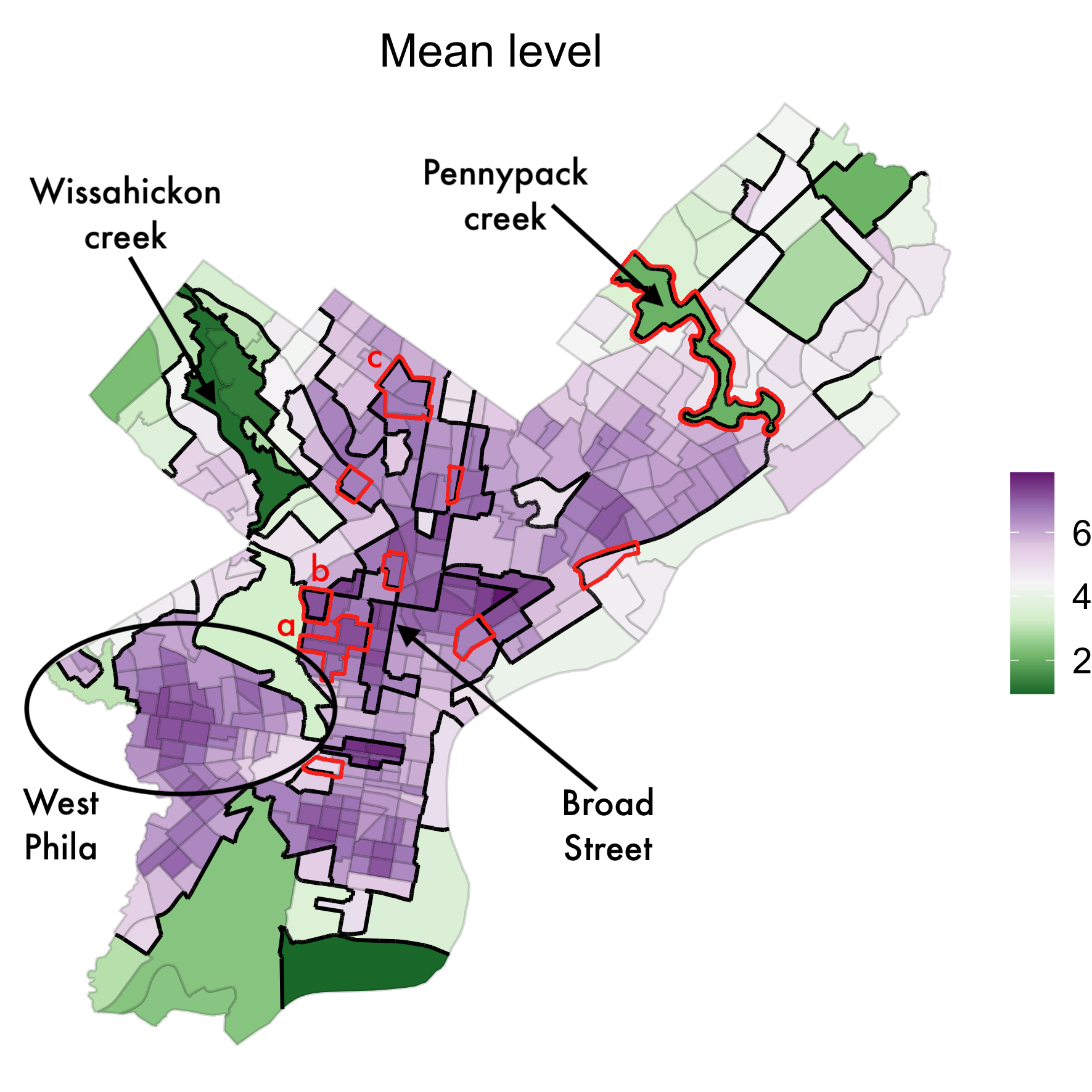}~
\includegraphics[width = 0.455\textwidth]{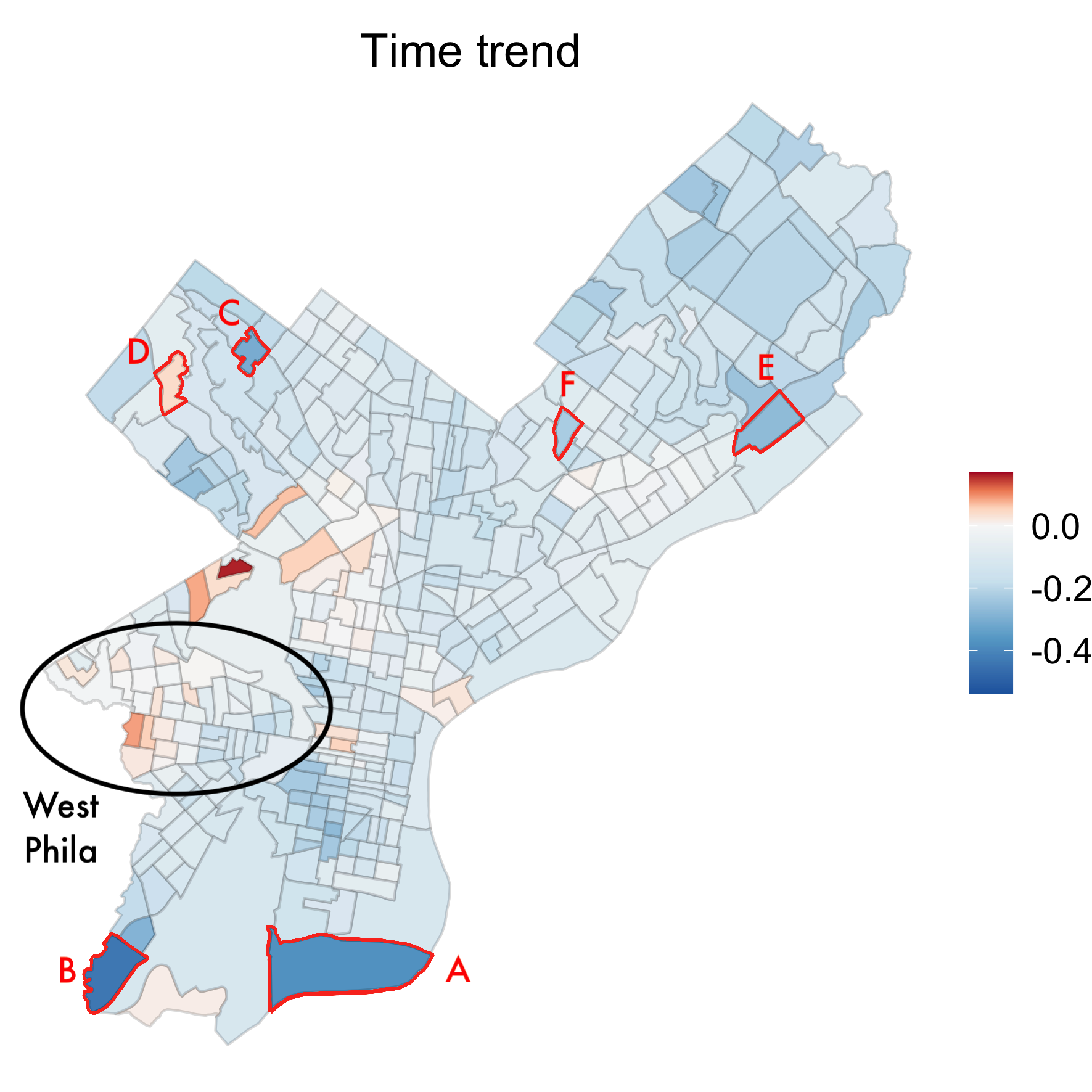}
\caption{{Top particle identified by \texttt{PartOpt}. The thick black lines delineate the borders between the clusters, and the color represents the posterior mean of $\balpha$ and $\bbeta$ given the top particle. Regions highlighted in red have different cluster assignment in the particle set.}}
\label{fig:best_part}
\end{figure}

\textbf{Identified particles}.
Figure~\ref{fig:best_part} shows the top particle recovered by \texttt{PartOpt}.
We highlighted the borders between clusters with thick black lines and each neighborhood's color correspond to the conditional parameter estimate given the partition.

Immediately we notice that while the top partition for the mean level of crime density $\gamma^{(\alpha)}$ contains many clusters, the top partition for the time trends $\gamma^{(\beta)}$ contains only one cluster.
We observe that it contains many neighborhoods which experienced small decreases in crime density and a handful of neighborhoods which experienced small increases in crime density.
We observe that the overall range of $\beta_{i}$'s is quite small, suggesting that the within-cluster variation is not large enough to overcome our prior's regularization towards a small number of large clusters. 
Indeed, as we discuss in more detail in Section~S4.2 of the Supplementary Materials, the posterior distribution of $\bgamma$ is somewhat sensitive to the choice of prior. 

The remaining nineteen particles identified by \texttt{PartOpt} are quite similar to the top particle shown in Figure~\ref{fig:best_part}.
In fact, the twelve particles with next highest importance weights had identical $\gamma^{(\beta)}$'s as the top particle.
The particles with second through fifth highest importance weight differs from the first particle in terms of the cluster assignment of the sets of neighborhoods labelled \textbf{(a)}, \textbf{(b)} and \textbf{(c)} in $\gamma^{(\alpha)}$, and the sixth through twelfth particles differ from the first in terms of the other neighborhoods highlighted in red.

The thirteenth through the twentieth particles have identical $\gamma^{(\alpha)}$'s, which differ from the partition shown in Figure~\ref{fig:best_part} in the assignment of \textbf{(a)}, but have different $\gamma^{(\beta)}$'s.
In the thirteenth particle, the neighborhood labelled \textbf{(A)} is removed from the large cluster and moved into a singleton cluster on its own.
The remaining particles respectively separate the neighborhoods \textbf{(B)}, \textbf{(C)}, \textbf{(D)}, \textbf{(E)}, and \textbf{(F)} into their own clusters.

\textbf{Analysis of the top particle}.
In Figure~\ref{fig:best_part}, we can also immediately recognize many interesting features of Philadelphia's built environment and geography.
The clusters labelled \textbf{Wissahickon creek} and \textbf{Pennypack creek} respectively correspond to the parks surrounding these two rivers.
In light of this, it is not especially surprising that these areas have somewhat lower baseline levels of crime than their surrounding neighborhoods.
We also observe several cluster borders coinciding along with sections of the major arterial road Broad Street (labelled \textbf{Broad Street}).

Figure~\ref{fig:best_part} also reveals an interesting pattern of crime dynamics in the West Philadelphia and University City region (circled in black and labelled \textbf{West Phila} in the figure).
This region contains both Drexel University and the University of Pennsylvania. 
For the most part, the region is characterized by relatively high crime density (darker shades of purple in the left panel of the figure) with three notable exceptions on the east and southeast areas of the region.
These three clusters, which are colored in lighter shades, are adjacent to university buildings and student housing. 
We also see a generally decreasing trend in crime in the vicinity of the universities and a slightly increasing trend further away from the universities. 
This finding aligns with previous reports of the positive impact of the University of Pennsylvania's West Philadelphia Initiatives aimed at improving the social and economic landscape around the university campus \citep{ehlenz2016neighborhood}.


\textbf{Predictive performance}. Table~\ref{table:outofsample} reports the root mean square error for predicting crime density in each neighborhood in 2018. 
We compare the performance of \texttt{PartOpt}, which averages over all of identified particles, with the MAP plug-in estimator. 
We compare it with the forecasts made by a method that does not impose any shrinkage or clustering and instead makes predictions based only on the maximum likelihood estimates of $\balpha$ and $\bbeta;$ we also compare it with the competitors considered in Section~\ref{sec:illustration}.

\begin{table}[H]
\centering
\caption{Out-of-sample RMSE for predicting the transformed crime density in 2018 based on models fit to data from 2006 -- 2017.} 
\label{table:outofsample}
\begin{tabular}{ccccccc}
\hline
MAP & \texttt{PartOpt} & \texttt{MLE} & \texttt{And} & \texttt{KM} & \texttt{SC} & \texttt{SCC} \\ \hline
0.2300 & 0.2299  & 0.2349 & 0.2395 & 0.2341 & 0.2401 & 5.970 \\ \hline
\end{tabular}
\end{table}

Recall that the twenty identified particles were all quite similar to the top particle shown in Figure~\ref{fig:best_part}.
Thus, it is not entirely surprising to see that MAP plug-in predictions are quite similar to the model averaged predictions made by \texttt{PartOpt}.
Nevertheless, we find that by averaging over more uncertainty about the latent partitions, \texttt{PartOpt} achieved slightly better predictive performance.
We find that \texttt{And} and \texttt{SC} perform worse than the simple MLE predictions.
Interestingly, \texttt{SCC} recovers a single cluster of time trends $\beta_{i}$, like the top particle shown in Figure~\ref{fig:best_part}.
However, \texttt{SCC}'s estimates of the mean levels $\alpha_{i}$ are substantially smaller than both the MLE and \texttt{PartOpt}'s estimates, yielding rather poor predictions.

\section{Discussion}
\label{sec:discussion}
Accurate estimation of the change in crime over time is a critical first step towards a better understanding of public safety in large urban environments. 
An especially important challenge to such estimation is the potential presence of sharp discontinuities, which may be smoothed over by naive spatial shrinkage procedures.
Focusing on the city of Philadelphia, we introduced a Bayesian hierarchical model that naturally identifies these discontinuities by partitioning the city into several clusters of neighborhoods and introduces spatial smoothness within but not between clusters.
In particular, we focused on recovering two latent spatial partitions, one for the approximate baseline level of crime over the twelve year period 2006--2017 and one for the approximate time-trend.

Rather than use a computationally prohibitive stochastic search, we identified partitions with highest posterior probability by solving a single optimization problem.
We showed that optimizing the proposed objective function is formally equivalent to finding a particular variational objective and introduced a local search strategy for solving this problem.
While our primary focus has been on crime in the city of Philadelphia, our ensemble optimization framework is more general and there are a number of areas of future development, which we discuss below. 

It is possible to run our particle optimization procedure at both higher and lower spatial resolutions. 
It would be interesting to run our procedure with data aggregated at the census block group level to reveal heterogeneity in crime dynamics within census tracts. 
Doing so would require considerably more computational resources, as there are 1336 block groups.

Though we have not done so in this paper, it is also possible to adjust for important neighborhood-level covariates within our framework.
For instance, one may extend Equation~\eqref{eq:tract_model} and model $y_{i,t} \sim \N(\alpha_{i} + \beta_{i}x_{t} + \bz_{i,t}^{\top}\theta, \sigma^{2})$ where $\bz_{i,t}$ is a vector of possibly time-varying covariates. 
Modifying our implementation of \texttt{PartOpt} for this purpose is relatively straightforward with a conditionally conjugate prior on $\theta$; it essentially amounts to writing a new function to compute the marginal likelihood $p(\by \vert \bgamma).$
Alternatively, one could replace the Ewens-Pitman priors on the latent partitions with priors that encourage neighborhoods with similar covariates to cluster together.
Many such covariate-dependent clustering priors have been introduced in the past \citep{ParkDunson2010, Muller2010, Wade2014} and it would be straightforward to include support for these priors in our implementation of \texttt{PartOpt}.
Profile regression \citep{molitor2010bayesian} incorporates covariates via direct regression adjustment and through the prior on clusters.
It does not, however, allow different regression parameters to cluster differently nor does it permit parameters to vary within clusters.
In spatial contexts, profile regression may not produce spatial clusters.


Although we have focused on identifying separate partitions of the mean levels and time trends, we believe that there are situations in which it is more desirable to identify clusters defined by both parameters.
For instance, policymakers may be interested in identifying clusters of neighborhoods which have both high mean levels of crime and increasing time trends.
In Section~S3.5 of the Supplementary Materials, we discuss how we can accomplish this with a small modification to our implementation of \texttt{PartOpt} that constrains $\gamma^{(\alpha)} = \gamma^{(\beta)}.$ 

Although linear approximations to the true regression functions $f_{i,0}$ were reasonable in our dataset, in datasets with stronger suggestions of non-linearity and more observations per census tract, it is reasonable to consider higher-order approximations of $f_{i,0}.$
With conditionally conjugate Gaussian priors on the tract-level parameters, it is still feasible to compute the marginal likelihood $p(\by \mid \bgamma)$ required by particle optimization.
We sketch such extensions in Section S4.1 of the Supplementary Materials

Finally, while our analysis has focused on modeling crime densities, it is natural to wonder whether \texttt{PartOpt} can be extended to modeling crime counts with Poisson or negative binomial regressions.
Our local search algorithm involves repeated calculation of the marginal likelihood $p(\by \mid \bgamma).$
Unfortunately, for Poisson and negative binomial regression with CAR--within--clusters prior, these likelihoods are not available in closed form.
Nevertheless, in small-scale experiments, we have found that running \texttt{PartOpt} with \textit{approximate} marginal likelihoods computed using Laplace approximations works rather well. 
We discuss these approximations and report the preliminary results of an approximate \texttt{PartOpt} for Poisson regression in Section S6 of the Supplementary Materials.

\bibliography{PartPhilly_bib}
\newpage
\begin{center}
{\Large {\bf Supplementary Materials for }}

\bigskip

{\Large {\bf ``Crime in Philadelphia: Bayesian Clustering with Particle Optimization"}}

\bigskip

\end{center}

\setcounter{equation}{0}
\setcounter{figure}{0}
\setcounter{table}{0}
\setcounter{page}{1}
\setcounter{section}{0}
\setcounter{theorem}{0}
\makeatletter
\renewcommand{\theequation}{S\arabic{equation}}
\renewcommand{\thefigure}{S\arabic{figure}}
\renewcommand{\bibnumfmt}[1]{[S#1]}
\renewcommand{\citenumfont}[1]{S#1}

\section{Proof of Proposition 1}
\label{app:proof}
In this Section 3.1 we state that we can find the set of $L$ particles with largest posterior by finding a variational approximation of the tempered posterior $\Pi_\lambda$. Here we restate Proposition 1 and provide the proof. 

Remember that we denote with $\Gamma_{L} = \{\bgamma^{(1)}, \ldots, \bgamma^{(L)}\}$ the set of $L$ particles with largest posterior mass, with $q(\cdot \mid \Gamma, \bw)$ the discrete distribution that places probability $w_{\ell}$ on the particle $\bgamma_{\ell}$ and with $\mathcal{Q}_{L}$ the collection of all such distributions supported on at most $L$ particles. 
Moreover, for each $\lambda > 0$, let $\pi_{\lambda}$ be the mass function of the tempered marginal posterior $\Pi_\lambda$, where $\pi_{\lambda}(\bgamma) \propto \pi(\bgamma \mid \by)^{\frac{1}{\lambda}}.$

\begin{proposition}
Suppose that $\pi(\bgamma \mid \by)$ is supported on at least $L$ distinct particles and that $\pi_{\lambda}(\bgamma) \neq \pi_{\lambda}(\bgamma')$ for $\bgamma \neq \bgamma'.$ 
Let $q^{\star}_{\lambda}(\cdot | \Gamma^{\star}(\lambda), \bw^{\star}(\lambda))$ be the distribution in $\mathcal{Q}_{L}$ that is closest to $\Pi_{\lambda}$ in a Kullback-Leibler sense:
$$
q^{\star}_{\lambda} = \argmin_{q \in \mathcal{Q}_{L}}\left\{\sum_{\bgamma}{q(\bgamma)\log{\frac{q(\bgamma)}{\pi_{\lambda}(\bgamma)}}}\right\}.
$$
Then $\Gamma^{\star}(\lambda) = \Gamma_{L}$ and for each $\ell = 1, \ldots, L,$ $w_{\ell}^{\star}(\lambda) \propto \pi(\bgamma^{(\ell)} | \by)^{\frac{1}{\lambda}}$
\end{proposition}

\begin{proof}

Denote the optimal particles $\Gamma^{\star}(\lambda) = \left\{\bgamma^{\star}_{1}, \ldots, \bgamma^{\star}_{L^{\star}}\right\}.$ 
Straightforward calculus verifies that $w^{\star}_{\ell}(\lambda) \propto \pi_{\lambda}(\bgamma^{\star}_{\ell}).$
We thus compute
$$
\text{KL}(q^{\star} \parallel \pi_{\lambda}) = \sum_{\bgamma}{q^{\star}(\bgamma)\log{\frac{q^{\star}(\bgamma)}{\pi_{\lambda}(\lambda) }}} 
= -\log{\Pi_{\lambda}(\Gamma^{\star}(\ell))}
$$

Since $\Pi_{\lambda}$ is supported on at least $L$ models, we see from this computation that if $\Gamma^{\star}$ contained fewer than $L$ particles, we could achieve a lower Kullback-Leibler divergence by adding another particle $\tilde{\bgamma}$ not currently in $\Gamma^{\star}$ that has positive $\Pi_{\lambda}$-probability to the particle set and updating the importance weights $\bw$ accordingly. 

Now if $\Gamma^{\star}$ contains $L$ models but $\Gamma^{\star}(\lambda) \neq \Gamma_{L},$ we know $\Pi_{\lambda}(\Gamma^{\star}(\lambda)) < \Pi_{\lambda}(\Gamma_{L}).$
Thus, replacing $\Gamma^{\star}(\lambda)$ by $\Gamma_{L}$ and adjusting the importances weights accordingly would also result in a lower Kullback-Liebler divergence.
\end{proof}

\section{Various hyper-parameter choices}
\label{app:hyper}

The main model described in Section 2 depends on several hyper-parameters, which need to be fixed by the practitioner: the parameters for the prior for $\sigma^2$ ($\nu_\sigma$ and $\lambda_\sigma$) and the multiplicative constants to specify within and between cluster variance ($a_1, a_2, b_1$ and $b_2$).
We will now describe the data-dependent approach to specify such values.

Let us consider each neighborhood separately and fit a linear regression model for each one: let $\hat{\alpha}_i$ and $\hat{\beta}_i$ be the least square estimates and $\hat{\sigma}^2_i$ be the estimated residual variance for neighborhood $i$. 

We can use the collection of $\hat{\sigma}^2_i$'s to specify the prior for $\sigma^2$: by matching mean and variance, we can recover $\nu_\sigma = 2 \frac{m^2}{v} + 4$ and $\lambda_\sigma =m (1-\frac{2}{\nu_\sigma})$, where we denote with $m$ and $v$ the empirical mean and variance of the $\hat{\sigma}^2_i$'s.

To specify the within and between cluster variance parameters we use a two-step heuristic: at a high level, we first find a temporary estimate of the hyperparameters $a_1, a_2, b_1$ and $b_2$, based on the MLE estimate of $\alpha_i$ and $\beta_i$ and on the expected number of clusters; we then recover the maximum a posteriori (MAP) partition under these values and find the empirical Bayes estimate of $a_1$ and $b_1$ given the MAP. 
These values are finally used to run our full Particle Optimization procedure, which can be initialized from the MAP partition recovered in the first step of the heuristic.

Specifically, we consider the least square estimates $\hat{\alpha}_i,\hat{\beta}_i$, which can be thought of as an approximation of $\alpha_i, \beta_i$ given the partition with $N$ clusters $\gamma_N$, since they do not incorporate any prior information or sharing of information; in fact under such configuration the coefficients are exchangeable and the only shrinkage induced is through the common variance parameter.
Given this, one heuristic desideratum is that the marginal prior on $\balpha \mid \gamma = \gamma_{N}$ should assign substantial probability to range of the $\hat{\alpha}_{i}$, assuming symmetry around zero.
Specifically, we will make sure that this conditional prior places 95\% of its probability over the range of the $\hat{\alpha}_{i}$'s.
Since $\balpha \mid \gamma = \gamma_{N} \sim N(0, \sigma^{2}(a_{1}/(1 - \rho) + a_{2})I_{n}),$ we constrain $a_{1}$ and $a_{2}$ so that
$$
\frac{a_1}{1-\rho}+a_2 = \frac{\max_i \vert \hat{\alpha}_i \vert^2}{4\hat{\sigma^2}}.
$$
When the MLE's are not symmetric around zero the prior probability on the range of $\hat{\alpha}_{i}$'s will be smaller than $95\%$, but at least each point in the range has prior density higher than $\phi(2).$

In order to determine each of $a_{1}$ and $a_{2},$ we need a second constraint.
To this end, consider the highly stylized setting in which we have $K$ overlapping clusters with equal variance $\sigma^{2}_{\rm cl}$ whose means are equally spaced at distance $2\sigma_{cl}.$
The idea of this second heuristic is to match such a stylized description to the observed distribution of $\hat{\alpha}_{i}.$
In essence, this involves covering the range of $\hat{\alpha}_{i}$ with $K+1$ ``chunks'' of length $2\sigma_{\rm cl}.$
While the exact value of $\sigma_{\rm cl}$ is unknown, we have found it useful to approximate it $a_{1}\sigma^{2}/(1 - \rho).$
This approximation tends to produce smaller values of $a_{1},$ which in turn encourages a relatively larger number of clusters.

With these two constraints we can find the temporary values:
\begin{align*}
a_1 &= \frac{(\max(\hat{\alpha}_i) - \min(\hat{\alpha}_i))^2}{4(K+1)^2\hat{\sigma}^2/(1-\rho)}\\
a_2 &= \frac{\max_i \vert \hat{\alpha}_i \vert^2}{4\hat{\sigma^2}} - \frac{a_1}{1-\rho}.
\end{align*}
Similarly for the $\hat{\beta}_i$'s we find:
\begin{align*}
b_1 &= \frac{(\max(\hat{\beta}_i) - \min(\hat{\beta}_i))^2}{4(K+1)^2\hat{\sigma}^2/(1-\rho)}\\
b_2 &= \frac{\max_i \vert \hat{\beta}_i \vert^2}{4\hat{\sigma^2}} - \frac{b_1}{1-\rho}.
\end{align*}

In order to operationalize these heuristics, we must specify an initial guess at $K.$
We have found in our experiments, setting $K = \lfloor \log{N}\rfloor$ works quite well.
It, moreover, accords with the general behavior of the Ewens-Pitman prior.


We now use these values to find the MAP partition $\bgamma^{(1)}$ (we can run our Particle Optimization procedure with $L = 1$) and find the Empirical Bayes estimates of $a_1$ and $b_1$ given $\bgamma^{(1)}$ and the other hyperparameter estimates, i.e. we find
$$
(\hat{a}_1,\hat{b}_1) = \argmax_{a_1,b_1} p(Y \vert a_1, b_1, a_2, b_2, \nu_\sigma,\lambda_\sigma,\bgamma^{(1)})
$$
using a numerical optimization algorithm.


Note that finding the MAP as part of our heuristic procedure does not increase the computational burden. In fact, even though it requires us to run the Particle Optimization algorithm twice (the first time with $L = 1$), the output from the first run can be used as a starting point in the initialization of the second run - together with the partitions recovered by running k-means on the maximum likelihood estimates.
We empirically see that the MAP recovered under the temporary hyperparameters has always higher posterior probability than other initializing  partitions.
Consequently, when we run our Particle Optimization procedure for the second time, we start from a point with high posterior probability, harnessing the work of the initial MAP search.

\section{Additional Synthetic Data Evaluation}
\label{app:additional_simulation_results}

\subsection{Synthetic data description}
\label{sec:sim:descr}

In Section 4, we generated several synthetic datasets based on a $20 \time 20$ grid of spatial units, given the true partitions $\tilde{\gamma}^\alpha$ and $\tilde{\gamma}^\beta$ and with various levels of cluster separations, as displayed in Figure~3. 

We now describe in details how the synthetic data was generated. Each cluster separation setting corresponds to a pair of values $(\Delta_\alpha, \Delta_\beta)$, which are set to $(\Delta_\alpha, \Delta_\beta) = (2, 1)$ in the high separation setting, to $(\Delta_\alpha, \Delta_\beta) = (1, 0.5)$ in the moderate separation setting and to $(\Delta_\alpha, \Delta_\beta) = (0, 0)$ in the low separation setting.
These values are used to construct the cluster-specific means, $\alphabar_{k} = 3.5 + \overline{a}_k \cdot \Delta_\alpha$ and $\betabar_{k} = \overline{b}_k \cdot \Delta_\beta$, where $\overline{a}_k \in \{-1, -0.5, 0, 0.5, 1\}$ and $\overline{b}_k \in \{-1, 0, 1\}$.
Within each cluster, we drew the $\alpha_{i}$'s (respectively $\beta_i$'s) from a CAR model centered at a specified cluster mean with $\rho = 0.95$, hyper-parameters $a_1 = b_1 = 0.125$ and standard deviation $0.25$.

Since the values of $\alphabar_k$ and $\betabar_k$ were artificially chosen and not sampled, there are no exact values for $a_2$ and $b_2$, but we can compute a lower bound for these hyperparameters so that the prior distributions of $\alphabar_k$ and $\betabar_k$ cover the values artificially chosen. If we require that the values of $\alphabar_k$ and $\betabar_k$ fall within the 0.025 and 0.975 quantiles of the prior distribution, we find that $a_2 \geq 9$ and $b_2 \geq 2$ for the high separation setting. In the moderate separation setting we find $a_2 \geq 8$ and $b_2 \geq 1$, while for the low separation setting $a_2 \geq 7$ and $b_2 \geq 0$. Note that the larger values of $a_2$ are due to the shift of $3.5$ in the construction of $\alphabar_{k}$, which is important to generate data that mimics the real data which has a positive mean level of crime. 

For each cluster separation setting we generated 100 pairs of vectors $(\balpha, \bbeta)$. For each set of parameters we generated the outcomes $y_{i,t} \sim N(\alpha_{i} + \beta_{i}x_{t}, \sigma^{2})$, where $x_t = (t - \mu_T)/\sigma_T$ is the time index standardized to have mean zero and unit variance. Since we used $t = 2006, \ldots, 2017$, we had $\mu_T = 2011.5$ and $\sigma_T = 3.6$.

We also considered a second data generating process, in which $c_{i,t} \sim {\rm Pois}(\exp(\alpha_{i} + \beta_{i}x_{t}))$.

\subsection{Additional cluster separation settings}

In Section~4, we compared the estimation, prediction and partition selection performance of our method to that of several competitors and we displayed results for the moderate separation setting. Specifically, for the estimation and prediction performance we displayed the root mean squared error (RMSE) for estimating the concatenated vector of parameters $(\balpha, \bbeta)$ and the RMSE for the vector of one-step-ahead observations $\by_{T+1}$ generated from the same model. For the partition selection performance we compared the adjusted Rand index between the true partitions $\tilde{\gamma}^\alpha$ and $\tilde{\gamma}^\beta$ and the recovered one.

\begin{figure}[H]
\centering
\includegraphics[width = 0.9\textwidth]{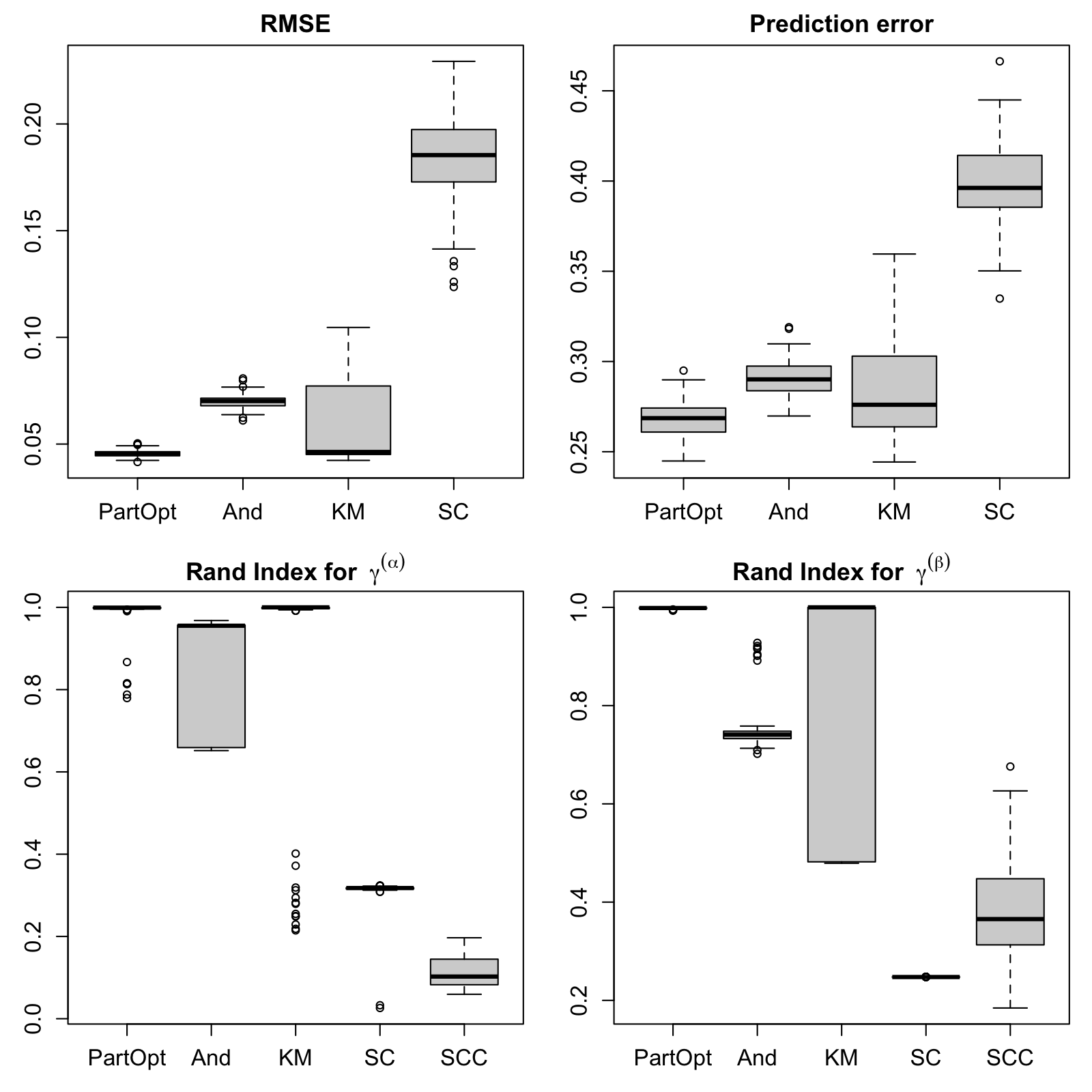}
\caption{The estimation and partition selection performance in the high cluster separation setting.}
\label{fig:sim_highsep_boxplot}
\end{figure}

\begin{figure}[H]
\centering
\includegraphics[width = 0.9\textwidth]{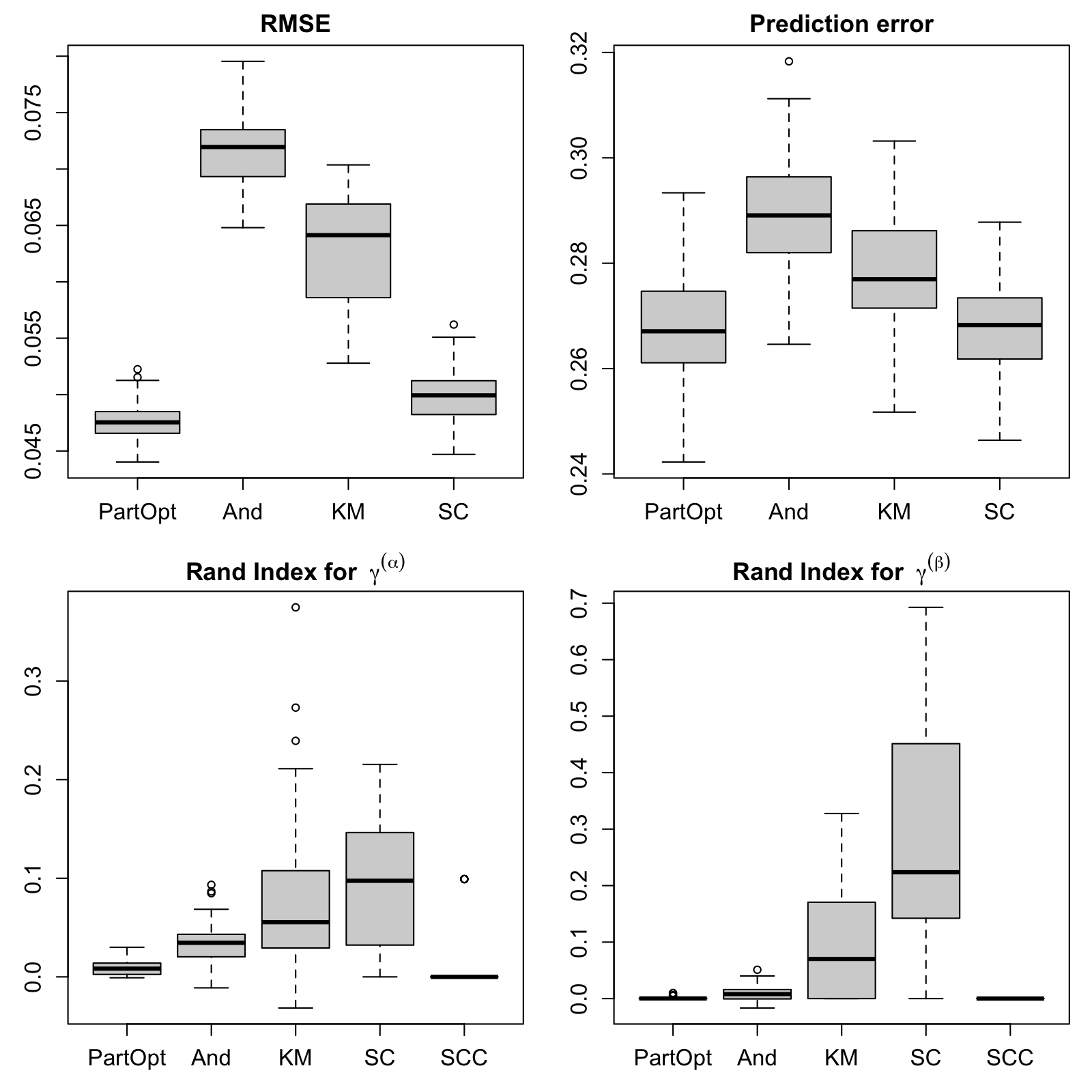}
\caption{The estimation and partition selection performance in the low cluster separation setting.}
\label{fig:sim_lowsep_boxplot}
\end{figure}

We now report the same measures for the additional cluster separation settings in Figure~\ref{fig:sim_highsep_boxplot} and Figure~\ref{fig:sim_lowsep_boxplot}. As in Section~4, we do not show the RMSE and Prediction error for the \texttt{SCC} method, since they were substantially greater than those of other methods (RMSE ranged between 3.9 and 4.3 in the high separation setting and between 2.4 and 3.7 in the low separation setting; prediction error ranged between 7.7 and 8.8 in the former and between 3.2 and 3.7 in the latter).

Similarly to the moderate cluster separation setting, we see that \texttt{PartOpt} performs better in terms of estimation and predictive performance, with no other method consistently achieving second best. In fact \texttt{KM} performs almost as well as \texttt{PartOpt} in the high separation setting, while \texttt{SC} achieves second best in the low separation setting.
In terms of selection performance, in the high separation setting, \texttt{PartOpt} recovers the true partitions almost always exactly, but \texttt{And} and \texttt{KM} also perform quite well; \texttt{SC} and \texttt{SCC} instead recover partitions that are quite different from the true one.
In the low separation setting instead, all method have low values for the adjusted Rand index; in fact, when there is no difference between the cluster means the true partitions $\tilde{\gamma}^\alpha$ and $\tilde{\gamma}^\beta$ lose meaning, and we expect the methods to recover the partitions with only one cluster.

\begin{figure}
\centering
\includegraphics[width = 0.95\textwidth]{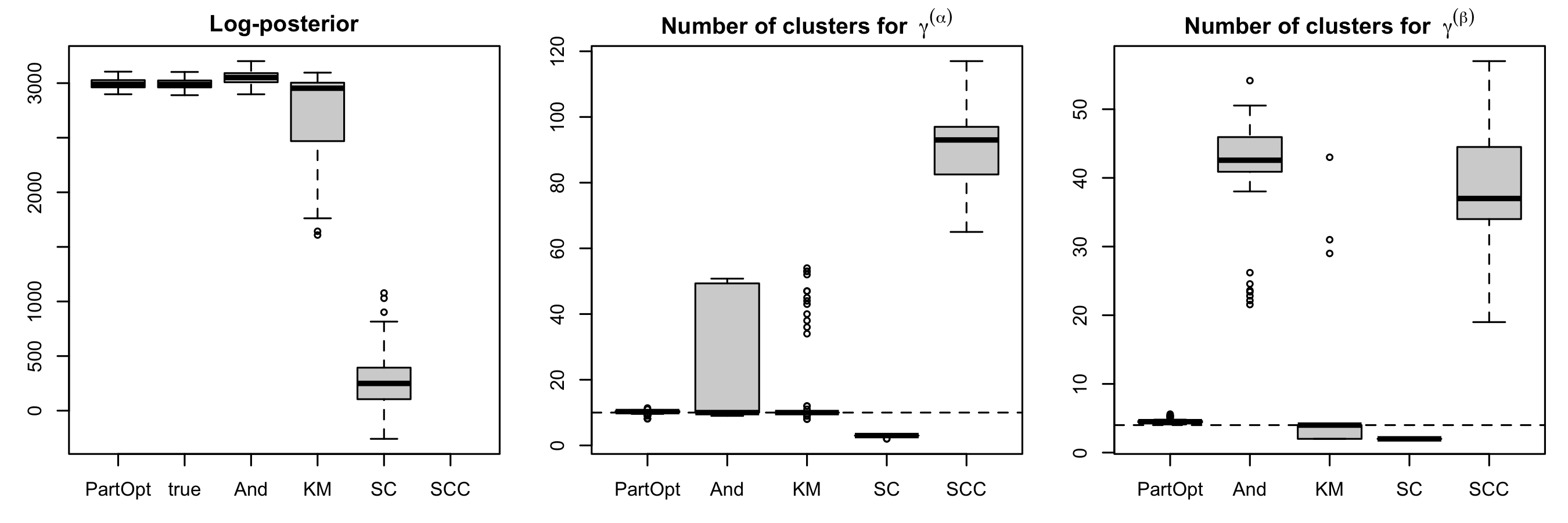}\\
\includegraphics[width = 0.95\textwidth]{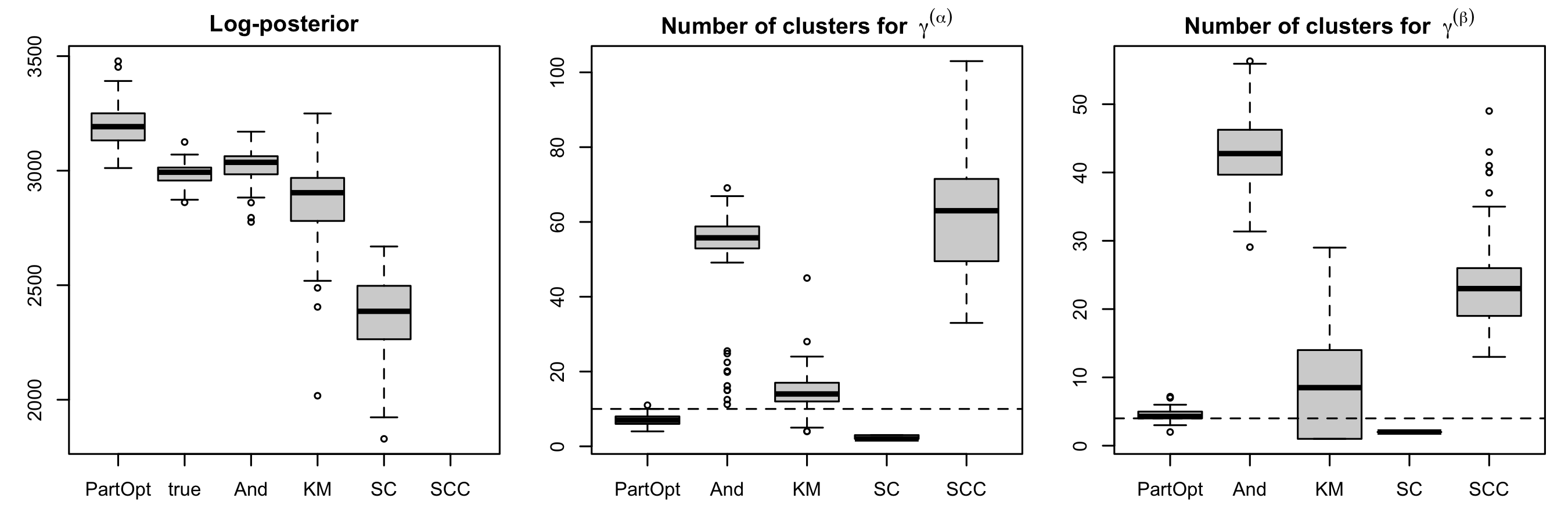}\\
\includegraphics[width = 0.95\textwidth]{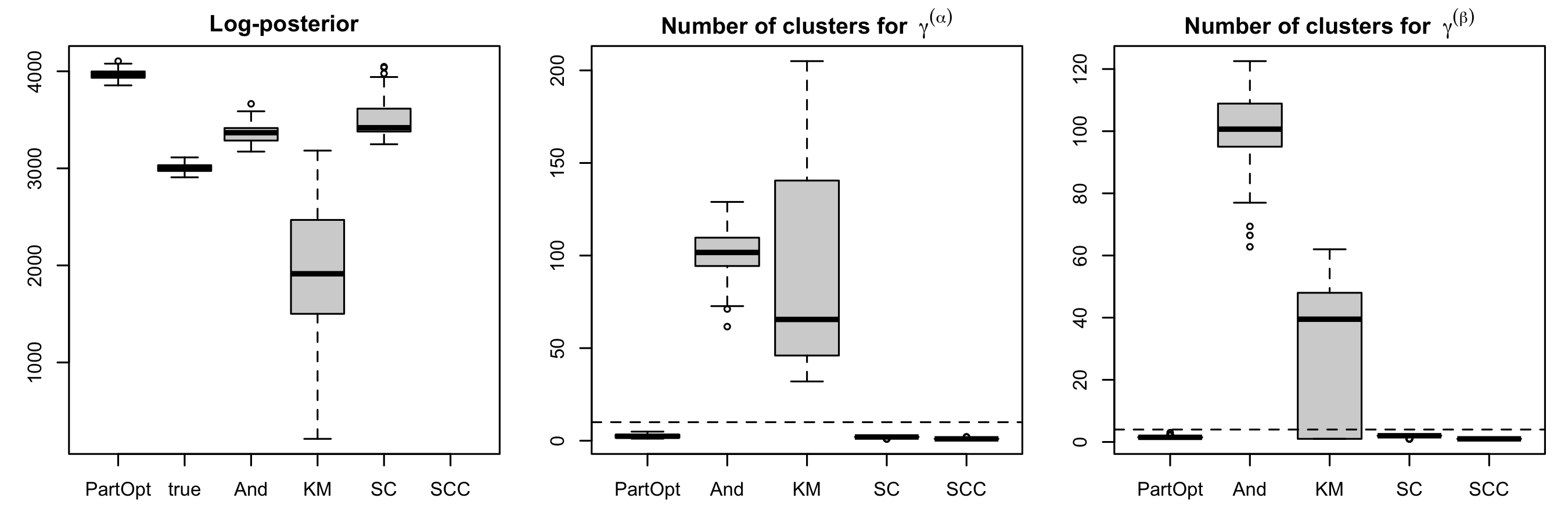}
\caption{Additional partition selection measures in several cluster separation settings. Top row: high cluster separation. Medium row: moderate cluster separation. Bottom row: low cluster separation.}
\label{fig:sim_boxplot2} 
\end{figure}

To better investigate some of these issues, we report in Figures~\ref{fig:sim_boxplot2} 
some additional measures of partition selection: the log-posterior of the recovered pair of partitions, computed under the model described in Section~2, the number of clusters for the recovered $\gamma^\alpha$ (K\_A) and $\gamma^\alpha$ (K\_B), together with the number of clusters of the true partitions, represented as horizontal dashed lines.

It's clear from these figures that even when the adjusted Rand index is low, the log-posterior of the partitions recovered by \texttt{PartOpt} is higher than the one of the true partitions. By examining the number of clusters of the recovered partitions, we see that \texttt{SC} always underestimate the number of clusters, suggesting the reason of the poor performance we had previously noticed. \texttt{SCC} and \texttt{And} instead often overestimate the number of clusters, but for different reasons. \texttt{And} in fact does not target spatial partitions, and we have to manually find the connected components, artificially inflating the number of clusters; however, while not spatial, the partitions recovered by \texttt{And} are not so distant from the true partitions and have relatively high log-posterior values. \texttt{SCC} instead targets spatial partitions, so no manual post-processing is necessary, but it's highly sensitive to the choice of spanning tree, resulting in low values of log-posterior.

\subsection{Second data generating process}

\begin{figure}
\centering
\includegraphics[width = \textwidth]{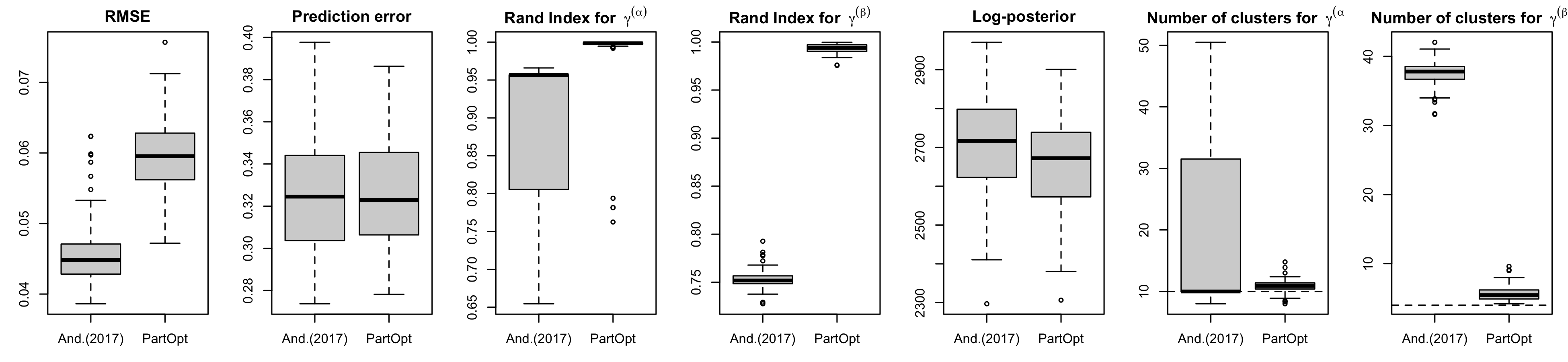}\\
\includegraphics[width = \textwidth]{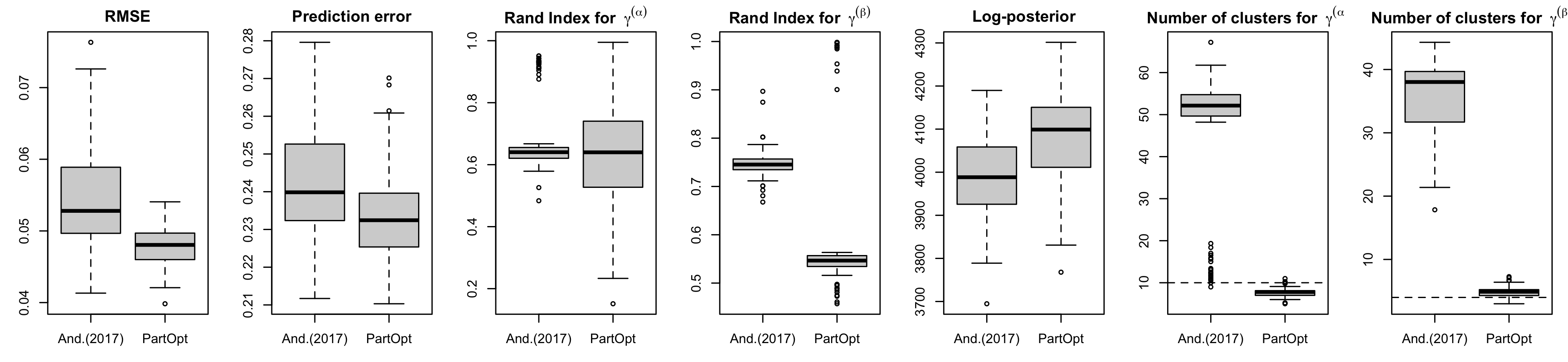}\\
\includegraphics[width = \textwidth]{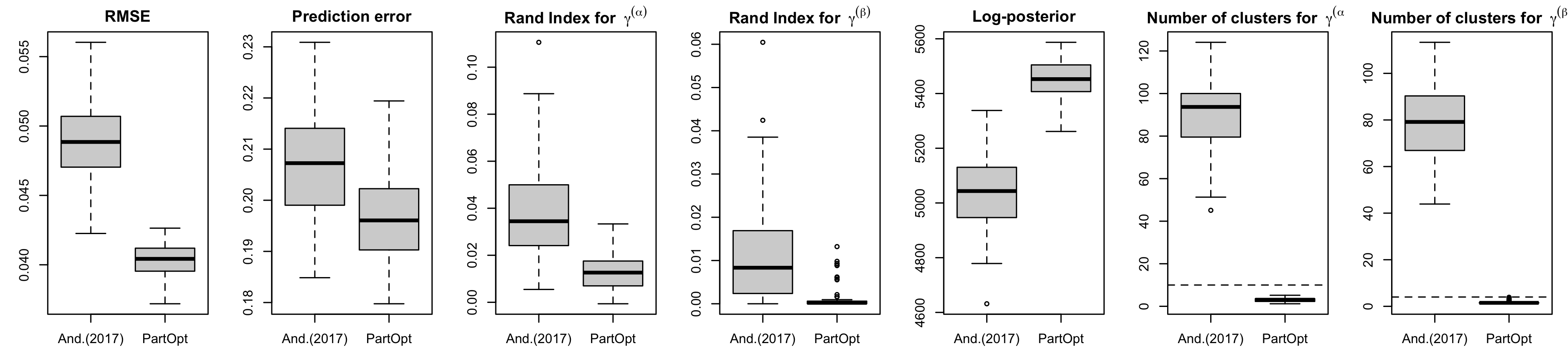}
\caption{\small{The estimation and partition selection performance under the second data generating process, for several cluster separation settings. Top row: high cluster separation. Medium row: moderate cluster separation. Bottom row: low cluster separation.}}
\label{fig:sim_boxplot_count} 
\end{figure}



So far we have compared the behavior of \texttt{PartOpt}, \texttt{And} and the other competitors under the data generating process suggested by our model in Section~2. However, the method by Anderson et al. (2017) is developed for count data, generated from a Poisson distribution. So we also considered a second data generating process, in which the count data $c_{i,t}$ is generated from a Poisson distribution with mean $\lambda_{i,t} = \exp(\mu_{i,t})$ and $\mu_{i,t} = \alpha_{i} + \beta_{i}x_{t}$, and the counts are transformed, as in equation (1): $y_{i,t} = \sinh^{-1} (c_{i,t}) - \log(2)$.
In Figure~\ref{fig:sim_boxplot_count} we report the results for the simulations under this second data generating process.

In high separation settings, \texttt{And} has better estimation performance than \texttt{PartOpt} (lower RMSE and higher log-posterior), even though the latter recovers the true partitions almost exactly. However, in moderate to low separation settings, \texttt{PartOpt} performs better than \texttt{And}, both in terms of estimation performance measures (lower RMSE and prediction error) and of log-posterior of the selected partitions, suggesting that \texttt{PartOpt} is robust to misspecification of the model.

\subsection{Sensitivity to hyperparameter choice}

In all our simulation settings and data analysis the spatial autocorrelation hyper-parameter $\rho$ is fixed and equal to $0.9$. This choice is motivated by our search for clusters that display a large spatial autocorrelation, without having to choose an improper prior for $\balpha$ and $\bbeta$.

We now explore how the results from our synthetic analysis change when we fix a different value for the hyperparameter $\rho$. In particular, we consider the moderate cluster separation setting and we separately fit our model with $\rho = 0.1, 0.5, 0.75, 0.95$ together with $\rho = 0.9$ which is the value we used in our main synthetic analysis.

\begin{figure}[H]
\centering
\includegraphics[width = 0.9\textwidth]{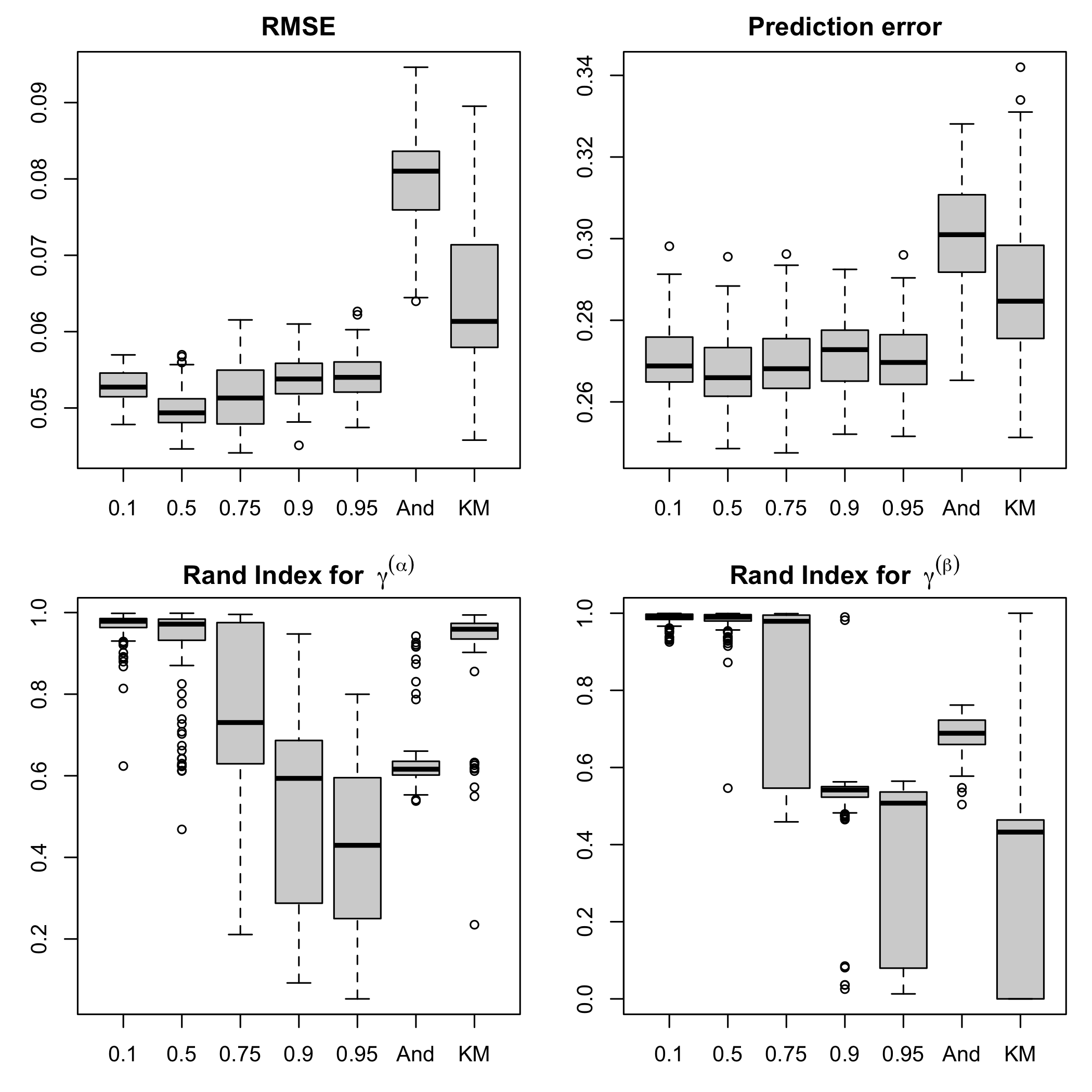}
\caption{The estimation and partition selection performance under the moderate cluster separation setting, for several specification of the hyperparameter $\rho$ in \texttt{PartOpt} (the labels report the value of $\rho$ for each specification). The performance of {\tt And} and {\tt KM} is also reported for reference.}
\label{fig:sim_boxplot_sensitivity} 
\end{figure}

Figure~\ref{fig:sim_boxplot_sensitivity} shows the estimation and partition selection performance of {\tt PartOpt} under the various values of $\rho$.
We first notice that there is quite some heterogeneity in the partition selection for different values of the hyperparameter $\rho$; in fact, for smaller values of $\rho$, such as $\rho=0.1$ and $\rho = 0.5$, \texttt{PartOpt} recovers partitions that are very close to the true partitions (with high adjusted Rand index values), while for larger values of $\rho$ it recovers partitions that are quite distant from the truth, similarly to what we discovered in our main synthetic analysis. 
Remember in fact that for each value of the hyperparameter the posterior distribution changes, and the particles that have largest posterior under different values of the hyperparameter will most likely not coincide.
However, it is reassuring that the particle recovered by \texttt{PartOpt} under each value of the hyperparameter has almost always larger posterior probability (under such value of $\rho$) than the particles recovered under different values of $\rho$ (results not shown).
Moreover, for all values of the hyperparameter, \texttt{PartOpt} achieves similarly good estimation and prediction performance, suggesting robustness with respect to the choice of $\rho.$

\subsection{Recovering equal partitions mean levels and time trends}

It is possible to adapt our procedure to the case where one is interested in recovering the same partition for the mean levels and the time trends, i.e. when $\gamma^\alpha = \gamma^\beta = \gamma$. In such case, our method approximates the posterior distribution $\pi(\gamma \vert \y)$ of one random partition that affects the distribution of both $\balpha$ and $\bbeta$.
We have implemented this version of PartOpt that constrains the two partitions to be equal (hereafter referred to as ``Equal Partition'' or \texttt{EqualPart}) and we have compared its performance to the unconstrained method that we have presented in the main manuscript, under the same synthetic data simulation described in Section \ref{sec:sim:descr}. Note that for this simulation study, there two true partitions used to generate the data were different. 

\begin{figure}[H]
\centering
\includegraphics[width = \textwidth]{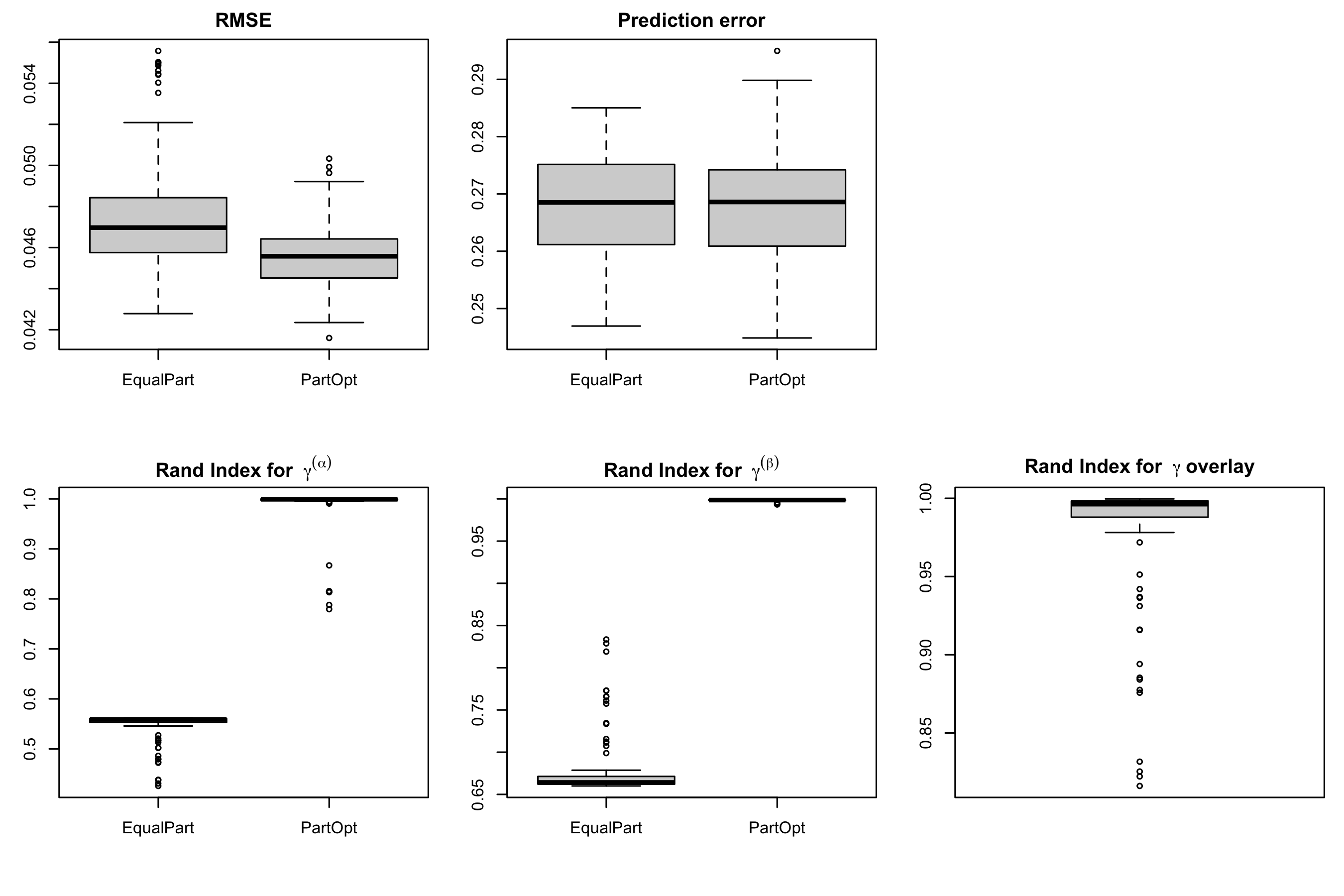}
\caption{The estimation and partition selection performance of \texttt{equalPart} under the high cluster separation setting. The performance of {\tt PartOpt} is also reported for reference. The panel ``Rand Index for $\gamma$ overlay'' plots the adjusted Rand index between the partition returned by \texttt{EqualPart} and the overlay of the two true latent partitions.}
\label{fig:sim_boxplot_equalpart} 
\end{figure}

Figure~\ref{fig:sim_boxplot_equalpart} reports the estimation and partition selection performance of \texttt{EqualPart} compared with the unconstrained \texttt{PartOpt} under the high cluster separation setting.

Interestingly, despite “Equal Partitions” being misspecified, we see that it has comparable estimation and prediction error as \texttt{PartOpt}. 
In terms of partition recovery, the adjusted Rand index between the partition estimated by \texttt{EqualPart} and the two partitions used to generate the data appears to be quite small. This is somewhat unsurprising:  our data was generated from a model with two latent partitions and ``Equal Partitions'' looks only for one. 
However, in virtually all of our simulation replications, the top partition recovered by \texttt{EqualPart} is quite close to the partition formed by ``overlaying'' the partitions in the top particle recovered by PartOpt. 
More specifically, this is the partition whose clusters are found as the pairwise intersection of clusters in the true partitions $\tilde{\gamma}^\alpha$ and $\tilde{\gamma}^\beta$. In the combinatorics literature this is known as the {\it meet} of the two partitions, which corresponds to the greatest lower bound of these partitions, under the partial order defined by the ``finer than'' relation.
The panel ``Rand Index for $\gamma$ overlay'' plots the adjusted Rand index between the partition returned by ``Equal Partitions'' and the overlay of the two true latent partitions. We see that the ``Equal Partitions'' routinely identified partitions close to the true overlay.

\section{Additional Results for Clustering in Philadelphia}
\label{app:additional_philly_results}


\subsection{Linearity of crime trends}

In our analysis of crime in Philadelphia, we model the change of crime over time for years between 2006 and 2017 with a linear trend (see Equation (2) of the main manuscript). 
While linearity might not perfectly characterize the trend over time, it is the most practical and common choice when using a relatively small number of time points (see Bernardelli et al., 1995 and Anderson et al., 2017). In fact, this simple model allows us to detect the general trend, i.e. whether crime is overall increasing or decreasing in a neighborhood.
However, the careful reader might worry about the validity of such assumption. We analyze here a representative sample of neighborhoods and their trend over time to check for strong non-linearities.

\begin{figure}[H]
\centering
\includegraphics[width = \textwidth]{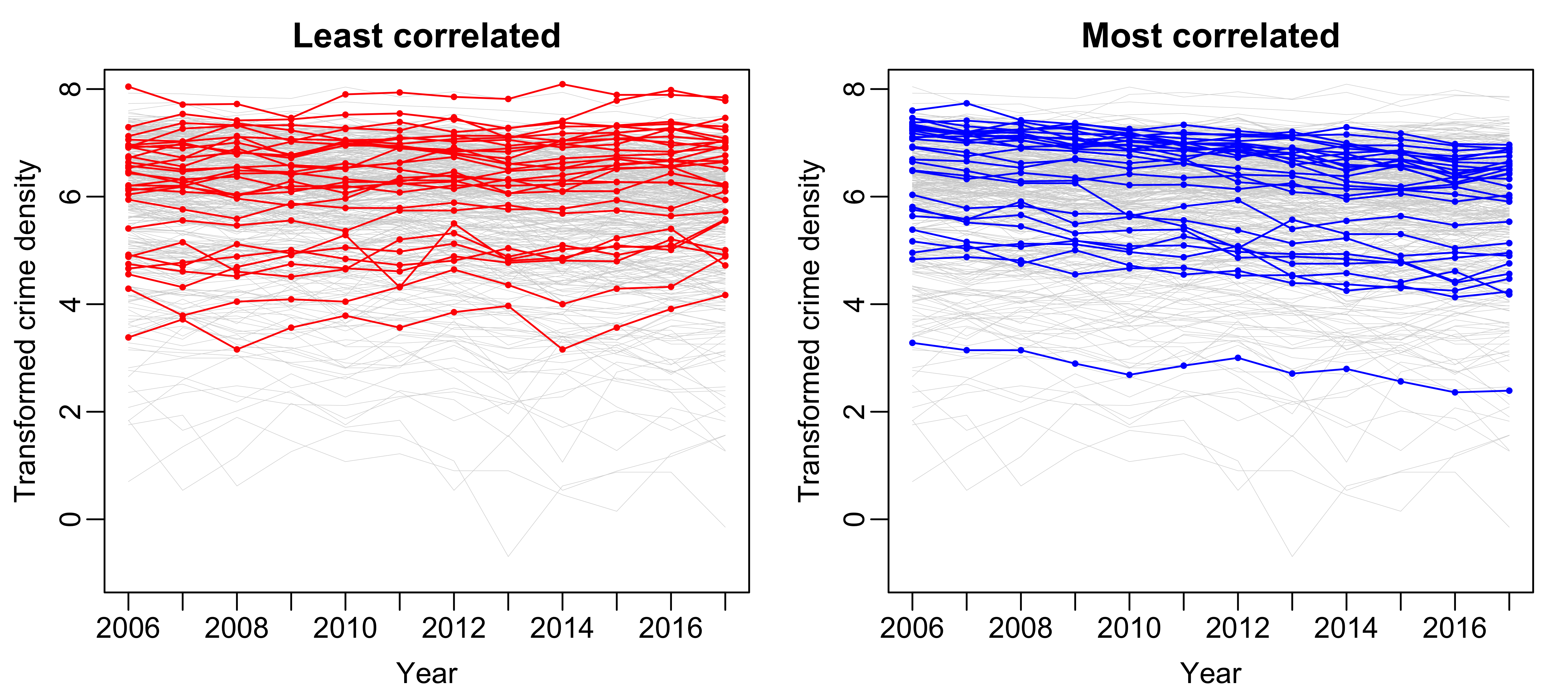}
\caption{Trace plots for the time trends of the thirty neighborhoods with lowest absolute value of the correlation coefficient (left panel) and with the highest value of the correlation coefficient (right panel).}
\label{fig:linearity30}
\end{figure}

To analyze the linearity of crime trends we computed the Pearson correlation coefficient between time and log crime density and examined the absolute value of their correlation. Neighborhoods characterized by a correlation coefficient close to 1 (in absolute value) have trends that are very close to linear. The ones with smaller values of the correlation coefficient (again, in absolute value) are either not changing over time, or could display non-linear trend.
We note that more than $50\%$ of the neighborhoods have correlation greater than $0.6$ (in absolute value) and more than $75\%$ greater than $0.4$. 
Figure \ref{fig:linearity30} shows the trends of the 30 neighborhoods with lowest correlation in absolute value (left panel) and the 30 neighborhoods with highest correlation in absolute value (right panel). While the low correlation neighborhoods display much more nonlinear variation than the high correlation ones, we still believe a linear trend is insightful in describing the time trend in such neighborhoods.

\begin{figure}[H]
\centering
\includegraphics[width = \textwidth]{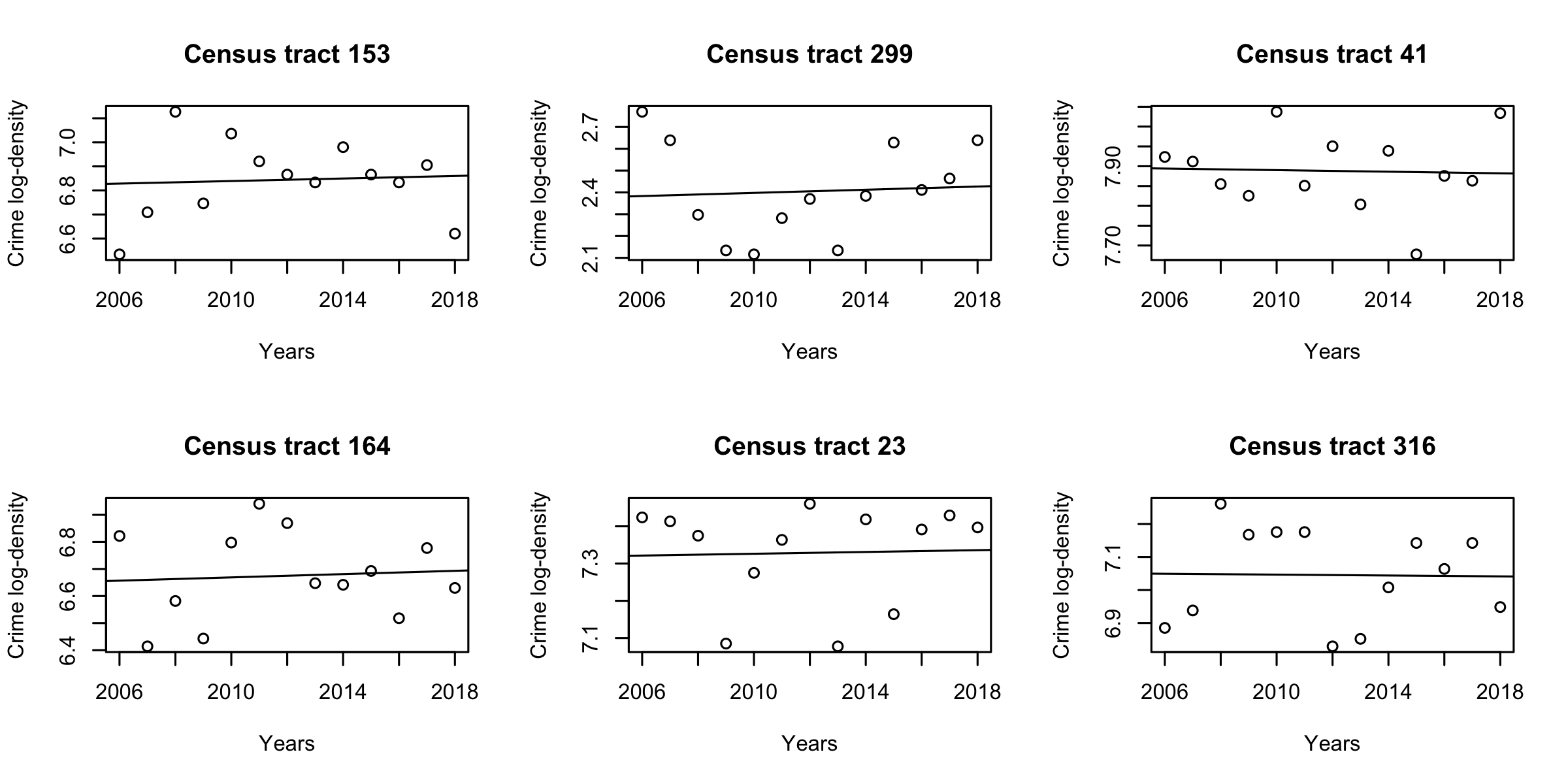}
\caption{Crime density over time for the six neighborhoods with lowest correlation in absolute value.}
\label{fig:linearity6}
\end{figure}

To study the neighborhoods with low correlation is more detail, Figure \ref{fig:linearity6} displays the time trends individually for the six neighborhoods with lowest correlation in absolute value, together with the least square line. Most of these neighborhoods do not display clear non-linear patters. The only neighborhood for which crime density seems to decrease and then increase is Census tract 299, where a quadratic term might better describe the trend.

\subsection{Extending \texttt{PartOpt} to accomodate non-linearities}

In our particular application, a first-order expansion of the expected transformed crime density was sufficient to characterize the general neighborhood-level trends in crime.
However, for datasets displaying strong non-linearities, such an approximate may not be appropriate.
We now describe how one might extend \texttt{PartOpt} to accommodate more flexible non-linear models.
To this end, consider a $D^{\text{th}}$ order expansion of the model in Equation 2 of the main text:
\begin{equation}
\label{eq:expanded_model}
y_{i,t} = \alpha_{i} + \sum_{d = 1}^{D}{\beta_{i,d}x_{t}^{d}} + \varepsilon_{it},
\end{equation}
where once again $x_{t}$ is the standardized time index.
We now must estimate the base-line transformed crime density $\alpha_{i}$ and a $D$-vector of regression coefficients$\bbeta_{i} = (\beta_{i,1}, \ldots, \beta_{i,D})^{\top}$ for each neighborhood $i.$

Doing so, however, requires us to make more modeling decision than we had to make with the first-order model in the main text.
Namely, we must decide how much heterogeneity we would like to allow in the spatial distributions of the coefficients in the expanded model.
At one extreme, we can introduce $D$ underlying partitions, one for each collection $\bbeta^{(d)} = (\beta_{1,d}, \ldots, \beta_{N,d}),$ that allows different spatial distributions for each coefficient.
We could then place conditionally independent CAR--within--clusters priors on each of these collections in a manner analogous to the priors on $\balpha$ and $\bbeta$ in the main text.
At the other extreme, we can introduce a single underlying partition, implicitly assuming that the spatial variability in, say, the quadratic coefficients is identical to the spatial variability in the linear coefficients.
We could then place conditionally independent \textit{multivariate} CAR \citep{GelfandVounatsou2003_mCAR} priors  on the vectors of regression coefficient for the neighborhoods in each cluster. 

Although there are many possibilities for specifying the latent cluster structure of the intercepts and regression coefficients in the $D^{\text{th}}$ order expansion, we may still run \texttt{PartOpt} so long as marginalize out the intercepts and regression coefficients.
Essentially, so long as we adopt conditionally conjugate priors within each cluster, we can still compute the log marginal likelihoods and conditional posterior expectations needed by \texttt{PartOpt}. 
We could, in fact, extend the model even further by using an alternative basis expansion in Equation~\eqref{eq:expanded_model}:
$$
y_{i,t} = \alpha_{i} + \sum_{d = 1}^{D}{\beta_{i,d}\phi_{d}(x_{t})} + \varepsilon_{i,t},
$$
where $\phi_{1}, \ldots, \phi_{D}$ are pre-specified basis functions.
Once again, so long as we maintain conditional conjugacy within cluster, we would be able to run \texttt{PartOpt}.

\subsection{Sensitivity to prior choice}


In Section 5, we analyzed the data on crime density in Philadelphia's census tracts, by running our Particle Optimization procedure on the model described in Section 2. The choice of the prior distribution and of some of the hyperparameters could affect the posterior estimates recovered. In this section, we will analyze sensitivity to prior and hyper-parameter choices.
In particular, we will compare results recovered under the Ewens-Pitman prior with different $\eta$ parameters and under the Uniform prior on the space of spatial partitions $\mathcal{SP}$. We additionally study the sensitivity under different values of the spatial autocorrelation parameter $\rho$.

\begin{figure}[t]
\centering
\includegraphics[width = \textwidth]{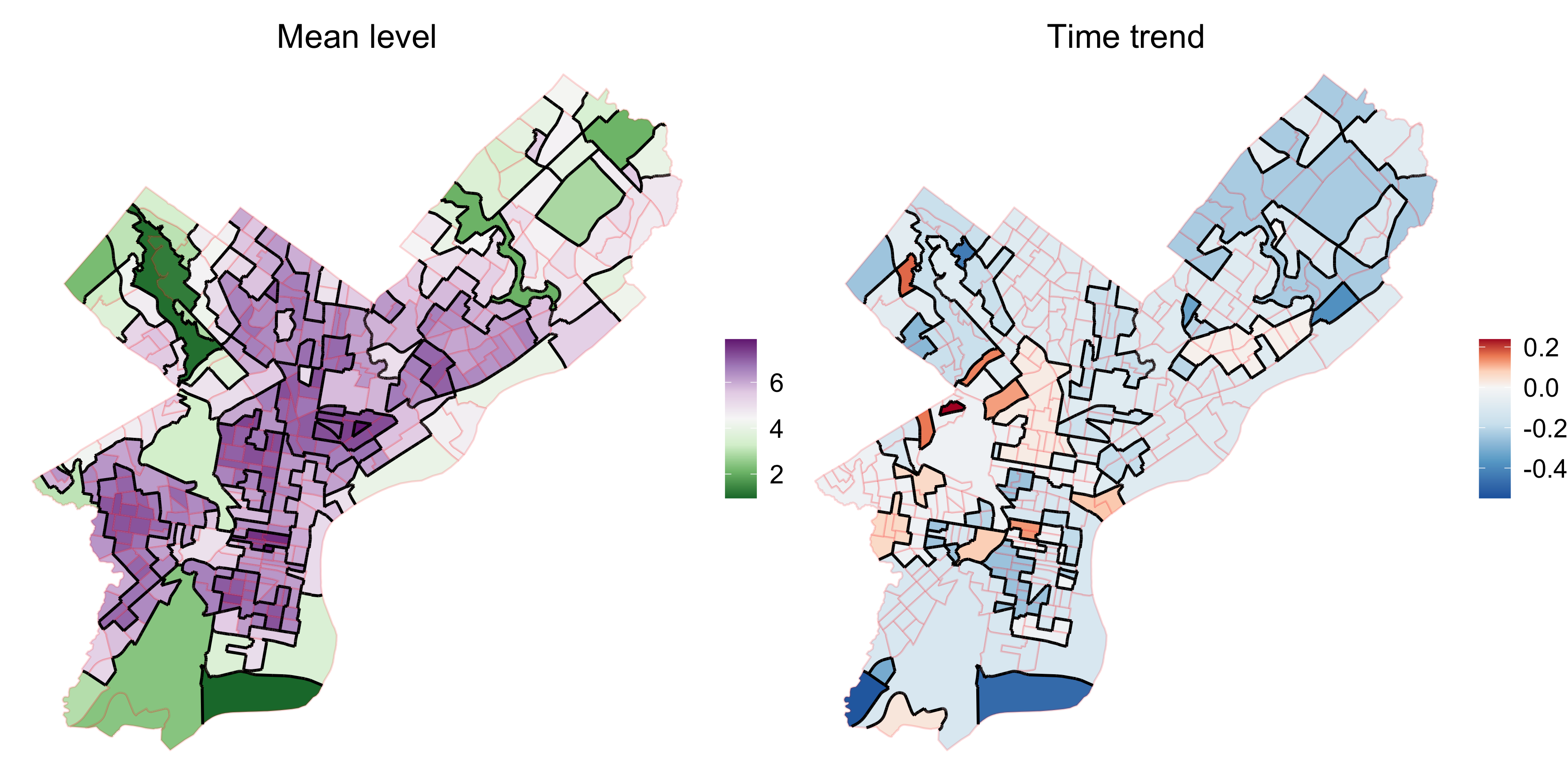}
\caption{Top particle identified by our procedure under the Uniform prior on $\mathcal{SP}$. The thick lines highlight the border between the clusters, and the color represents the posterior mean of the parameters $\balpha$ and $\bbeta$ conditional on the displayed particle.}
\label{fig:best_part_unif}
\end{figure}

We start by reporting the top particle recovered by our procedure under the Uniform prior in Figure~\ref{fig:best_part_unif}. We first notice that under this prior, the partition of the time trends $\gamma^\beta$ presents several clusters, identifying many moderately sized clusters that display a range of time trends, both increasing and decreasing. Moreover, the parameter estimates under the uniform prior are almost constant within cluster, in contrast to the ones under Ewens-Pitman prior that showed much larger levels of within-cluster variation.
Interestingly, though we recover more clusters in the mean level partition $\gamma^{(\alpha)}$ with a uniform prior, the estimates of $\alpha_{i}$ arising from both priors show little substantive difference. While it might not be obvious from visually comparing the two plots, we can easily check by analyzing the linear correlation between the estimates of the vector of crime trends $\balpha$ under the two priors, which is equal to 0.999. The same correlation measure on the estimates of $\bbeta$ under the two models is instead 0.931, which suggests the estimates are somewhat different.

While the partition under this prior can be seen as more interpretable, it is associated with worse predictive performances: it's out of sample predictive error is $0.2344$, which is larger than the error under the Ewens-Pitman prior, but smaller than the one achieved by running \texttt{And} or by finding separate MLE's (see Table~1).

\begin{figure}[t]
\centering
\includegraphics[width = \textwidth]{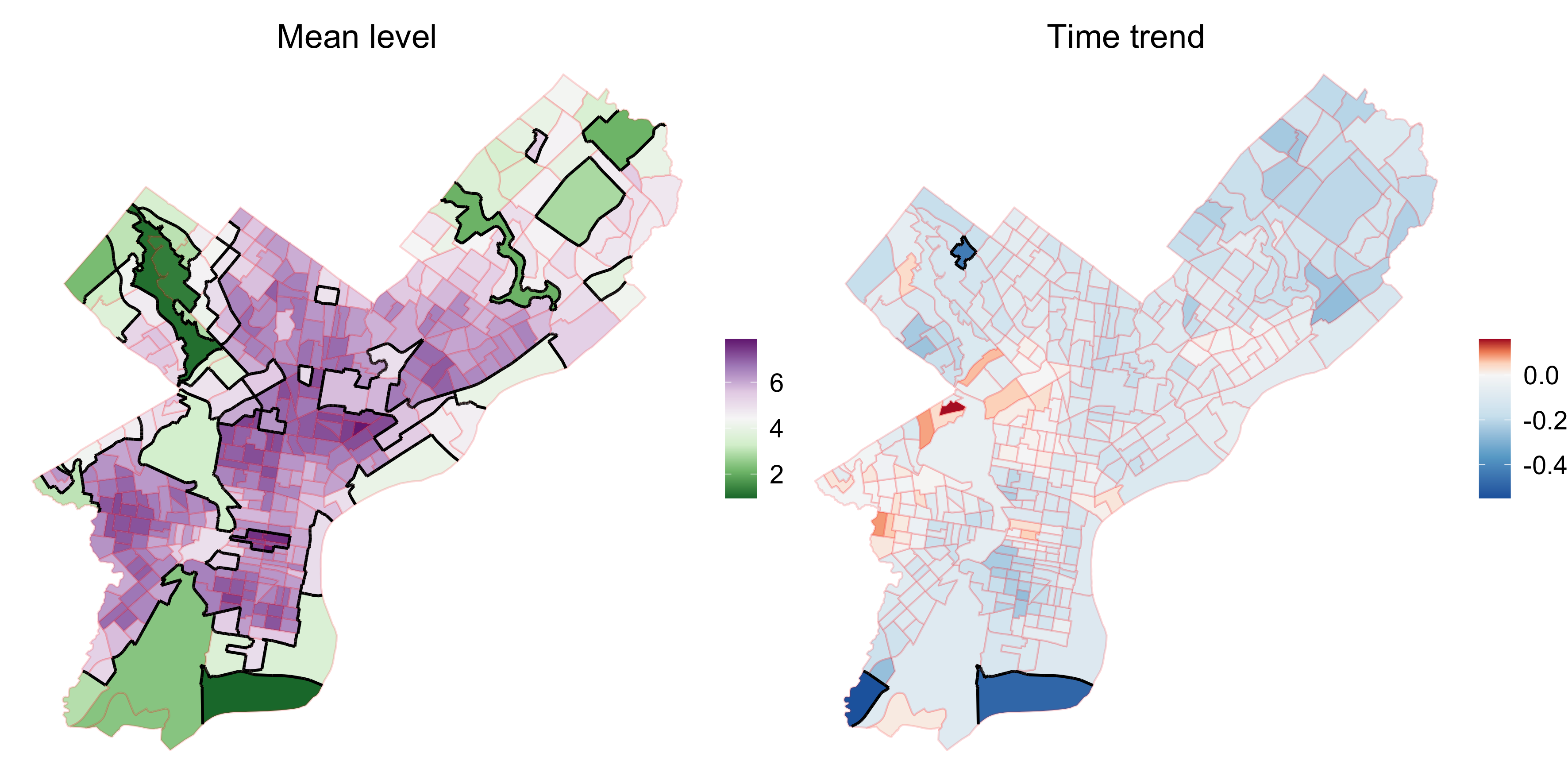}
\caption{Top particle identified by our procedure under the Ewens-Pitman prior with concentration hyperparameter $\eta = 5$.}
\label{fig:best_part_eta5}
\end{figure}

We now analyze the particles recovered by our procedure under the Ewens-Pitman prior with different values for the concentration hyperparameter $\eta$. This hyperparameter regulates the probability of a units joining a new cluster, with higher values inducing a larger number of clusters in expectation. Specifically we compared values $\eta = 1, 3, 5$. We find that the partition of the mean level $\gamma^\alpha$ in the top particle differs for different values of $\eta$, with the partition recovered under $\eta = 5$ displayed in Figure~\ref{fig:best_part_eta5} and the one under $\eta = 3$ being very similar to it. We notice instead that the partition of the time trends $\gamma^\beta$ does not change substantially, still showing one cluster but also recovering several singleton clusters for neighborhoods with extreme values of the time trend $\beta_i$.
The change in the partition $\gamma^\alpha$ is likely caused by the local nature of our algorithm, which can get stuck in local modes.
In fact, we found that the particle recovered under $\eta = 5$ has a larger posterior probability under the model with $\eta = 1$, compared to the top particle recovered under $\eta = 1$. 

\begin{figure}[t]
\centering
\includegraphics[width = \textwidth]{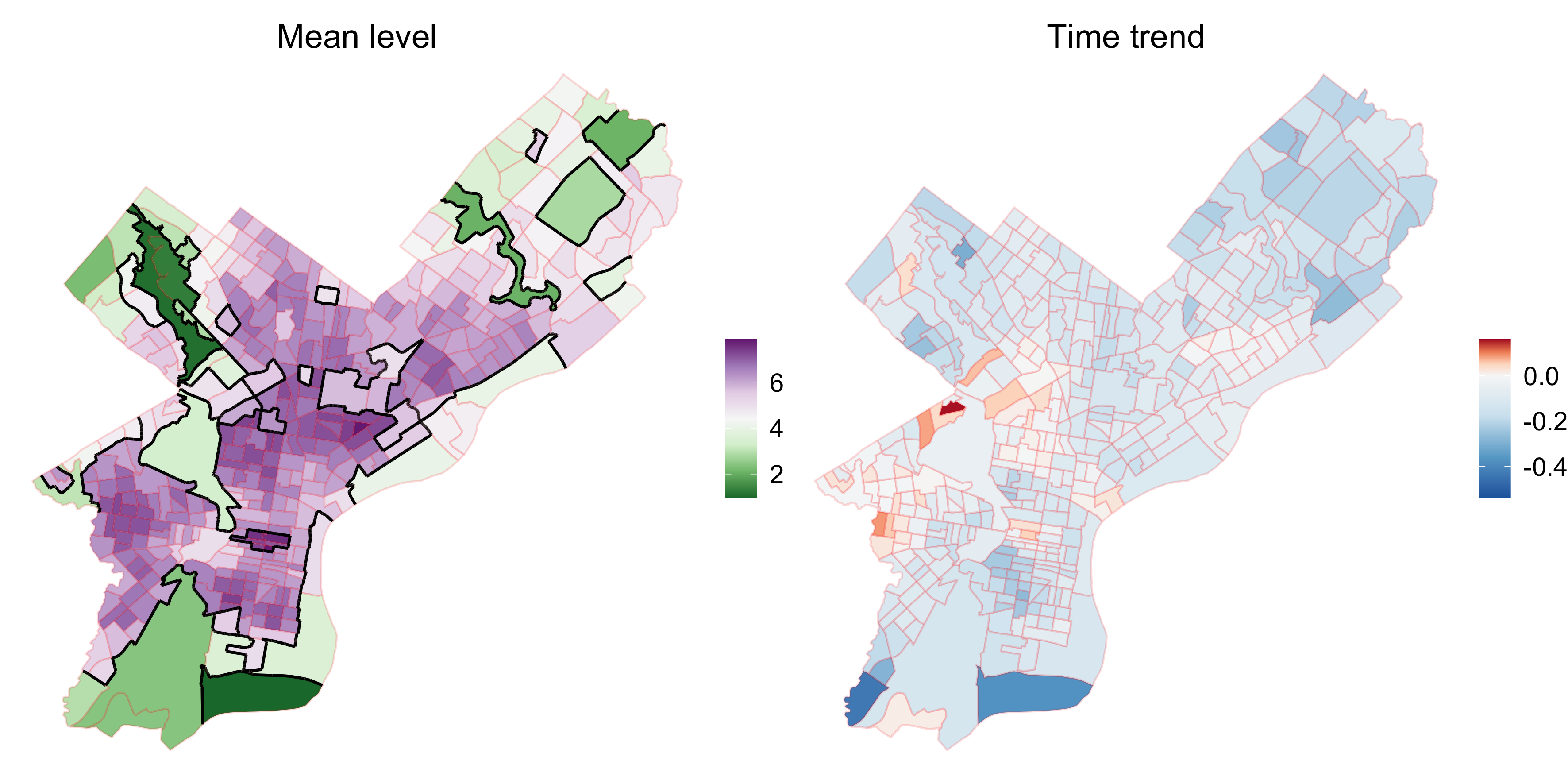}
\caption{Top particle identified by our procedure when the spatial autocorrelation hyperparameter $\rho = 0.99$. Many of the top particles recovered in our sensitivity analysis for other values of $\rho$ looked similar to this.}
\label{fig:best_part_rho99}
\end{figure}

Finally, we considered sensitivity to the choice of the spatial autocorrelation hyperparameter $\rho$. In our analysis and simulation, it was chosen equal to 0.9, which induces a strong degree of spatial autocorrelation, without causing the prior distributions of $\balpha$ and $\bbeta$ to be improper (this happens when $\rho = 1$). 
To test the sensitivity of our procedure to this value, we ran our method with different values of $\rho$, ranging from low prior autocorrelations with $\rho = 0$ to very strong prior autocorrelation with $\rho = 0.99$.
We analyzed the top particle recovered by our procedure. For the mean levels of crime, our procedure recovered two sets of similar partitions for different values of $\rho$. One set corresponds to partitions that look like the top particle displayed in Figure~5 of our main manuscript, and it was recovered for $\rho=0.9$ and for $\rho=0.85$. The other set of partitions (recovered for $\rho=0.99, 0.95,  0.80, 0.75, 0.50, 0.25, 0$) is similar to the partition reported in Figure~\ref{fig:best_part_rho99}, which corresponds to the top particle for $\rho=0.99$.
Similarly to what we discussed previously, 
the difference between these two sets is likely caused by the local nature of our algorithm, which can get stuck in local modes. This seems to be the case in this example, as the partition recovered by our procedure with $\rho=0.99$ has a higher posterior probability under the model with $\rho = 0.9$, suggesting that the algorithm with $\rho = 0.9$ and $\rho = 0.85$ got stuck in a local mode.
While this is not ideal, the positive aspect is that the estimate for the $\balpha$ values is robust to these changes, and does not show differences; in fact, the linear correlation between any pair of estimates under different $\rho$ values is always greater than 0.99.

Instead, the partition of the time trends recovered in the top particle is always equal to the partition with one large cluster (with one exception where a singleton cluster is recovered too). In this case the estimate of the $\bbeta$ values changes slightly, but not substantially: the linear correlation between the estimate found under $\rho = 0$ and the one under $\rho = 0.99$ is 0.95 but is greater than 0.98 for any two pairs of parameters found with positive values of $\rho$. This effect is not surprising, as we expect the CAR prior to have stronger effects when the partition recovered is formed by only one large cluster, rather than when it is formed by many smaller clusters, as it's the case for $\gamma^\alpha$.

\section{Derivation of Closed Form Expressions}


\label{app:two_partitions_derivation}

Recall from Section~2 that our full model is: 
\begin{align*}
\gamma^{(\alpha)}, \gamma^{(\beta)} &\sim \text{EP}(\eta; \mathcal{SP}) \\
\sigma^{2} &\sim \text{IG}\left(\frac{\nu_{\sigma}}{2}, \frac{\nu_{\sigma}\lambda_{\sigma}}{2}\right) \\
(\alphabar_{k})_{k} &\overset{iid}{\sim} N(0, a_{2}\sigma^{2}) \\
(\betabar_{k'})_{k'} &\overset{iid}{\sim} N(0, b_{2}\sigma^{2}) \\
(\balpha_{k})_k &\overset{ind}{\sim} \text{CAR}(\alphabar_{k}, a_{1}\sigma^{2}, W^{(\alpha)}_k) \\
(\bbeta_{k'})_{k'} &\overset{ind}{\sim} \text{CAR}(\betabar_{k'}, b_{1}\sigma^{2}, W^{(\beta)}_{k'})\\
(y_{i,t})_{i,t} &\overset{ind}{\sim} \text{N}(\alpha_{i} + \beta_{i}x_{t}, \sigma^{2})
\end{align*}

We exploit the conditional conjugacy present in this model in several places.
First, we have closed form expressions for the conditional posterior means $\E[\balpha \mid \by, \bgamma]$ and $\E[\bbeta \mid \by, \bgamma],$ which we use in our particle optimization procedure to propose new transitions.
Second, we can compute the marginal likelihood $p(\by \mid \bgamma)$ in closed form, which we use to evaluate the optimization objective and pick between multiple transitions.
Below, we carefully derive these closed form expressions, noting that in several places, we can avoid potentially expensive matrix inversions.
In particular, the choice to center the time variable, thereby ensuring an orthogonal design matrix within each neighborhood, facilitates rapid likelihood evaluations.

\paragraph{Distribution of $\balpha_k$}
Let us first consider the vector of parameters $\balpha_{k}$ in cluster $S^{(\alpha)}_k$ given $\sigma^2$: by marginalizing the distribution of the grand cluster mean $\alphabar_k$, we find that its distribution is a multivariate normal with covariance matrix $\sigma^2 \Sigma^{(\alpha)}_k$, where $\Sigma^{(\alpha)}_k = a_1\Sigma^{(\alpha)}_{k,\rm{CAR}}+a_2\1\1^{\top} = a_1 \left[ \rho (W^{(\alpha)}_k)^* + (1-\rho)\I \right]^{-1}+a_2\1\1^{\top} $.
Note that its precision matrix can be computed using Woodbury's formula without having to invert any matrix:
\begin{align*}
(\Sigma^{(\alpha)}_k)^{-1} &= a_1^{-1}\Omega^{(\alpha)}_{k,\rm{CAR}} - a_1^{-1} \Omega^{(\alpha)}_{k,\rm{CAR}} \1 \left(a_1^{-1} \1^{\top}  \Omega^{(\alpha)}_{k,\rm{CAR}} \1 + a_2^{-1} \right)^{-1} \1^{\top} a_1^{-1} \Omega^{(\alpha)}_{k,\rm{CAR}} =\\
&= a_1^{-1}\Omega^{(\alpha)}_{k,\rm{CAR}} - \frac{a_1^{-2} (1-\rho)^2}{a_1^{-1} n_k (1-\rho) + a_2^{-1}}  \1  \1^{\top} 
\end{align*}
where $\Omega^{(\alpha)}_{k,\rm{CAR}} = \left( \Sigma^{(\alpha)}_{k,\rm{CAR}} \right)^{-1} = \rho (W^{(\alpha)}_k)^* +(1-\rho)\I$; the second line follows from noticing that $\1$ is both a left and right eigenvector of $\Omega^{(\alpha)}_{k,\rm{CAR}}$ with eigenvalue $1 - \rho$.
Similarly this holds for the distribution of $\bbeta_{k'}$.

\paragraph{Distribution of $\balpha$}
Next, we can write the distribution of the whole vector $\balpha$ given $\sigma^2$ and $\gamma^{(\alpha)}$: by combining the distributions of the cluster specific parameters $\balpha_k$'s, and using the independence between different clusters, we find that the distrubution of $\balpha$ given $\sigma^2$ and $\gamma^{(\alpha)}$ is a multivariate normal with mean zero and covariance matrix that can be found by combining the $\Sigma^{(\alpha)}_k$'s.
Because of the independence between clusters, \textit{there exists an ordering of the indices of $\balpha$} so that the covariance matrix of $\Balpha \vert \gamma_{\alpha}, \sigma^2$ has a block-diagonal structure. 
We denote such permutation of the indices with $\pi^{(\alpha)}$, and it can be constructed by mapping the first $n_1$ elements to the indices in the first cluster ($\{ \pi^{(\alpha)}(1), \ldots, \pi^{(\alpha)}(n_1) \} = S^{(\alpha)}_1$), the following $n_2$ elements to the indices in the second cluster ($\{ \pi^{(\alpha)}(n_1+1), \ldots, \pi^{(\alpha)}(n_1+n_2) \} = S^{(\alpha)}_2$), and so on. With such ordering, the $k$th diagonal block of the covariance matrix is $\sigma^2 \Sigma^{(\alpha)}_k$.
Similarly, we can find a (potentially different) permutation $\pi^{(\beta)}$ for $\Bbeta$ and derive the distribution of $\bbeta_\pi \vert \sigma^2,\gamma^{(\beta)}$. 

\paragraph{Notation} 
To describe the distributions of interest we can represent our model in the form of a unique linear model, by combining all the observations in a vector $Y$, combining the reodered coefficients in a unique vector $\btheta = (\Balpha_\pi, \Bbeta_\pi)$ and appropriately constructing the covariate matrix $X$. In the next paragraphs we will provide with the details on how we constructed such vectors and matrix.

To build the column vector $Y$ we stack the vectors $\y_i$ with $i= 1, \ldots, N$: $Y$ is a vector of length $N \cdot T$ and each block of $T$ rows corresponds to a particular neighborhood; in particular, the $((i-1)T + t)$th entry of $Y$ corresponds to $y_{i,t}$.

The vector of coefficients $\btheta$ is found by concatenating the reordered $\balpha_\pi$ and $\bbeta_\pi$: for $i = 1, \ldots, N$, elements $\theta_i = \alpha_{\pi^{(\alpha)}(i)}$ and  $\theta_{N+i} = \beta_{\pi^{(\beta)}(i)}$.

The matrix of covariates $X$ then has dimensions $NT \times 2N$; each block of $T$ rows corresponds to a neighborhood and each column corresponds to an element of $\btheta$: the first $N$ columns correspond to the elements of $\balpha_\pi$ and the second $N$ columns to $\bbeta_\pi$. 
The rows of $X$ corresponding to neighborhood $i$ (rows $(i-1)T + t$ with $t = 1, \ldots T$) have an element equal to 1 in the $(\pi^{(\alpha)})^{-1}(i)$th column, an element equal to $x_{it} = (t - \overline{t})/sd(\bt)$ in the $(N+(\pi^{(\beta)})^{-1}(i))$th column, and zero elsewhere. 
With such construction, the $(i-1)T + t$ row of the equation $Y = X\btheta$ corresponds to $y_{i,t} = \theta_{(\pi^{(\alpha)})^{-1}(i)} + x_{it} \theta_{N+(\pi^{(\beta)})^{-1}(i)} = \alpha_i  + x_{t} \beta_i$.


\paragraph{Marginal likelihood $Y \vert \gamma^{(\alpha)}, \gamma^{(\beta)}$}
To recover the marginal likelihood $p(Y \vert \gamma^{(\alpha)}, \gamma^{(\beta)})$ we compute 
\begin{align*}
&\int \left[ \int p(Y \vert \Balpha,\Bbeta,\sigma^2)p(\Balpha \vert \gamma^{(\alpha)},\sigma^2) p(\Bbeta \vert \gamma^{(\beta)},\sigma^2) d \Balpha d\Bbeta \right] p(\sigma^2)d \sigma^2=\\
=&\int \left[ \int p(Y \vert \Balpha_\pi,\Bbeta_\pi,\sigma^2)p(\Balpha_\pi \vert \gamma^{(\alpha)},\sigma^2) p(\Bbeta_\pi \vert \gamma^{(\beta)},\sigma^2) d \Balpha_\pi d\Bbeta_\pi \right] p(\sigma^2)d \sigma^2=\\
=&\int \left[ \int p(Y \vert \btheta,\sigma^2)p(\btheta \vert \gamma^{(\alpha)},\gamma^{(\beta)},\sigma^2)  d \btheta \right] p(\sigma^2)d \sigma^2.
\end{align*}
Let us first compute $p(Y \vert \sigma^2,\gamma^{(\alpha)},\gamma^{(\beta)})  = \int p(Y \vert \btheta,\sigma^2)p(\btheta \vert \gamma^{(\alpha)},\gamma^{(\beta)},\sigma^2)  d \btheta $. Using the notation for linear regression we can write $p(Y \vert \Btheta, \sigma^2) = N(X\Btheta, \sigma^2 \I)$.
The prior for $\Btheta$ is a normal distribution with mean zero and block covariance matrix $\Sigma_\theta$: the first $n\times n$ block corresponds to the covariance matrix of $\Balpha$ and the second to the one for $\Bbeta$.

By integrating out $\Btheta$, $p(Y \vert \gamma^{(\alpha)},\gamma^{(\beta)},\sigma^2) =N\left(\0, \sigma^2 \Sigma_Y \right)$ where $\Sigma_Y = \I + X \Sigma_\theta X^{\top}$.
Its precision matrix can be computed using Woodbury's formula again:
$ \Sigma_Y^{-1} = \I - X(\Sigma_\theta^{-1} + X^{\top} X)^{-1} X^{\top}$. Note that $X^{\top} X$ is a diagonal matrix, and we derive its form towards the end of this section.

The marginal likelihood can now be derived by integrating out $\sigma^2$: 
\begin{align*}
p(Y \vert  \gamma^{(\alpha)},\gamma^{(\beta)}) &= \int p(Y \vert \sigma^2,\gamma^{(\alpha)},\gamma^{(\beta)}) p(\sigma^2) d\sigma^2 =\\
&=   \pi^{-nT/2} {\rm det}(\Sigma_Y)^{-1/2} \frac{(\nu_\sigma \lambda_\sigma/2)^{\nu_\sigma/2}}{\Gamma(\frac{\nu_{\sigma}}{2})}\int(\sigma^2)^{-\frac{NT+\nu_\sigma}{2}-1} { e}^{-\frac{Y^\top \Sigma_Y^{-1} Y + \nu_\sigma \lambda_\sigma}{2\sigma^2} }d\sigma^2 =\\
&=  \pi^{-nT/2} {\rm det}(\Sigma_Y)^{-1/2} \frac{\Gamma(\frac{NT+\nu_{\sigma}}{2})}{\Gamma(\frac{\nu_{\sigma}}{2})} \left( \frac{\nu_\sigma \lambda_\sigma}{2} \right)^{\nu_\sigma/2}  \left( \frac{\nu_\sigma \lambda_\sigma+Y^\top \Sigma_Y^{-1} Y}{2} \right)^{-(NT+\nu_\sigma)/2}=\\
&= \pi^{-nT/2} {\rm det}(\Sigma_Y)^{-1/2} \frac{\Gamma(\frac{NT+\nu_{\sigma}}{2})}{\Gamma(\frac{\nu_{\sigma}}{2})}   \left(\frac{\nu_\sigma \lambda_\sigma}{2} \right)^{-NT/2} \left( 1 + \frac{Y^\top \Sigma_Y^{-1} Y}{\nu_\sigma \lambda_\sigma} \right)^{-(NT+\nu_\sigma)/2}.
\end{align*}
Note that if $\lambda_{\sigma} = 1$, this is multivariate t-distribution with $\nu_\sigma$ degrees of freedom.

For this we need to compute the quadratic form 
$$ Y^{\top} \Sigma_Y^{-1} Y =  Y^{\top}Y - Y^{\top} X(\Sigma_\theta^{-1} + X^{\top} X)^{-1} X^{\top}Y.
$$
Because of the block diagonal structure of $\Sigma_\theta^{-1} + X^{\top} X$ we can write this as a sum over the clusters of the two partitions.
Consider the column vector $X^{\top}Y$ of length  $2N$: the first $N$ elements correspond to the summary statistics related to the $\alpha_{\pi(i)}$'s and we will denote the ones corresponding to cluster $S^{(\alpha)}_k$ with $(X^{\top}Y)_{k}^{(\alpha)}$, while the second $N$ elements are for the $\beta_i$'s and we denote with $(X^{\top}Y)_{k'}^{(\beta)}$ the ones for cluster $S^{(\beta)}_{k'}$ . Now we can write 
\begin{align*}
Y^{\top} X(\Sigma_\theta^{-1} + X^{\top} X)^{-1} X^{\top}Y &= \sum_{k=1}^{K^{(\alpha)}} (X^{\top}Y)_{k}^{(\alpha)\top} ((\Sigma^{(\alpha)}_k)^{-1}+ T\I)^{-1}  (X^{\top}Y)_{k}^{(\alpha)}\\
&+\sum_{k'=1}^{K^{(\beta)}} (X^{\top}Y)_{k'}^{(\beta)\top} ((\Sigma^{(\beta)}_{k'})^{-1} +\textstyle \sum x_t^2 \I)^{-1}  (X^{\top}Y)_{k'}^{(\beta)}
\end{align*}
where $(\Sigma^{(\alpha)}_k)^{-1} + T\I$ is the diagonal blocks of $\Sigma_\theta^{-1} + X^{\top} X$ corresponding to cluster $S^{(\alpha)}_k$ and $(\Sigma^{(\beta)}_{k'})^{-1} +\textstyle \sum x_t^2 \I$ corresponds to $S^{(\beta)}_{k'}$; each of them can be inverted using methods for symmetric positive definite matrices.

To compute the marginal likelihood we are left we calculating the determinant of $\Sigma_Y$, where we can use the reciprocal of the determinant of its inverse
$$
{\rm det}(\Sigma_Y^{-1}) = {\rm det}(\I - X(\Sigma_\theta^{-1} + X^{\top} X)^{-1} X^{\top}) = {\rm det}(\I- (\Sigma_\theta^{-1} + X^{\top} X)^{-1} X^{\top}X)
$$
where the last equality is given by Sylvester's formula, and allows us to compute the determinant of a smaller dimensional matrix. Moreover, because of its block diagonal structure, we can compute the determinant block-wise.


\paragraph{Posterior mean of $\balpha, \bbeta$}
The calculations for the posterior mean of $\Balpha, \Bbeta$ are very similar: using the same notation and the results for linear regression, we can find 
$$
\E \left[ \Btheta \vert Y, \gamma^{(\alpha)},\gamma^{(\beta)}, \sigma^{-1} \right] = \left( X^{\top}X + \Sigma_\theta^{-1} \right)^{-1} X^{\top} Y
$$
and since this does not depend on $\sigma^2$, it coincides with $\E \left[ \Btheta \vert Y, \gamma^{(\alpha)},\gamma^{(\beta)} \right]$. Because of the block diagonal structure of the matrices involved, we can compute the estimate of the parameter for each cluster independently.
Moreover, note that the inverse of $X^{\top}X + \Sigma_\theta^{-1}$ is computed in the likelihood calculation, so it can be stored and does not need to be computed two times.

\paragraph{Derivation of $X^{\top} X$}
Since in our formulation the covariates are orthogonal, i.e. $\sum_{t=1}^{T} x_{it}=0$ for all $i$, $X^{\top} X$ is a diagonal matrix. Note that column $X_{(\pi^{(\alpha)})^{-1}(i')}$ contains $T$ 1's in rows $t + (i'-1)\times$T and zeros elsewhere; similarly column $X_{N+(\pi^{(\beta)})^{-1}(i')}$ contains elements $(x_{i't})$ in rows $t + (i'-1)\times T$ and zero's elsewhere. Thus, when we compute $(X^{\top} X)_{ij}$ we consider the cross product of columns $X_i$ and $X_j$. Depending on the value of $i$ and $j$, we have the following cases:
\begin{itemize}
\item if $i = j \leq N$, then $(X^{\top} X)_{ij} = T$,
\item if $i = j \geq N$, then $(X^{\top} X)_{ij} = \sum_t x_{\pi^{(\beta)}(j-N),t}^2$, 
\item if $i \leq N$ and $j = N + i$, then $(X^{\top} X)_{ij} = \sum_t x_{\pi^{(\beta)}(i),t}$ = 0, 
\item if $j \leq N$ and $i = N + j$, then $(X^{\top} X)_{ij} = \sum_t x_{\pi^{(\beta)}(j),t}$ = 0, 
\item for any other $i,j$, $(X^{\top} X)_{ij} = 0$.
\end{itemize}
Thus the matrix $X^{\top} X$ is a diagonal matrix:
the first $n\times n$ diagonal block is $T\I$, and the second diagonal block is a diagonal matrix whose entries are $\sum_{t=1}^{T} x_{it}^2$; when we have fixed design, $x_{it} =x_t= (t - \overline{t})/sd(\bt)$, then
 $\sum_{t=1}^{T} x_{it}^2=\sum_{t=1}^{T} ((t - \overline{t})/sd(\bt))^2$ is constant, so the second diagonal block is $\sum x_{it}^2 \I$. Because of the orthogonality of the covariates, the upper-right and lower-left blocks are zero matrices, since $\sum_{t=1}^{T} x_{it}=0$.

\paragraph{Note on cluster-wise update of calculations.}

In our greedy search when we perform a move only one or two clusters in only one partition is changed: in a \textit{split} move for $\gamma^{(\cdot)}$, a cluster is divided into two sub-clusters, and the original cluster replaced by the first, while the second creates an additional cluster; in a \textit{merge} move, one of two clusters is deleted and the other is replaced to the merge of the two original clusters. In each case, we need to update the value of the marginal likelihood, of the prior for $\gamma^{(\cdot)}$ and of the estimate of the parameters. 

Because of the block structure given by orthogonality of covariates and by the reordering of the parameters, changing the structure of some clusters does not affect the parameter estimates for other clusters that are not involved in the move. This implies that updates for updates to $S^{(\alpha)}_k$ do not affect the parameter estimates $\balpha_{h}$ for $h\neq k$ or $\bbeta_{k'}$ for any $k'$. 
Similarly, since the quadratic form $ Y^{\top} \Sigma_Y^{-1} Y$ can be written as sum of cluster-specific quadratic forms, we can update only the quadratic form of the clusters affected and we can compute the determinant of the blocks of $\Sigma_Y$ corresponding to the modified clusters.

This allows us to invert matrices that scale like the size of the clusters, reducing the computational costs dramatically.


\section{Extension to Non-Conjugate Models}
\label{app:laplace}

The proposed particle optimization strategy relies on the ability to compute the marginal likelihood $\pi(\y \vert \Bgamma)$. 
This is often straightforward with conjugate models, such as the one considered for our application.
However, it may be challenging or impossible to compute $\pi(\y \vert \Bgamma)$ in more complicated non-conjugate settings.
We can, nevertheless, \textit{approximate} the marginal likelihood using, e.g., a Laplace approximation and deploy an approximate particle optimization strategy.
Below, we outline how this works for Poisson regression.

For example, consider a Poisson regression model for count data, $c_{it} \sim \textrm{Pois}\left(\exp\left\{\alpha_{i} + \beta_{i}x_{t}\right\}\right)$, with separate CAR-within-clusters priors on the $\alpha_{i}$'s and $\beta_{i}$'s.
That is, similar to model (3) in the main text, we model
\begin{align}
\label{eq:poisson_model}
\begin{split}
\gamma^{(\alpha)}, \gamma^{(\beta)} &\overset{iid}{\sim} \mathcal{T\mbox{-}EP} \\
\sigma^{2} &\sim \text{IG}\left(\frac{\nu_{\sigma}}{2}, \frac{\nu_{\sigma}\lambda_{\sigma}}{2}\right) \\
\alphabar_{1}, \ldots, \alphabar_{K_{\alpha}} \mid \gamma^{(\alpha)}, \sigma^{2} &\overset{iid}{\sim} N(0, a_{2}\sigma^{2}) \\
\betabar_{1}, \ldots, \betabar_{K_{\beta}} \mid \gamma^{(\beta)}, \sigma^{2} &\overset{iid}{\sim} N(0, b_{2}\sigma^{2}) \\
\balpha_{k} \mid \alphabar_{k}, \sigma^{2}, \gamma^{(\alpha)} &\sim \text{CAR}(\alphabar_{k}, a_{1}\sigma^{2}, W^{(\alpha)}_{k}) \quad \text{for $k = 1, \ldots, K_{\alpha}$} \\
\bbeta_{k'} \mid \betabar_{k'}, \sigma^{2}, \gamma^{(\beta)} &\sim \text{CAR}(\betabar_{k'}, b_{1}\sigma^{2}, W^{(\beta)}_{k'}) \quad \text{for $k' = 1, \ldots, K_{\beta}$} \\
c_{i,t} \mid \balpha, \bbeta &\sim \textrm{Pois}(\exp(\alpha_{i} + \beta_{i}x_{t}))
\end{split}
\end{align} 
where $x_{t}$ corresponds to the time index standardized to have mean zero and unit variance. 

In the main text, we defined a particle to be the pair of partitions $(\gamma^{(\alpha)}, \gamma^{(\beta)}).$
Now, we extend the definition of particle to include $\sigma^{2}$; that is, let $\Bgamma = (\gamma^{(\alpha)}, \gamma^{(\beta)}, \sigma^2)$.
To approximate the posterior distribution $\pi(\Bgamma \vert \y),$ we first approximate the marginal likelihood $\pi(\y \vert \Bgamma)$. 
Using a Laplace approximation, we can compute 
\begin{align}
\label{eq:map_approximation}
\pi(\Bgamma \vert \y) &= \frac{\pi(\Bgamma, \Balpha, \Bbeta \vert \y)}{\pi(\Balpha, \Bbeta \vert \y, \Bgamma)} \simeq \left. \frac{\pi(\Bgamma, \Balpha, \Bbeta \vert \y)}{\hat{\pi}(\Balpha, \Bbeta \vert \y, \Bgamma)}\right|_{(\Balpha, \Bbeta) = (\hat \Balpha,\hat \Bbeta)}
\end{align}
where $\hat{\pi}(\Balpha, \Bbeta \vert \y, \Bgamma)$ is the density of a multivariate normal distribution whose mean is equal to the conditional MAP estimate $(\hat \Balpha,\hat \Bbeta) = \argmax_{\Balpha,\Bbeta} \pi(\Balpha, \Bbeta \vert \y, \Bgamma)$ and whose covariance matrix is equal to the the inverse Hessian of the conditional log-density $\log \pi(\Balpha, \Bbeta \vert \y, \Bgamma)$ evaluated at the conditional MAP $(\hat \Balpha, \hat \Bbeta).$
The expression on the right-hand side of ~\eqref{eq:map_approximation} is evaluated at $(\hat \Balpha,\hat \Bbeta).$
Computing the MAP and the relevant Hessian is straightforward using standard optimizers. 
Note further that the density in the numerator of~\eqref{eq:map_approximation}, $\pi(\Bgamma, \Balpha, \Bbeta \vert \y)$ can be computed up to normalizing constant (which does not depend on $\Bgamma$ or $\Balpha, \Bbeta$) as $\pi(\y \vert \Balpha, \Bbeta) \pi(\Balpha, \Bbeta \vert \Bgamma) \pi(\Bgamma)$.

The approximate particle optimization algorithm is extremely similar to the one described in the main text.
Namely, we sequentially update each particle in the particle set $\BGamma.$
To update a single particle, we sequentially update $\gamma^{(\alpha)}$ and $\gamma^{(\beta)}$ using the same suite of coarse and fine transitions.
To update $\sigma^{2}$ we maximize the conditional posterior $\pi(\sigma^2 \vert \hat \alpha, \hat \beta, \gamma^{(\alpha)}, \gamma^{(\beta)})$, where $\hat \alpha, \hat \beta$ is the conditional MAP corresponding to $\Bgamma$.
In these updates, we approximate $\pi(\Bgamma \vert \y)$ for every candidate particle $(\gamma^{(\alpha)}, \gamma^{(\beta)},\sigma^{2})$ that we consider. 

\subsection{Illustration on synthetic data}

\begin{figure}[h]
\centering
\includegraphics[width = 0.5\textwidth]{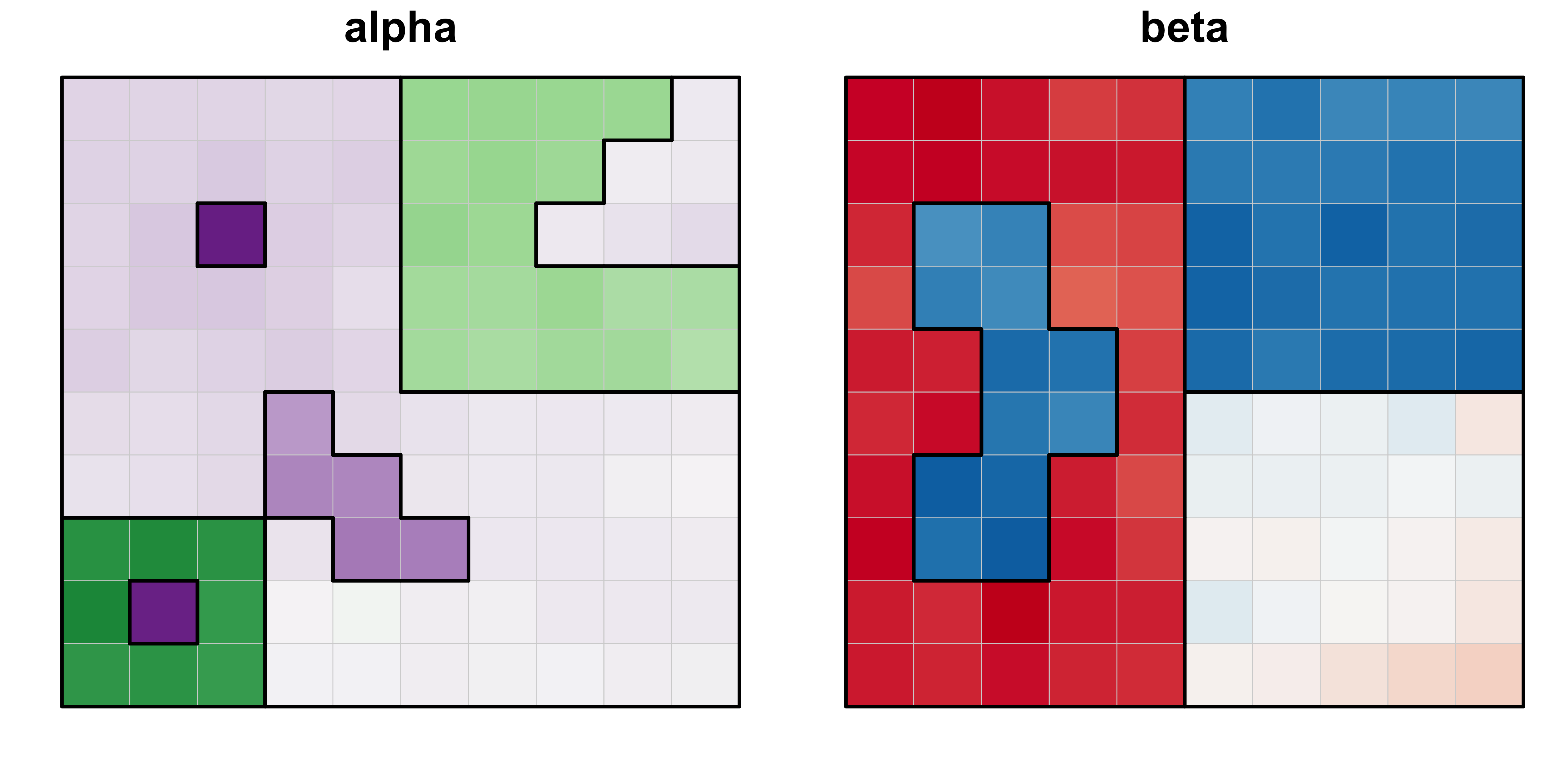}\\
\includegraphics[width = 0.5\textwidth]{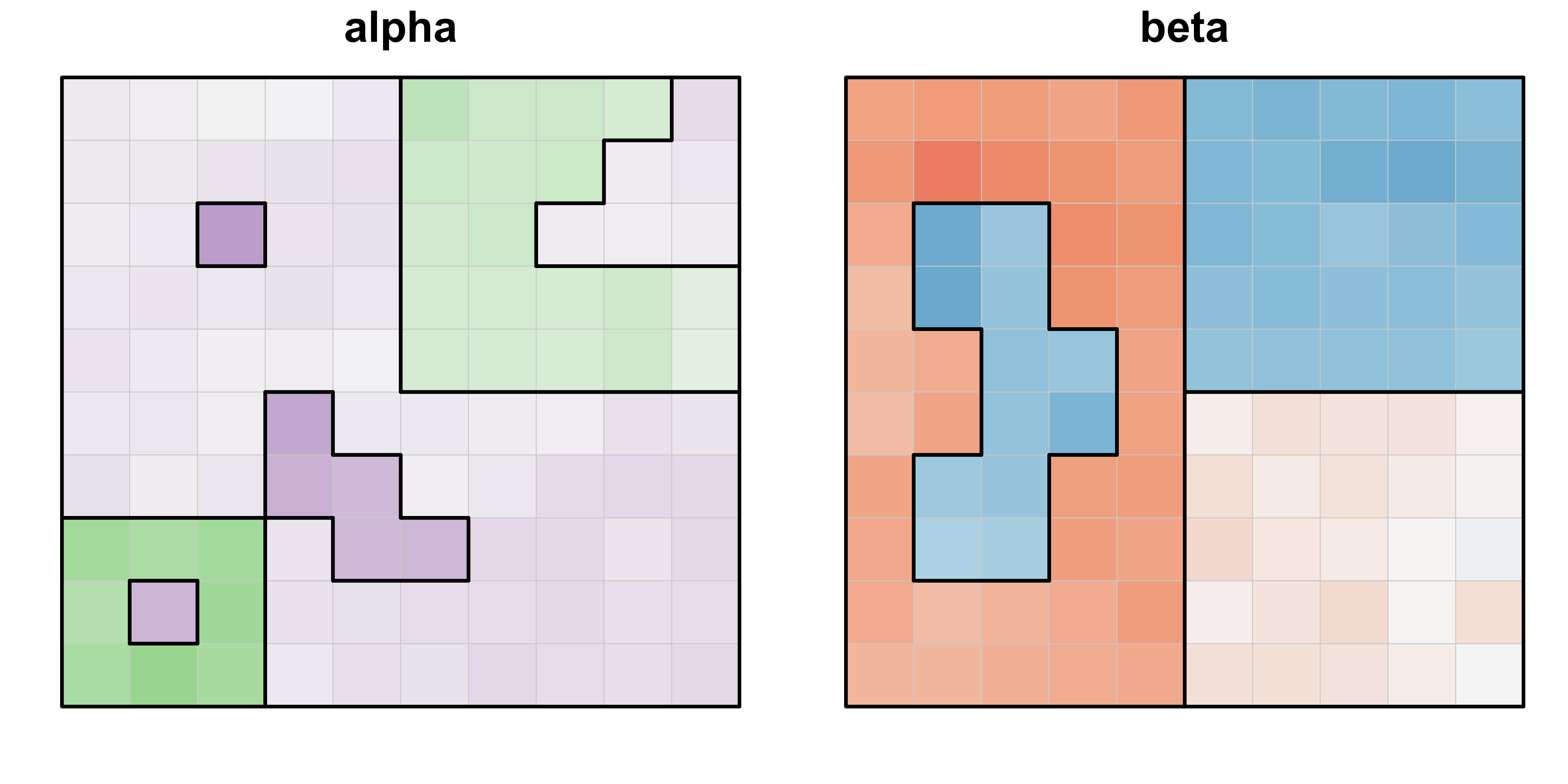}\\
\includegraphics[width = 0.5\textwidth]{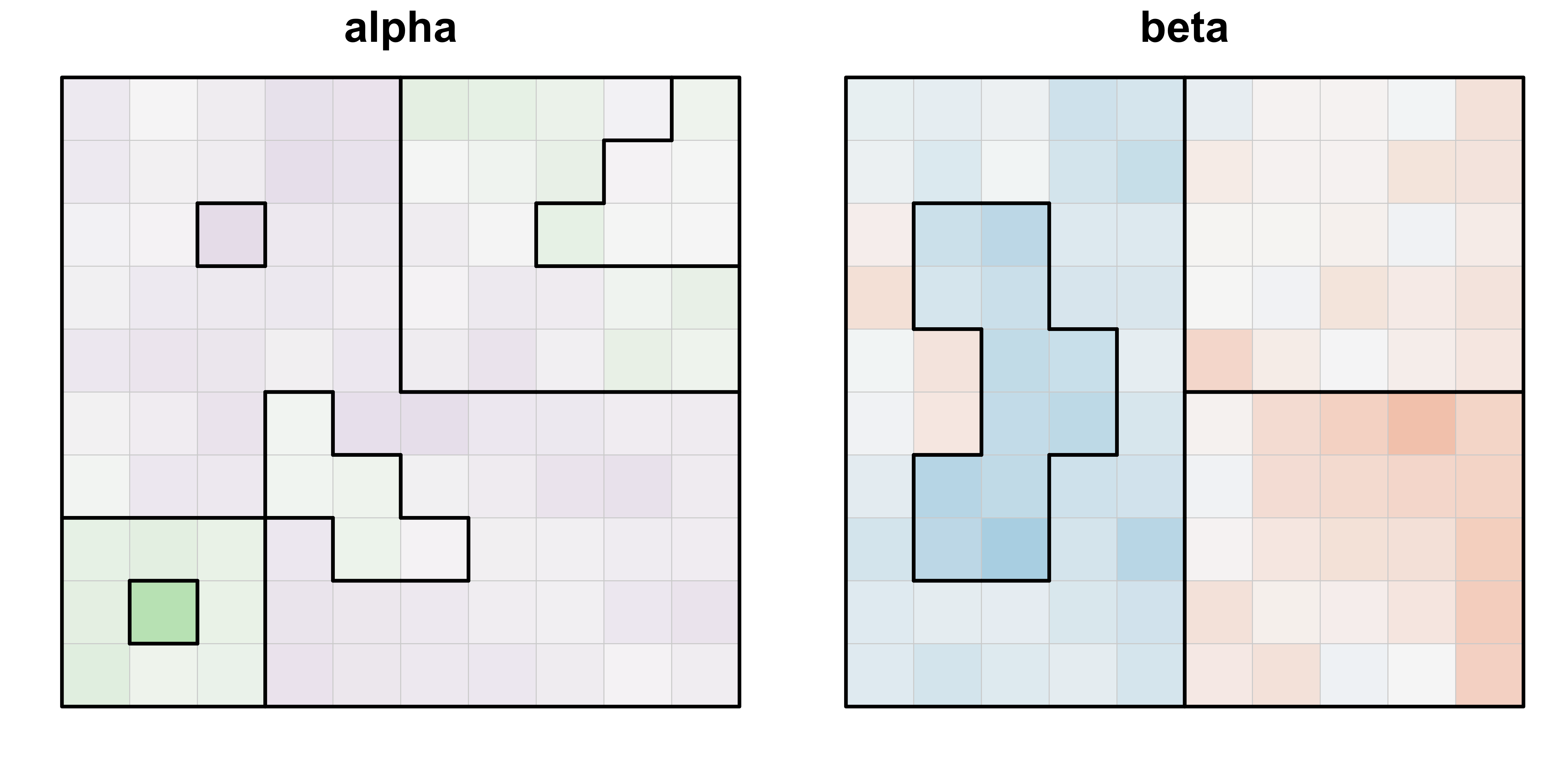}
\caption{The true partition $\gamma^{(\alpha)}$ and $\gamma^{(\beta)}$ used to generate the synthetic data to test the extension of Particle Optimization to non-conjugate models. First row: high cluster separation configuration. Second row: moderate cluster separation configuration. Third row: low cluster separation configuration.}
\label{fig:sim_laplace_true}
\end{figure}

To understand the performance of our approximate \texttt{PartOpt}, we performed a simulation study very similar to that described in Section~\ref{sec:sim:descr}. Specifically, we generated observations $c_{it} \sim \textrm{Pois}(\exp\{\alpha_i + \beta_i x_t\})$ where the $\alpha_i$'s and $\beta_i$'s were drawn from a CAR-within-cluster distribution. We considered three settings of the $\alpha_i$'s and $\beta_i$'s, one where the grand cluster means were highly separated (first row of Figure~\ref{fig:sim_laplace_true}), one where the grand cluster means were moderately separated (second row of Figure~\ref{fig:sim_laplace_true}), and one where the grand cluster means were very close in value (third row of Figure~\ref{fig:sim_laplace_true}).
We ran our approximate \texttt{PartOpt} with $L = 10$ particles and penalties $\lambda = 1, 10,$ and $100.$

\begin{figure}[H]
\centering
\includegraphics[width = 0.75\textwidth]{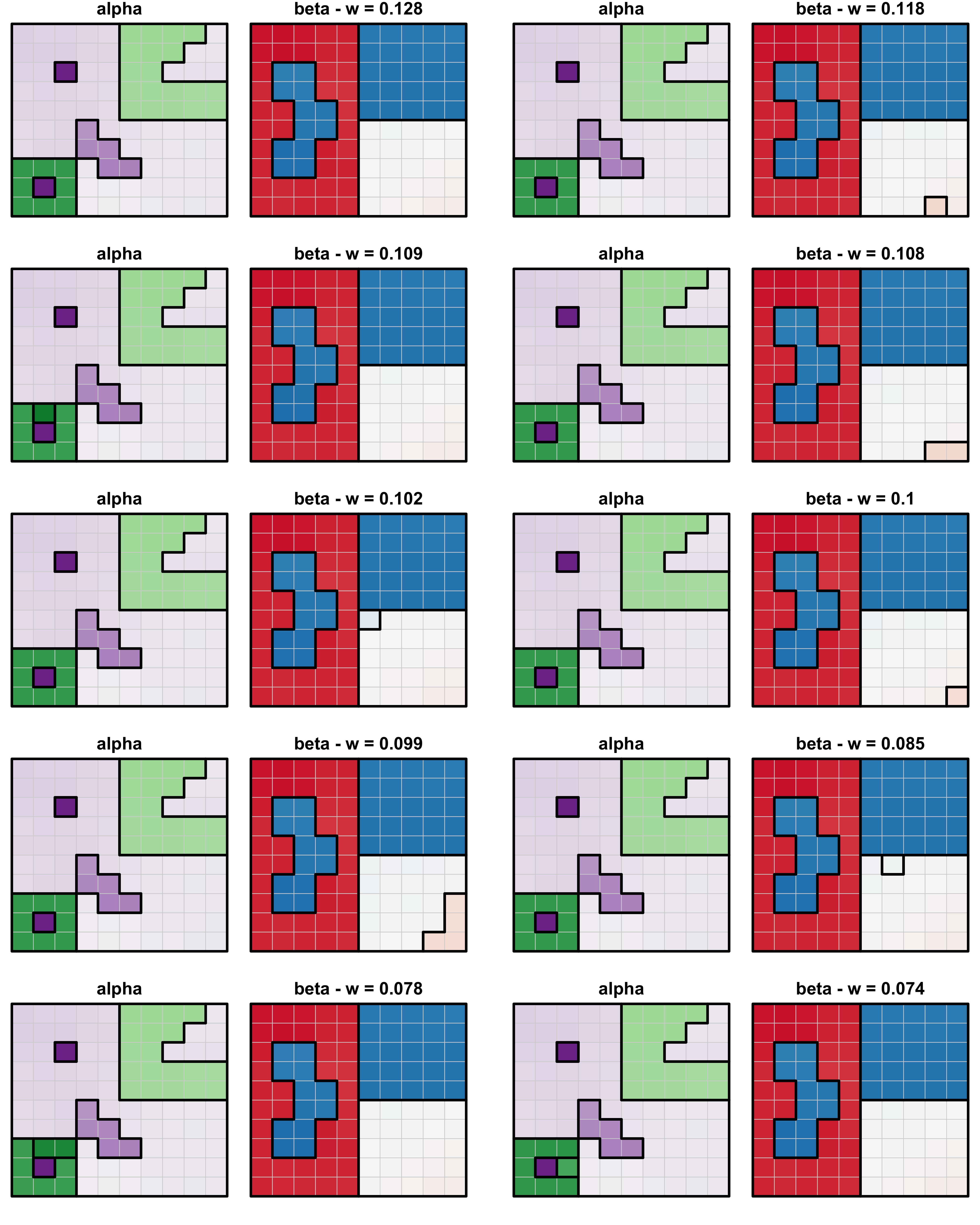}
\caption{The estimate particles $(\gamma^{(\alpha)},\gamma^{(\beta)})$ recovered for $\lambda = 10$ in the high separation setting, and the weight associated to each one.}
\label{fig:sim_laplace_lam10_hs}
\end{figure}

\begin{figure}[H]
\centering
\includegraphics[width = 0.75\textwidth]{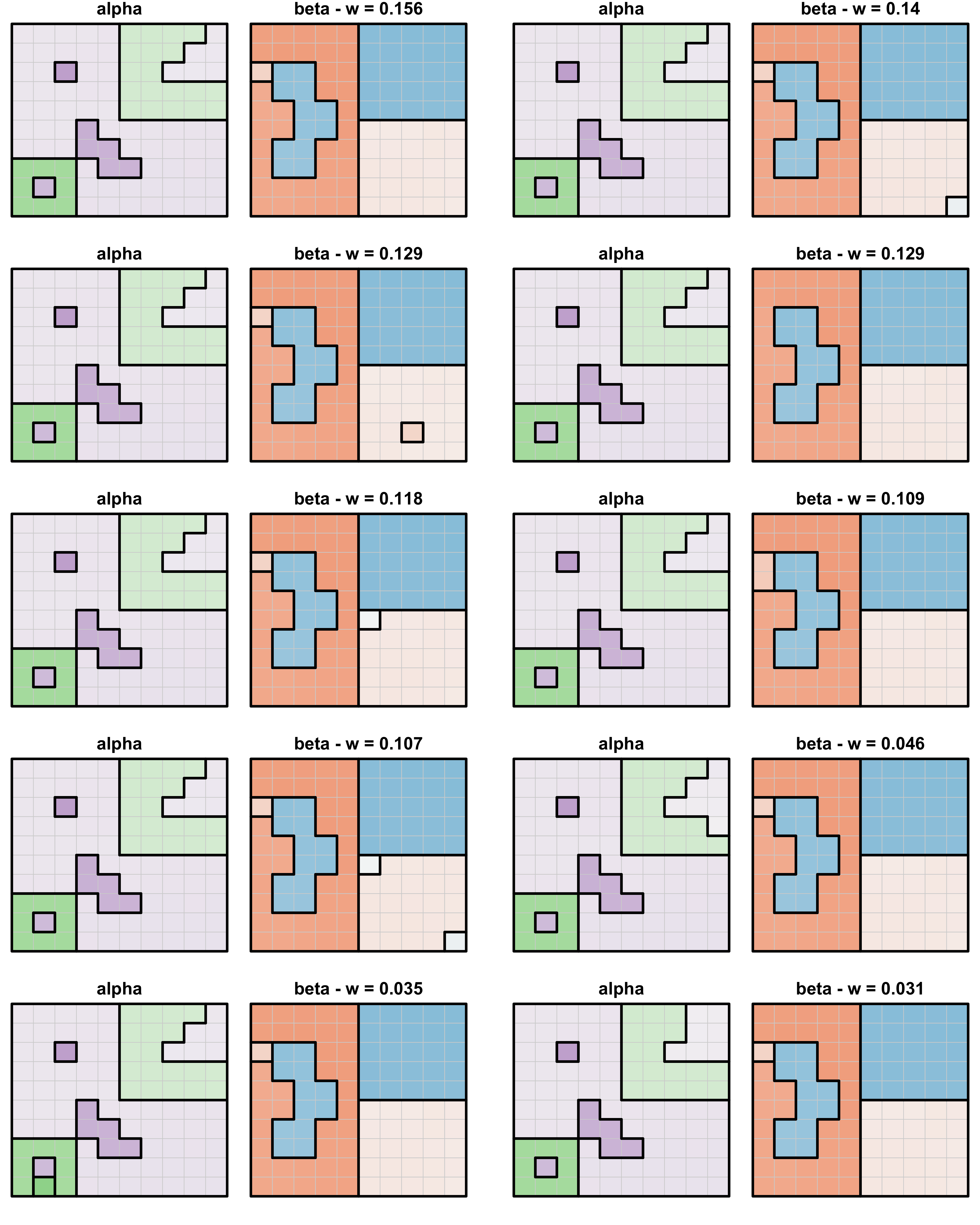}
\caption{The estimate particles $(\gamma^{(\alpha)},\gamma^{(\beta)})$ recovered for $\lambda = 10$ in the moderate separation setting, and the weight associated to each one.}
\label{fig:sim_laplace_lam10_ms}
\end{figure}

\begin{figure}[H]
\centering
\includegraphics[width = 0.75\textwidth]{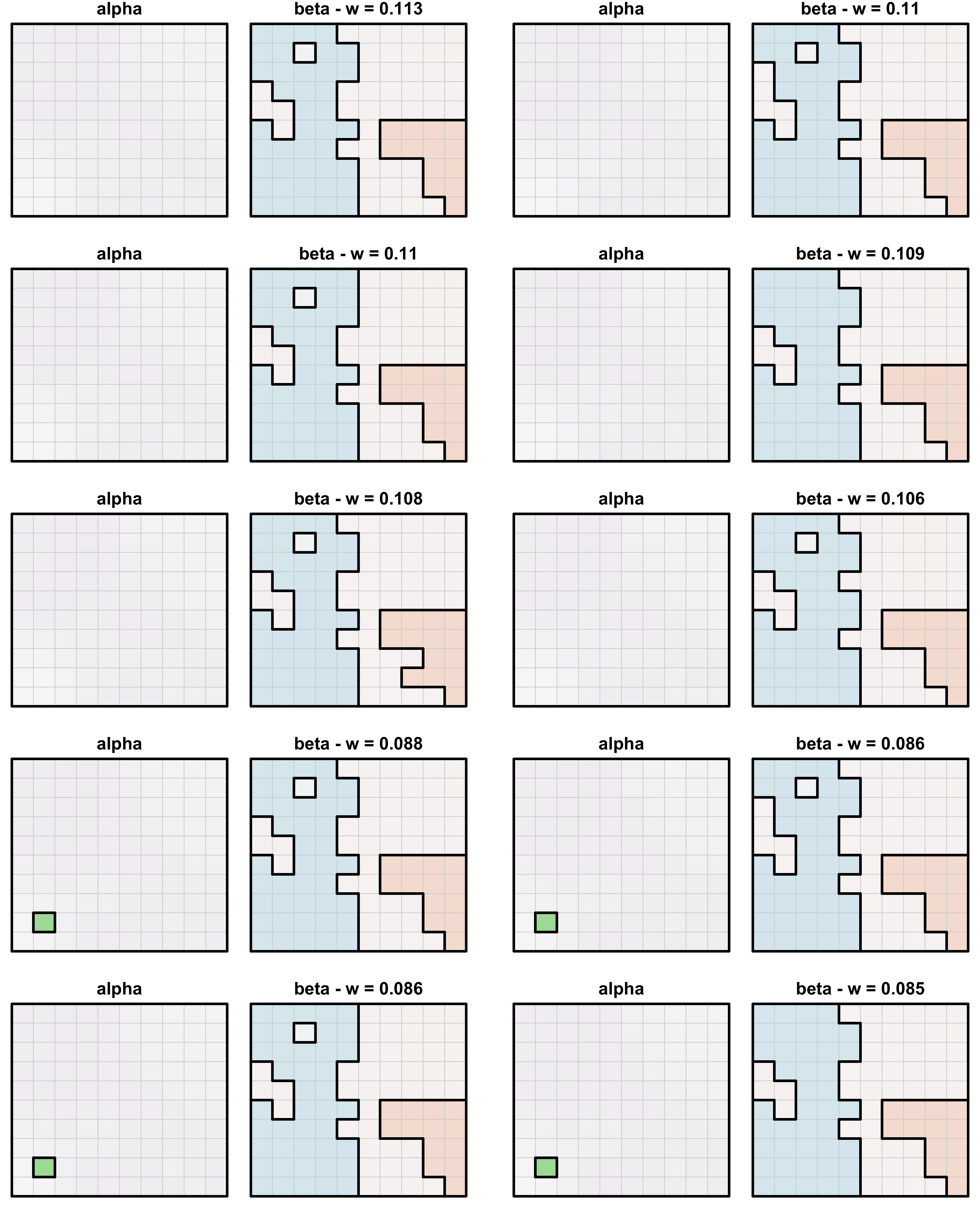}
\caption{The estimate particles $(\gamma^{(\alpha)},\gamma^{(\beta)})$ recovered for $\lambda = 10$ in the low separation setting, and the weight associated to each one.}
\label{fig:sim_laplace_lam10_ls}
\end{figure}

For all values of $\lambda$, the particle set contained the true partitions $\gamma^{(\alpha)}$ and $\gamma^{(\beta)}$ shown in Figure~\ref{fig:sim_laplace_true}. 
Like the simulations reported in the main manuscript for the Gaussian regression setting, we found that when $\lambda = 1$, many particles collapsed to the same point. 
However, we recovered a much more diverse particle set when we set $\lambda = 10$ and $\lambda = 100.$
Figure~\ref{fig:sim_laplace_lam10_hs} shows the $L = 10$ particles recovered when we ran the approximate \texttt{PartOpt} on data generated from~\eqref{eq:poisson_model} with the $\alpha_{i}$'s and $\beta_{i}$'s from the high separation setting, shown in the top panel Figure~\ref{fig:sim_laplace_true}.
Similarly, Figure~\ref{fig:sim_laplace_lam10_ms} shows the recovered particles for the moderate separation setting, when $\lambda = 10$. Finally, Figure~\ref{fig:sim_laplace_lam10_ls} shows the recovered particles for the low separation setting, when $\lambda = 10$.
In the high and moderate cluster separation settings, all of the estimated partitions are extremely close to the true partitions and the corresponding conditional posterior means of $\alpha_{i}$ and $\beta_{i}$ are quite close to the true values. 
In the low separation setting, the particle set recovered partitions quite different from the ones used to generate the data. 
The behavior of the approximate \texttt{PartOpt} in the low separation setting is not surprising: when there is little between-cluster variation in parameter values, the posterior strongly favors a single large cluster instead of several smaller clusters that all containing similar parameter values. 


\end{document}